\documentclass[aps,prd,superscriptaddress,showpacs,nofootinbib,eqsecnum]{revtex4-1}
\usepackage{epsfig}
\usepackage{natbib}
\usepackage{mciteplus}
\usepackage{color}

\usepackage{graphicx}

\graphicspath{{images/}}
\def\be{\begin{equation}}
\def\ee{\end{equation}}
\def\bea{\begin{eqnarray}}
\def\eea{\end{eqnarray}}
\def\tb{\tan\beta}

\def\t{\tilde }
\def\mhalf{m_{1/2}}
\def\mgrav{m_{3/2}}
\newcommand{\unit}[1]{\; \mbox{#1}}
\newcommand{\pfrac}[2]{\left( \frac{#1}{#2} \right)}
\newcommand{\lsim}{\raisebox{-0.13cm}{~\shortstack{$<$ \\[-0.07cm] $\sim$}}~}
\newcommand{\gsim}{\raisebox{-0.13cm}{~\shortstack{$>$ \\[-0.07cm] $\sim$}}~}

\begin{document}
\baselineskip .5cm

\title{ Revisiting No-Scale Supergravity Inspired Scenarios:\\ Updated Theoretical and  
Phenomenological Constraints\\}
\author{Amine Benhenni}
\author{Jean-Lo\"{\i}c Kneur}
\author{Gilbert Moultaka}
\affiliation{CNRS, Laboratoire Charles Coulomb UMR 5221, F-34095, Montpellier, France}
\affiliation{Universit\'e Montpellier 2, Laboratoire Charles Coulomb UMR 5221, F-34095, Montpellier, France}
\author{Sean Bailly}
\affiliation{Laboratoire de Physique Th\'eorique LAPTH, Universit\'e de Savoie, CNRS
(UMR 5108)\\
9 chemin de Bellevue, BP 110, F-74941 Annecy-Le-Vieux Cedex, France}
\begin{abstract}
We consider no-scale inspired supergravity  scenarios, where the gravitino mass and
related soft supersymmetry-breaking parameters are determined  dynamically by radiative corrections 
to an essentially flat tree-level potential in the supersymmetry breaking hidden sector.
We examine the theoretical and phenomenological viability of such a mechanism, when including up-to-date 
calculations of the low energy sparticle spectrum and taking into account the latest LHC results and other
experimental constraints. 
We (re)emphasize the role of the scale-dependent vacuum energy contribution to the 
effective potential, in obtaining realistic no-scale electroweak minima, examining carefully the impact 
of boundary conditions and of variants of the minimization procedure. 
We also discuss and implement the $B_0$ (soft breaking Higgs mixing parameter) 
input boundary condition at high scale, therefore fixing $\tan\beta(B_0)$
at low scales. For general high scale boundary conditions with $B_0, m_0,\cdots \ne0$, 
our analysis provides theoretical correlations among the  supersymmetric, soft and vacuum energy parameters
and related phenomenological consequences at the LHC. For instance, a zero vacuum energy at the GUT scale 
would lead to a decoupled supersymmetric spectrum, together with a light standard model-like Higgs boson
at the electroweak scale.
Given the experimental exclusion limits, a substantial class of the boundary conditions, and in particular 
the strict no-scale with $m_0=A_0=B_0=0$, are only compatible 
with a stau being the lightest MSSM particle. Then an enlarged allowed parameter space emerges
when assuming a gravitino LSP to account for the observed dark matter relic density.     

\end{abstract}


\maketitle



\section{Introduction}

Supersymmetry (SUSY) has imposed itself as the most popular `beyond the standard model" scenario for
many good reasons. In addition to appealing extended symmetry principles, it has the potential to  
solve some of the problems raised by the standard model, 
even though it was not originally introduced for this purpose. 
It solves the hierarchy problem by protecting the scalar sector from 
unnaturally large radiative corrections, provided that the superpartners lie in the TeV
range~\cite{Witten:1981nf,Sakai:1981gr,Dimopoulos:1981zb,Kaul:1981hi}.
It also predicts the gauge coupling 
unification~\cite{Ellis:1990wk,Amaldi:1991cn,Langacker:1991an,Giunti:1991ta} at a high scale consistent with 
experimental constraints, and provides 
very plausible particle candidates for the dark matter~\cite{Goldberg:1983nd,Ellis:1983ew,dmrev}.  
Last but not least, the very structure of the Minimal Supersymmetric Standard Model (MSSM) (and
of many non-minimal extensions) leads generically to the
radiative electroweak symmetry breaking (REWSB) 
mechanism~\cite{Ibanez:1982fr,Ibanez:1982ee,Ellis:1982wr,AlvarezGaume:1983gj}. 

The remaining open questions concern mainly the precise mechanism underlying the supersymmetry breaking 
itself. 
Most present viable scenarios assume that a dynamical or
spontaneous symmetry breaking occurs in a hidden sector. The supersymmetry breaking 
is then transmitted to the visible low energy sector via different mechanisms depending on the 
models. One of the most popular such scenario is when supersymmetry breaking is transmitted 
essentially via the gravitational interaction, in the gravity-mediated models. 
In an unbroken supergravity (SUGRA) 
model~\cite{Chamseddine:1982jx,Barbieri:1982eh,Hall:1983iz,Cremmer:1982vy,Ohta:1982wn}, the graviton and its superpartner, 
the gravitino, both have a vanishing mass. Once supersymmetry is broken, 
only the gravitino gets a mass via the super-Higgs mechanism. 
Therefore the breaking of local supersymmetry is directly linked to the non-vanishing gravitino mass. 
In a standard SUGRA scenario with a canonical K\"ahler 
potential, when SUSY breaking is communicated gravitationally to the visible sector 
the soft parameters are roughly of the same order as $\mgrav \sim M_{SUSY}^2 / M_P$, itself expected to be of
order the electroweak (EW) scale.  
In this way one ends up with the correct hierarchy between the Planck scale $M_P$, the SUSY breaking scale 
$M_{SUSY}$ and the EW breaking scale  $M_{EW}$, 
although the requirement of the vanishing tree-level potential is somewhat ad hoc. \\
In the no-scale models, the basic idea is that the vanishing of the tree-level potential in the
hidden sector direction can be automatic for an appropriately chosen form of the K\"ahler potential.
 Moreover,  the value of    
$\mgrav$ can be fixed dynamically by radiative correction stabilization, and is simply related 
to other soft SUSY-breaking parameters.
This no-scale approach has emerged quite early~\cite{Cremmer:1983bf,*Ellis:1983sf,*Ellis:1983ei,noscalerev}, and since 
then it has been regularly claimed to be ruled out and resurrected in different forms several times. 
However, in most of nowadays phenomenological studies, `no-scale' 
is often a name for just the
specific and very restricted boundary  conditions on the SUGRA parameters, namely $m_0 = A_0 = 0$
where $m_0$ and $A_0$ are respectively the universal GUT scale values of the scalar mass
and trilinear soft SUSY-breaking parameters, or 
$B_0= m_0 = A_0 = 0$ in the strict no-scale model ($B_0$ being the soft SUSY-breaking Higgs mass-mixing parameter).
Although well-known to the no-scale model aficionados, it is worth emphasizing here 
that an essential feature of the original no-scale program is the possible dynamical determination of the
gravitino mass and other related soft SUSY-breaking parameters, that may be realized
with the above boundary conditions but also possibly with more general SUGRA ones. 
More precisely the basic framework \cite{Cremmer:1983bf,*Ellis:1983sf,*Ellis:1983ei} is to first assume a specific 
K\"ahler potential such that there exists a flat direction (moduli) at tree-level, thus also ensuring
automatically a (tree-level) vanishing cosmological constant. The SUSY-breaking order parameter  is the
gravitino mass but  is not determined at the tree-level, i.e. the gravitino mass is `sliding'. 
Then the flatness of the potential can be lifted by (non-gravitational) radiative corrections originating
from the strong and electroweak sectors at the electroweak (EW) symmetry breaking scale.  These corrections may trigger, 
under appropriate circumstances, a non-trivial minimum of the potential as a function of the gravitino mass, thus 
fixing the latter. In principle, this picture does not forbid having weak quantum corrections to the vacuum energy, 
as long as those are of order $\sim m^4_{3/2}$ \cite{vacplanck}. 
So actually one has rather an `almost flat' moduli direction of the K\"ahler potential. 
Overall the mechanism is somewhat similar to 
the REWSB mechanism, but provides an even more direct (and calculable)
link between the EW and SUSY-breaking scales. In this way the no-scale scenario
relates those two scales more dynamically, explaining naturally the hierarchy $m_{SUSY} \ll M_P$. 
The vacuum thus corresponds to a minimum of the potential with respect to the two Higgs fields and a hidden sector 
field $z$, whose vacuum expectation values (vev) determine respectively the weak scale masses and the gravitino mass
$\mgrav$. Thus, the occurrence of a minimum in the $z$ direction is a consequence of the loop-improved effective 
potential at the EW scale where the tree-level flatness is lifted.    
For the whole picture to work, one must furthermore
assume that there are no stronger, purely gravitational 
(quantum) corrections near the Planck scale $M_P$, with dangerous vacuum energy contributions of the 
form $\Lambda^2 Tr M^2$, where $\Lambda \sim {\cal O}(M_P)$ is an appropriate cutoff  
beyond which quantum gravitational effects are non negligible, and $M$ generically the relevant masses of the high 
scale hidden sector. In fact specific superstring models
compatible with the no-scale boundary conditions, are known to avoid this problem \cite{vacplanck}. 
On more phenomenological grounds one may always assume that those issues will ultimately be solved
by a fully consistent superstring framework, an assumption no more (nor less) problematic than the standard
{\em minimal} SUGRA picture with (assumed) universal soft parameters at high scale. 

On the one hand, most analyses in the past aiming to determine possible no-scale $\mgrav$ minima
were conducted with definite approximations in the effective potential and sparticle spectrum calculations.  
Typically, those studies mostly used (one-loop) RGE analytical solutions restricted to low $\tan\beta$ values, 
and also typically neglecting the non-dominant couplings, non-RG radiative corrections, 
and other non-dominant terms in the one-loop effective potential, etc, 
with the legitimate aim of determining (semi)-analytical solutions.  
On the other hand, the situation since those early
days of no-scale models has drastically changed concerning the elaboration level of (s)particle 
spectra calculation, so that such approximations are hardly considered satisfactory nowadays.
There have been of course  
numerous more recent studies of the viability of the mSUGRA subspace defined 
by the specific no-scale initial conditions with more 
elaborate particle spectra calculations, and updated phenomenological constraints. However, most of  
those more recent studies are generally not addressing the existence of $V_{eff}(\mgrav)$ minima.  
Actually, to the best of our knowledge a systematic study of the occurrence of non-trivial
no-scale $\mgrav$  minima of a well-defined (RG improved) effective potential $V_{full}(v_u,v_d; m_{3/2})$,
taking into account its full one-loop radiative corrections, 
has not been done (though a number of special cases, other models, or partial studies of those aspects 
have been examined in the past or 
recently \cite{Dutta:2007xr, *Maxin:2008kp, *Ellis:2010jb, *Li:2010mi, *Li:2011xu},~\cite{Li:2010ws,*Li:2011dw}). 
In addition, in spite of the above mentioned superstring
motivations and different appealing scenarios, in the following we will
essentially consider a more phenomenological approach. 
More precisely we shall consider 
the standard mSUGRA parameter models, with more specific boundary conditions like in the strict 
no-scale models, but also more general ones, for which we examine 
the conditions for the emergence of non-trivial $\mgrav$ minima at the EW scale. This is 
thus a `no-scale inspired' but more `bottom-up' framework, with the goal of determining what kind of
high scale parameter relations can emerge from our analysis. 

The paper is organized as follows: section 2 gives a short reminder of the basic no-scale 
supergravity scenario. In section 3 we specify our procedure for minimizing the loop-improved effective 
potential, examining some salient features to take into account. We emphasize in particular the role
played in the $V_{eff}(\mgrav)$ minimization 
by the necessary scale-dependent vacuum energy contribution to the effective potential. 
This point was indeed raised earlier~\cite{Kounnas:1994fr,Kelley:1994ab,nsvac2}, 
and in fact our practical procedure is closely related to the latter work. (However    
at the time those analyses used semi-analytical approximations essentially 
similar to the ones mentioned above, while we will 
perform more complete numerical studies based on the available present SUSY spectrum calculators).  
In section 4 we discuss in some details important generic properties and results for the minimization 
of the loop-improved effective potential with respect to the extra soft-breaking parameters.  
We emphasize also the differences in choice of $\tan\beta$ or $B_0$ as input parameter, the latter
being the consistent choice in no-scale scenarios, which has non-trivial
technical as well as phenomenological consequences. Section 5 examines 
a few different representative parameter cases, either 
in the strict no-scale models or its generalizations, and following
well-defined prescriptions for a loop improved effective potential.
The main phenomenological constraints 
affecting the viability of the theoretical results when confronted with collider and other 
experimental limits are illustrated. We also explore the constraints given
by the dark matter relic density, either in the standard neutralino LSP scenario, or 
considering alternatively that the LSP is the gravitino, which can be a priori assumed  
in a generalized no-scale scenario.
Finally we give some conclusions in section 6.

\section{basics of no-scale supergravity} \label{basics}
For completeness, we review very sketchily in this section the essential features
of the original no-scale models, referring for more details to the pioneering literature~\cite{Cremmer:1983bf,
*Ellis:1983sf,*Ellis:1983ei}, \cite{Ellis:1984kd,*Ellis:1984bs},~\cite{noscalerev}. 
The $N=1$ supergravity Lagrangian is fully determined by the K\"ahler potential and superpotential. 
For simplicity, we focus here only  on the matter chiral superfield dependence (referring only implicitly 
to the gauge vector superfields and gravitino supermultiplet sectors).   
The gauge kinetic function $f_{a b}(\phi_i)$, the K\"ahler potential $K(\phi_i, \bar \phi_i)$ and the superpotential 
$W(\phi_i)$ are 
specified in terms of the chiral superfields $\phi_i$ and their complex conjugate $\bar \phi_i$. Here $\phi_i$
denotes generically all visible and hidden sector superfields, possibly transforming under some gauge groups. 
In terms of the K\"ahler function 
\begin{eqnarray}
G(\phi_i,\bar \phi_i)= {K(\phi_i,\bar \phi_i) \over m_{p}^2} + \ln{\frac{|W(\phi_i)|}{m_{p}^6}^2}
\end{eqnarray}
where $m_{p}$ is the reduced Planck scale $m_{p} \equiv {M_{P} / \sqrt{8 \pi}}$,
one obtains the $F$-term part of the scalar potential
\begin{eqnarray}\label{Vf}
V_F=m_{p}^4 \; e^G \left( G_i G^{i \overline{j}} G_{\overline{j}}-3 \right)
\end{eqnarray}
where 
\begin{eqnarray}
G_i \equiv \frac{\partial G}{\partial \phi_i},~ ~ G_{\overline{i}} \equiv \frac{\partial G}{\partial \overline{\phi}_i},
~ ~ G_{i \overline{j}} \equiv \frac{\partial^2 G}{\partial \phi_i 
\partial \overline{\phi}_j }, ~ ~ G^{i \overline{j}} \equiv (G^{-1})_{i \overline{j}} \;.  
\end{eqnarray}
Due to local supersymmetry, the vacuum energy deduced from (\ref{Vf}) is in general non vanishing even before 
SUSY breaking. After SUSY breaking the gravitino acquires a mass $\mgrav$ given by
 \begin{equation}
\mgrav =  m_{p}  \; e^{\langle G \rangle/2} 
\label{m32}
\end{equation}
where $\langle G \rangle$ is a function of the  vevs of the scalar components of a sub-class  
$\phi_k$ of the chiral superfields responsible for the SUSY-breaking, whatever the underlying breaking mechanism may be. 
The corresponding non-vanishing F-term vevs $F_k = m_p^3 \langle  e^{G/2} G_k \rangle$ yield the SUSY-breaking 
mass scale
\begin{equation}
M_{SUSY} = (F_k G^{k \overline{j}} F_{\overline{j}})^{1/4}
\label{Msusy}
\end{equation}
Fine-tuning $\langle V_F \rangle$ to $0$ to keep the 
(tree level) cosmological constant around its observed value   
fixes $\langle G_k G^k \rangle$ uniquely, leading to the well-known relation
\begin{eqnarray}
\mgrav & = & { M_{SUSY}^2 \over \sqrt{3} m_{p} }\;.
\end{eqnarray}
If SUSY breaking is communicated gravitationally to the visible sector, one expects {\sl generically} the soft
parameters $m_{soft}$ to satisfy 
\begin{eqnarray}
 m_{soft} & = & {\cal O}(1) \times \mgrav  
\end{eqnarray}
and $\mgrav$ not far from the electroweak scale. Moreover, the generic magnitude 
$\langle V_F \rangle \sim {\cal O}(\mgrav^2 m_p^2)$ (see (\ref{Vf}) and (\ref{m32}))
exacerbates the fine-tuning of the vacuum energy to zero for $\mgrav$ of order the EW scale.
No-scale supergravity~\cite{Cremmer:1983bf,*Ellis:1983sf,*Ellis:1983ei},
~\cite{Ellis:1984bm},~\cite{noscalerev} was introduced to ensure naturally, 
through a suitable choice of the K\"ahler potential, a vanishing 
potential at the tree-level in the hidden sector scalar fields directions at every value of these
fields. 
Since on the one hand the ${\cal O}(\mgrav^2 m_p^2)$ magnitude in $V_F$ is now flattened, and on the other the 
vevs of the hidden sector fields, and thus  $\mgrav$, are undetermined
at the tree-level, $\mgrav$ will be fixed at the loop level through (generalized) REWSB 
which takes place in the observable sector, with $\mgrav \sim {\cal O}(m_Z)$ as a natural outcome. This holds only if 
a large mass scale $M$ that can be present in the observable sector, such as a GUT scale, does not contribute to the 
potential by quantities of ${\cal O}(\mgrav^2 M^2)$.    

Assuming for simplicity that the hidden sector contains only one chiral superfield $z$ (say $\phi_1$ of the 
above set of $\phi_i$'s), the simplest K\"ahler potential realizing the above potential
flatness, entailing a (non-compact) $SU(1,1)$ symmetry, is given by 
\begin{equation}
K=-3 \ln (z+\bar z) \; .
\label{simplens}
\end{equation}
This K\"ahler potential has to be supplemented in realistic models by other K\"ahler potential and superpotential
parts depending on the visible sector fields. A generalization of the corresponding K\"ahler
function $G(z, \phi_k)$
\begin{equation}
G = -\frac{3}{m_{p}^2} \ln (z+\bar z - f(\phi_k, \bar \phi_k)) + \ln{\frac{|W(\phi_k)|}{m_{p}^6}^2}
\end{equation} 
where the $\phi_k$'s (with $k \ge 2$) are all in the visible sector, was found to have very nice properties,
either {\sl i)}  when $f$ is an arbitrary function but $W$ trivial 
(e.g. $W= m_{p}^3$), or {\sl ii)} when $W$ is an arbitrary superpotential (e.g. that of the MSSM or of an extended GUT 
model) but $f$ taking the special form $\displaystyle f(\phi_k, \bar \phi_k) = \sum_{k \ge 2} |\phi_k|^2$. 
In case {\sl ii)} the $SU(1,1)$ symmetry of (\ref{simplens}) is extended to $SU(n,1)$ where $n-1$ 
is the number of fields in the observable sector.
A key point for both {\sl i)} and {\sl ii)} cases is that SUSY-breaking in the hidden sector leaves the visible 
chiral superfield sector supersymmetric.
This is welcome particularly in case {\sl ii)} where a full GUT sector can be accommodated, including the MSSM
as the low energy effective theory, since the non-transmission of SUSY-breaking will protect the visible sector from 
the large ${\cal O}(\mgrav^2 M_{GUT}^2)$ effects mentioned previously. However, for the same reason contributions 
${\cal O}(\mgrav^2 m_W^2)$ will not be present either, thus preventing the usual radiative EW symmetry breaking
and the ensuing dynamical determination of $\mgrav$. All soft breaking squark, slepton and Higgs
masses and couplings are thus vanishing at all scales, and in particular the universal parameters at the GUT scale,

\begin{eqnarray}
m_0 = A_0 = B_0 =0
\label{strict}
\end{eqnarray}
The only source left for SUSY-breaking is in the gauge/gaugino sector through the inclusion of  a $z$-field dependent 
non-canonical gauge kinetic functions $f_{a b}(z,...)$ 
which are essentially free in a general supergravity 
framework~\cite{Chamseddine:1982jx,Barbieri:1982eh,Hall:1983iz,Cremmer:1982vy,Ohta:1982wn}. 
The ensuing soft breaking gaugino mass terms take then the form 
$\frac{1}{4} \mgrav \langle G_{\bar z} / G_{\bar z \bar z} \partial f_{a b}/\partial z \rangle$. One should still
assume that $f_{a b}$ is chosen such that the heavy GUT gaugino soft masses remain vanishing so that again 
large unwanted contributions ${\cal O}(\mgrav^2 M_{GUT}^2)$ are not present. The remaining MSSM gaugino masses are
proportional to $\mgrav$ as just noted. Within the universal gaugino mass assumption 
we rewrite this relation for later phenomenological use at the GUT scale as 
\be
m_{3/2} =  c_{3/2} m_{1/2}   \label{mgravmhalf}
\ee

Equations (\ref{strict}, \ref{mgravmhalf}) define the boundary conditions of the strict no-scale scenario.
It is now possible to imagine variants to these boundary conditions. For instance supplementing case {\sl i)}
with a superpotential in the visible sector implies a non vanishing $A_0$. Another variant is to consider
$B_0 \neq 0$ keeping the boundary conditions  
\begin{equation}
m_0 = A_0 = 0
\label{nsinit}
\end{equation}

Analogous relations arise naturally in low-energy effective models for some specific string theories~\cite{vacplanck,
Brignole:1993dj,Munoz:1995bi}, 
leading to  generalized boundary conditions depending on the type of the string and the compactification mechanism.
For instance, in the dilaton-dominated SUSY breaking scenario, the conditions become
\begin{eqnarray}
m_0={1 \over \sqrt{3}} \mhalf \ , \ A_0=- \mhalf \ , \
B_0= {2 \over \sqrt{3}} \mhalf \ , \ m_{1/2} = \sqrt{3} m_{3/2}
\label{stringbc}
\end{eqnarray}
with thus non-zero values of $m_0$, $B_0$ and other high-scale parameters, but all related to the
unique SUSY-breaking scale, $\sim m_{1/2}$. In contrast, 
note that the strict no-scale relations in Eq.(\ref{strict}) 
are equivalent to the so-called moduli-dominated SUSY breaking superstring model. 
It is now possible to study the electroweak potential and to predict values for $\mhalf$ instead
 of $\mgrav$ by an extra minimization in addition to the ordinary EW minimization driven by
the REWSB mechanism. Using a simplified analysis with approximations 
allowing analytical handling of expressions~\cite{Cremmer:1983bf,*Ellis:1983sf,*Ellis:1983ei},~\cite{noscalerev}, 
the preferred values appeared to be of order 
\begin{eqnarray}
\mhalf \sim {\cal O}(m_{3/2}) \sim {\cal O}( m_Z)\;.
\end{eqnarray}
As mentioned before, strict no-scale supergravity is characterised by the specific boundary 
conditions at GUT scale given by Eq.(\ref{strict}). 
We are left with only $\mhalf$ as a free parameter, parameterizing the supersymmetry breaking, 
and driving the other parameters through renormalization group evolution (RGE) effects. 

In the following we consider 
a more phenomenological approach, essentially string model-independent,
motivated by the fact that the ultimate superstring framework, and even more how it is linked 
to the GUT scale is not yet fully established.   
We will thus assume the most general standard mSUGRA high scale parameters and boundary conditions, 
but study also the  special cases of  no-scale (\ref{nsinit}) and strict no-scale models (\ref{strict}).  
This more phenomenological approach aims to  
concentrate more on the conditions for the emergence of a third non-trivial minimum of $V_{eff}$
 with respect to $\mhalf$ at the EW scale, without
too strong prejudice on the high scale models. We will assume the following generic form of the boundary conditions
\be
B_0= b_0 m_{1/2},\;\; m_0= x_0 m_{1/2},\;\; A_0=a_0 m_{1/2}
\label{softbc}
\ee 
where $b_0$, $a_0$, $x_0$ are mass independent constants taken as input parameters. A further input information about 
the supersymmetric $\mu$-parameter is also needed, even though
the value of $\mu$ at the electroweak scale will be as usual eventually fixed (up to a sign) by the REWSB conditions.  
Indeed since $m_{1/2}$ will be determined dynamically from the potential, it is important to know beforehand
whether $\mu$ has a functional dependence on $m_{1/2}$, for instance through
its boundary value $\mu_0$ at the high scale. For reasons which will become clear in the sequel, we will adopt 
throughout the paper the boundary condition
\be
\mu_0 =c^0_\mu m_{1/2} 
\label{muscale}
\ee
with $c^0_\mu$ an independent constant. Although $\mu$ is a supersymmetric parameter, such an assumption is 
well-motivated as there are various mechanisms where it can be related to the SUSY breaking 
order parameter $m_{3/2}$ within supergravity. Typically, this can occur through a non-minimal term in the 
K\"ahler potential involving the two Higgs superfields and some gauge singlet superfield whose F-term triggers
supersymmetry ~\cite{Giudice:1988yz, Soni:1983rm}, or alternatively through a minimal term in the K\"ahler potential
and the addition of an R-symmetry breaking constant in the superpotential~\cite{Bagger:1994hh},
(see also for instance \cite{Munoz:1995yp} for an early review of other possible mechanisms including the superstring 
induced ones).

We close this section by stressing that the usual free parameters of mSUGRA ($m_0, m_{1/2}, A_0, \tan \beta,
{\rm sign}(\mu)$) have been traded here for $b_0$, $a_0$, $x_0$ and (sign of) $c^0_\mu$, thus with one less free 
parameter, $m_{1/2}$, to be determined dynamically, hence a more constrained scenario. Furthermore, in the spirit of 
no-scale, the dimensionless parameters  $b_0$, $a_0$, $x_0$ and $c^0_\mu$ are expected to be of ${\cal O}(1)$,
or else strictly vanishing. These features are useful criteria distinguishing the no-scale scenario from the less 
constrained mSUGRA, even if both can lead to similar low energy MSSM spectra.

\section{Present situation vs early no-scale model analysis}\label{present}
In this section we first review some rather generic and important 
features of the no-scale mechanism, by which a dynamical determination of the soft parameters is provided
via the extra minimization of the effective potential.  
Consider the familiar MSSM effective potential, for the moment just at tree-level for simplicity, which reads
\be
V_{tree} = m^2_1 |H_u|^2 + m^2_2 |H_d|^2 -B \mu H_u.H_d +\frac{g^2+g^{'2}}{8} (|H_u|^2 -|H_d|^2)^2
\label{Vtree}
\ee
in the relevant electrically neutral Higgs field directions
\begin{equation}
H^0_u =
\left(\begin{array}{c}
0 \\
h_u \\
\end{array} \right)  \qquad 
H^0_d =
\left(\begin{array}{c}
h_d \\
0 \\
\end{array} \right)
\end{equation}
where $m^2_1\equiv m^2_{H_u} +\mu^2$, $m^2_2\equiv m^2_{H_d} +\mu^2$, $g, g'$ denote respectively the $SU(2)_L$ and
$U(1)_Y$ gauge couplings and the `.' denotes the $SU(2)$ scalar product.  
Once the EW symmetry breaking mechanism occurs, the Higgs fields develop
non-vanishing vacuum expectation values $\langle h_u\rangle = v_u$, $\langle h_d\rangle = v_d$, and 
the EW extremum is characterized by
\be
\frac{\partial V}{\partial H_u|_i} |_{H_u= \langle H^0_u\rangle, H_d= \langle H^0_d\rangle} = 0\;,\:\;\;  
\frac{\partial V}{\partial H_d|_i} |_{H_u= \langle H^0_u\rangle, H_d= \langle H^0_d\rangle}=0\;
\label{ewsb0}
\ee
where $i=1,...4$ runs over the four field components of the $H_u$ and $H_d$ doublets, of which only the two conditions
\be
\frac{\partial V}{\partial h_u} = 0\;,\:\;\;  \frac{\partial V}{\partial h_d} =0\;,
\label{ewsb}
\ee
are not trivially satisfied, and allow to determine $v_d$ and $v_u$ in terms of $m_1^2, m_2^2, B \mu$ and $g^2 + g'^2$.
Since the gauge invariant point $v_d= v_u=0$ is also a solution of (\ref{ewsb}), one should require the 
consistency condition
\be
m^2_1 m^2_2 \le (B\mu)^2\;
\label{ewsbc}
\ee
to assure that this point is not a minimum so that the electroweak symmetry is indeed broken.
Note that one has to require as well 
\be
m^2_1 + m^2_2 \ge 2 |B\mu| 
\label{ewsbs}
\ee
to guarantee the tree-level stability of the potential.\footnote{There are connections between these two
consistency conditions. If (\ref{ewsbc}) is violated then (\ref{ewsbs}) is necessarily fulfilled. This implies
that the scale at which EWSB occurs is always higher than the scale at which the potential becomes unstable.
Furthermore, when (\ref{ewsbs}) is satisfied together with (\ref{ewsb}) then the EWSB extremum is guaranteed to be
a minimum and (\ref{ewsbc}) is automatically satisfied. These connections are typical of the MSSM Higgs sector 
potential and are not valid in a general two Higgs doublet model. 
They are also modified by loop corrections to the effective potential \cite{LeMouel:1997tk}.}
Of course one should further consider one-loop (and possible higher order) corrections to the effective potential and 
other related radiative corrections to the sparticle masses without which the simple tree-level 
analysis is not sufficiently reliable.  
Now essentially the main additional 
feature of no-scale models is to seek for an extra 
non-trivial minimum: 
\be
\frac{\partial V}{\partial m_{3/2}} = 0
\label{ns}
\ee
or with $m_{3/2}$ replaced by $m_{1/2}$ in the above described case of a unique SUSY-breaking scale that can be
conveniently parameterized in terms of $m_{1/2}$, as discussed in the previous section. 
The parameters in Eq.~(\ref{Vtree}) depend 
non-trivially on $m_{1/2}$ via the high scale boundary conditions, and follow RGE from
high scale values e.g. $m_1(GUT)$ down to $m_1(EW)$ values at the EW scale, where the 
minimization equations (\ref{ewsb}) and (\ref{ns}) are required. Before examining in 
more details the extra minimization (\ref{ns}), let us first examine some important aspects
in defining the actual expression for the effective potential to be minimized. 
As mentioned before, in the early days of no-scale model 
analyses~\cite{Cremmer:1983bf,*Ellis:1983sf,*Ellis:1983ei},~\cite{noscalerev} or even a bit more recently 
\cite{Kounnas:1994fr,Kelley:1994ab,nsvac2}, a number of
approximations were used in order to get analytic expressions with a rather 
transparent picture for the behavior of
the effective potential and its possible minima. 
While those approximations were legitimate at the time, 
clearly the situation will change substantially with an up-to-date
analysis, potentially affecting the existence and location 
of possible $m_{1/2}$ minima of the effective potential. 
We list below some of the important features to be taken into account. 
Rather than giving somewhat blind final results 
we find instructive to disentangle and discuss the different (tree-level, loop level) 
contributions as much as possible in order to pinpoint what contributions are actually responsible 
for the occurrence of non trivial $m_{1/2}$ minima. 
\begin{itemize}
\item Perhaps the most relevant point concerns the RG invariance of the effective potential:
in principle one would expect that the existence of minima is not strongly dependent on the choice
of the EW and renormalization scales. However, this is a non trivial issue since even the  
one-loop improved effective potential for the MSSM exhibits a rather important 
scale-dependence in general, due to its intrinsic non-RG invariance unless one subtracts a scale-dependent 
vacuum-energy-like term. Although the necessity of including in general such a term
was established since the work in refs \cite{Kastening:1991gv,Bando:1992wz,Bando:1992wy,rginv2}, 
we stress that it has a drastic influence on the specific $m_{1/2}$ minimization results, 
as was indeed pointed out earlier in refs. \cite{Kounnas:1994fr,Kelley:1994ab,nsvac2}.   
This will deserve a more detailed discussion below.
\item In the standard REWSB mechanism, the occurrence of a non trivial EW minimum
at some scale $Q_{EW}$ is strongly determined by the driving of $m^2_{H_u}$ towards its EW scale value, 
characterized (very roughly) by $m^2_{H_u}(Q_{EW}) <0$.   
Indeed the (one-loop) RGE for $m_{H_u}$ reads:
\be
8 \pi^2 \frac{d}{d\,\ln Q} m^2_{H_u} = 3 Y^2_t (m^2_{Hu} +m^2_{\t Q_L}+m^2_{\t t_R}+A^2_t) - {g'}^2 M^2_1
-3g^2 M^2_2
\ee 
where for sufficiently large $Y_t$ the first term on the RHS largely dominates. 
[Note that the trace term `$Tr Y m^2$' is absent since we assumed mSUGRA boundary conditions at the GUT scale.]
One might expect similarly that the occurrence of a non-trivial extra minimum of 
$V_{eff}(m_{1/2})$ resembles the REWSB mechanism, therefore 
relying mostly as a first approximation 
on the running properties and $m_{1/2}$ dependence of the relevant Higgs sector parameters entering
(\ref{Vtree}). This is, however, not the case, the detailed mechanism triggering possible $m_{1/2}$ minima 
being quite more subtle: within the initial conditions (\ref{softbc}, \ref{muscale}) and even when including
the RGE running of the tree-level potential parameters, (\ref{Vtree}) does not lead to non trivial 
$m_{1/2}$ extrema satisfying simultaneously (\ref{ewsb}) and (\ref{ns}), [except possibly at $m_{1/2}=0$, but
where the extremum is a maximum!]. Adding merely a vacuum energy term will already allow for local $m_{1/2}$ minima. 
More generally, as we shall examine later on, the occurrence of (phenomenologically relevant) non trivial $m_{1/2}$ 
minima will result from the interplay between (\ref{Vtree}), the vacuum energy term {\sl and} the one-loop corrections
to the effective potential.

\item Further influence on the precise location of $m_{1/2}$ minima 
comes from the necessary non-logarithmic radiative corrections to (s)particle masses, 
i.e. that are not determined only from RG properties and can indirectly affect the effective potential 
dependence on $m_{1/2}$. 
Among those are the field-dependent contributions coming from the one-loop part of the effective potential,
most conveniently included, at the EW minima, in the form of tadpole contributions~\cite{tad}. 
Though these are naively 
reasonably moderate corrections with respect to a tree-level analysis, 
they can have in fact a strong influence on some crucial relations such as those involving 
$\mu$, $m^2_{H_i}$ and the $m_Z$ mass at the EW minimum.
As a result their global effect may shift substantially the $m_{1/2}$ minima 
with respect to a simple tree-level analysis.  
In addition, other non-RG radiative corrections are important especially for 
the top and bottom Yukawa couplings and (to a lesser extent) for the gauge couplings. For example
in the standard procedure where the top (pole) mass is input, one extracts the Yukawa coupling
values at some chosen input scale $Q_{in}$ (typically $m_Z$ or EW) from the relation
\be
m^{pole}_{top} =  Y_t(Q_{in}) v_u(Q_{in}) (1+\delta^{RC}_y(Q_{in})+\cdots) \; .
\label{Ythresh}
\ee
Here the one-loop non-logarithmic (SUSY and SM) corrections $\delta^{RC}_y$ 
relate the pole mass to the running mass, and ellipsis stands for higher order corrections. 
Now, the supersymmetric
and standard model contributions to $\delta^{RC}_y$ are quite large and positive in most of the mSUGRA parameter 
space, such that the extracted value of $Y_t(Q_{EW}) $ is substantially smaller
than what it would be in a pure RGE approximation neglecting $\delta^{RC}_y$. 
Then the precise value of $Y_t$ has an important impact on the $m^2_{H_u}$ running
among other things, thus also on the subsequent determination of other relevant parameters
$B$, $\mu$ via the EW constraints.  This emphasizes the importance of controlling  
all sources of radiative corrections
for a better determination of the no-scale $m_{1/2}(m_{3/2})$ minima. 

\item Finally there are other minor differences between up-to-date standard SUSY spectrum calculations
and the above mentioned approximations, like the fact that the RGEs are solved numerically for
any $\tan \beta$, rather than analytically for a restricted range of small $\tan\beta$ values. 
In the following we also mainly examine the influence of considering consistently the soft breaking Higgs 
mixing parameter, $B_0$, to be an input at the high scale, which is quite different from considering 
$\tan\beta\equiv v_u/v_d$ input (at low scale).
\end{itemize}
\subsection{RG invariance and the effective potential}

The tree-level potential (\ref{Vtree}) is known to have in general an unwelcome  scale-dependence. 
A first step to improve this situation is to consider the  one-loop improved effective potential 
\cite{CWV} defined in the $\overline{DR}'$ scheme \cite{V2Martin, Martin:2001vx} as 
\be
V_{eff} \equiv V_{tree} +V_{1-loop} = V_{tree}[H_u,H_d](Q)+ \frac {1}{64\pi ^2}
\sum_n (-1)^{2n}  M_n^4(H_u,H_d)(\ln  \frac{ M_n^2(H_u,H_d)}{Q^2}-\frac 32)
\label{Veff}
\ee  
where $M_n$ are (field-dependent) mass eigenvalues and 
the summation runs over all (s)particle species and possible degeneracies due to color, flavor, etc. 
In the general MSSM  the one-loop term in (\ref{Veff}) takes explicitly the 
form~\footnote{One should in principle also include in (\ref{V1loop}) the one-loop contribution
of the gravitino, $-4 h(\tilde G_{3/2})$. 
This contribution  makes, however, very little numerical
differences in our analysis, at least as long as $m_{3/2}\lsim m_{1/2}$, 
since it has a rather small weight relative to the total sum over
all contributions to (\ref{V1loop}). For generic input values and $m_{3/2}=m_{1/2}$  
it is typically an ${\cal O}(1\%)$ effect.},
\bea
V_{1-loop} &= &\sum_{\phi^0} h(\phi^0) +2\sum_{\phi^+} h(\phi^+)+2\sum_{\tilde f} h(\tilde f)
-2\sum_{i=1,...4} h(N_i)-4\sum_{i=1,2} h(C_i)-16 h(\tilde g)\nonumber \\
& &-12 (h(t)+h(b))-4h(\tau) +3h(Z)+6h(W)
\label{V1loop}
\eea 
in the notations of \cite{V2Martin}, where the name of each particle denotes its squared (field-dependent)
mass and
\be
h(x) \equiv \frac{x^2}{64\pi^2} (\ln \frac{x}{Q^2}-\frac{3}{2})\;.
\ee
The expression of RG invariance is formally
\be
\left[\frac{\partial}{\partial \ln Q} +\sum_i\beta_i(\lambda_i)\frac{\partial}{\partial \lambda_i} 
-\gamma_{H_u} v_u\frac{\partial}{\partial v_u}
-\gamma_{H_d} v_d\frac{\partial}{\partial v_d}
\right] V_{eff} =0
\label{RG}
\ee
where $\lambda_i$ designates generically all relevant couplings or masses, with $\beta_i$ their
corresponding beta functions, and $\gamma_{H_i}$ are the anomalous dimensions of the Higgs fields. 
As is well known the practical cancellations of scale-dependence implied by RG invariance 
 generally occurs among terms of different perturbative orders. 
Thus the (one-loop level) cancellation of the scale dependence    
would be expected to occur between the relevant one-loop beta functions parts in Eq.~(\ref{RG})
 acting on the tree-level parameters in $V_{tree}$ and the explicit $Q$-dependence in the one-loop
term above, up to higher order (two-loop) remnant terms. However, in general in the presence of massive
fields, and in particular in the MSSM due to the SUSY-breaking terms,
this does not work so, because the effective potential in the form (\ref{Veff}) is not 
a proper RG-invariant physical quantity, so that there are remnant terms of one-loop order that do not cancel. 
Perhaps rather curiously, apart from early hints \cite{rginv0} this fact was not fully appreciated 
until the early nineties, where different detailed prescriptions were 
proposed \cite{Kastening:1991gv,Bando:1992wz,Bando:1992wy,rginv2}, all pointing out the necessity to include a
`vacuum energy' piece in addition to the above one-loop effective potential (\ref{Veff}). 
One simple prescription is to subtract the field-independent zero-point (vacuum) energy \cite{rginv2}:
\be
V_{full}\equiv  V_{tree} +V_{1-loop}-V_{1-loop,sub} 
\label{Vfullsub}
\ee
with
\be
V_{1-loop,sub} \equiv V_{1-loop}(H_u,H_d=0)
\label{sub}
\ee  
which can easily be shown to have the same one-loop RG-running as the remnant part from 
Eq.~(\ref{Veff}), therefore cancelling the scale-dependence in the latter 
up to higher (two-loop) order terms.
This subtraction is by construction similar to the supertrace in Eq.(\ref{Veff})
but with a spectrum involving only soft terms and the supersymmetric $\mu$ parameter~\footnote{For 
$v_u = v_d = 0$ all particle masses originating from EW symmetry breaking are thus vanishing.
The chargino and neutralino masses are respectively
$|M_2|$, $|\mu|$ and  $|M_1|$, $|M_2|$, $|\mu|$ with two degenerate ones. Similarly the sfermion and scalar sector 
eigenmasses depend only on the soft terms and the $\mu$ parameter.}.
However, this subtraction is only one possible prescription, sufficient for RG
invariance properties at this one-loop order, but having 
some limitations and unwelcome features \cite{rginv2}, such as that the subtracted potential may become 
complex~\footnote{This happens in particular in the MSSM case, where for $v_u=v_d=0$ the (would be lightest) neutral
and charged Higgs states become tachyonic whenever the EWSB conditions (\ref{ewsbc}) are satisfied.
Note however that these problems may  be avoided by subtracting at other values of the Higgs fields.} 
A more general convenient prescription consists in adding a 
running vacuum energy to the potential:
\be
V_{full} \equiv V_{tree}(Q) +V_{1-loop}(Q) +\tilde{\Lambda}_{vac}(Q)
\label{Vfull}
\ee
where the running of $\tilde{\Lambda}_{vac}$ is determined by requiring $V_{full}$ to satisfy (\ref{RG}), 
which leads to the (one-loop) RGE equation,
\be
Q\frac{d}{d\,Q}\tilde{\Lambda}_{vac}(Q) = \frac{1}{32\pi^2} \sum_n (-1)^{2n} {M}_n^4(H_u, H_d=0).
\label{rgevac}
\ee
A reasonably tractable expression at the two-loop level is also available~\cite{V2Martin}. 
The RGE of $\tilde{\Lambda}_{vac}(Q)$ has no direct influence on other RG parameters, and $\tilde{\Lambda}_{vac}(Q)$ 
behaves at the one-loop level qualitatively exactly
like the above defined subtraction $V_{1-loop,sub}$, thus canceling the remnant non RG-invariant terms from 
$V_{tree}+V_{loop}$ in (\ref{Veff}). Moreover, the running $\tilde{\Lambda}_{vac}(Q)$ will be uniquely fixed
once a choice is made of its boundary value at some arbitrary initial scale. In view of later discussions where
the boundary conditions (\ref{softbc}, \ref{muscale}) will be assumed, we adopted here a notation with  
the tilde to indicate that the boundary value $\tilde{\Lambda}^0_{vac}$ can be in principle quite general,
 including the possibility that the dimensionless quantity $\tilde{\Lambda}^0_{vac}/m^4_{1/2}$ be $m_{1/2}$ 
{\sl dependent}, while a $\Lambda$ without a tilde will implicitly indicate that the boundary conditions 
$\Lambda^0_{vac}/m^4_{1/2}$ are chosen $m_{1/2}$ ($m_{3/2}$) {\sl independent}. We will refer to the latter as the
untwiddled prescription'. This distinction is important insofar as we are interested in the minima of the potential
with respect to $m_{1/2}$, given that one can always add to $\tilde{\Lambda}_{vac}(Q)$ 
any arbitrary $Q$-independent function of $m_{1/2}$ without altering the RGE properties while modifying
the structure of the $m_{1/2}$ minima.  Throughout the paper we adopt mainly the `untwiddled prescription' version 
of Eq.~(\ref{Vfull}), but also occasionally illustrate the subtraction prescription (\ref{Vfullsub}),
in particular in section \ref{sub-presc}. Obviously, the latter prescription is but
a special case of the general twiddled prescription (\ref{Vfull}), and (\ref{sub}) a special solution of the RGE 
(\ref{rgevac}) with a specific boundary condition at some scale $Q=Q_0$. Indeed, taking 
\be
\tilde{\Lambda}^0_{vac} = -V_{1-loop,sub}(H_u, H_d=0,Q_0)
\label{initialsub}
\ee
as a boundary condition, ensures through (\ref{rgevac}) that
\be
\tilde{\Lambda}_{vac}(Q) = -V_{1-loop,sub}(H_u, H_d=0,Q)
\label{solsub}
\ee
for all $Q$. It is also obvious from the form of $V_{1-loop,sub}$ that the boundary condition (\ref{initialsub}) is of 
the twiddled type, i.e. $\tilde{\Lambda}^0_{vac}/m_{1/2}^4$ is a function of $m_{1/2}$ when 
(\ref{softbc}, \ref{muscale}) are assumed. It will prove phenomenologically useful to compare the subtraction
prescription (\ref{Vfullsub}, \ref{sub}) with the untwiddled prescription of (\ref{Vfull}). Clearly, these are two 
different prescriptions from the point of view of no-scale, since they differ in the $m_{1/2}$ dependence of the 
boundary conditions, and thus lead to different $V_{full}(m_{1/2})$ potentials.  

The vacuum energy being field-independent by definition, has no influence
on the EW minimization of the effective potential, Eqs. (\ref{ewsb}), so that it can be safely omitted
in all related issues. But 
it can have a definite influence on the fate of eventual $V(m_{1/2})$ minima, contributing non trivially
to Eq.~(\ref{ns}), as we will see in more detail later on. 
 In a top-down approach, the running vacuum energy allows
to choose different boundary conditions for $\tilde{\Lambda}_{vac}(Q)$. In particular we shall consider
different choices of $\tilde{\Lambda}_{vac}(Q_{GUT})$ or $\tilde{\Lambda}_{vac}(Q_{EW})$ 
and explore the consequences on the existence of $m_{1/2}$ minima
for the full effective potential. This point was already noted and studied 
earlier~\cite{Kounnas:1994fr,Kelley:1994ab,nsvac2}, as mentioned
in the introduction, but those studies relied on semi-analytical
expressions within approximations similar to the ones mentioned above.
Following these authors, we 
parameterize the vacuum energy contribution at an arbitrary scale $Q$ 
in terms of 
$m_{1/2}$ (or equivalently $m_{3/2}$), in the most general twiddled context,  as
\be
\tilde{\Lambda}_{vac}(Q)\equiv \tilde{\eta}(Q) m^4_{1/2}
\label{Lambdaeta}
\ee
where the running of $\tilde{\eta}(Q)$ at one-loop is determined by Eq. (\ref{rgevac}) together with 
the boundary condition defined e.g. at the GUT scale as $\tilde{\eta}(Q_{GUT})\equiv \tilde{\eta}_0$.
We stress here that, on top of the $Q_{GUT}$ dependence, $\tilde{\eta}(Q)$ can in general depend also on $m_{1/2}$ 
(or equivalently $m_{3/2}$). 
[we will come back to this point later on when discussing equation (\ref{nsapprox}).] 

\begin{figure}[htb]
\begin{center}
\epsfig{figure=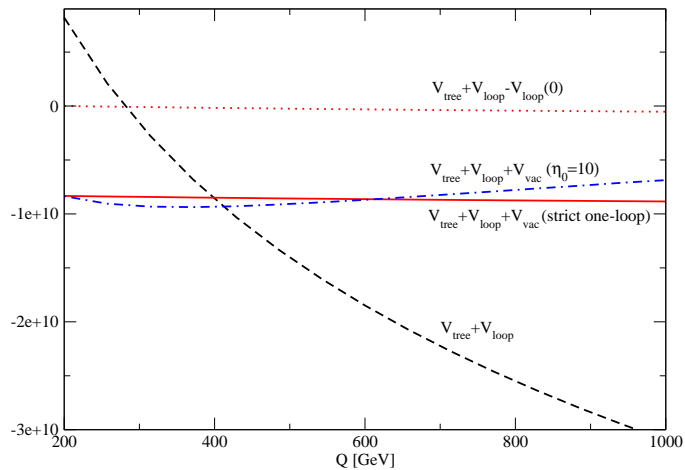,width=8cm,angle=-90}
\caption[long]{\label{Qdep} The scale dependence of the effective potential, with or without the
vacuum energy contributions, for representative input values $B_0=.2\mhalf$, $m_0=A_0=0, m_{1/2}=300$GeV and
 $m_{top}=$173 GeV. 
Dashed line: without vacuum energy contribution; full line:
$V_{full}$ from Eq.~(\ref{Vfull}) with running vacuum energy and $\eta_0=10$ at the strict one-loop order
(i.e. no running of parameters inside one-loop expressions); 
dash-dotted line: same as the previous case but with all running parameters at one-loop, showing moderate 
spurious scale-dependence to be cancelled by two-loop contributions;
dotted line: same with subtracted vacuum contribution from Eq.~(\ref{sub}).}
\end{center}
\end{figure}

The improvement in scale (in)dependence of $V_{full}$ (\ref{Vfull}), as compared to 
$V_{tree}+ V_{loop}$ alone, is illustrated in Fig.~\ref{Qdep} for both the untwiddled prescription
with $m_{1/2}$-independent $\eta_0$ values, and the subtraction prescription that corresponds
to an $m_{1/2}$-dependent initial condition $\tilde{\eta}_0$. 
Of course the absolute value of $V_{full}$ depends
very much now on the initial condition for $\eta$, but this is a constant shift as far as the $Q$ dependence
is concerned. 
Note that, strictly speaking,  to ensure the RG invariance at one-loop level one should not consider 
the running of parameters within the one-loop expressions in (\ref{Vfull})~\cite{rginv2,rginv3}, since those
induce formally two-loop order terms.  
Indeed, as we have checked explicitly, the scale independence of the full effective potential
at one-loop is almost perfect when freezing
the running of all relevant parameters entering the different one-loop contributions, while the
formally higher order terms induced from those runnings produce a remnant but rather moderate scale 
dependence visible in Fig.~\ref{Qdep}. We have checked however that such spurious 
effects remain reasonably small in all cases of our subsequent analysis. In particular
they influence only very moderately  the location of $m_{1/2}$ minima, whenever those exist.

\subsection{The fate of $\mathbf m_{1/2}$ minima} \label{presentB}
Having a well-defined (one-loop) RG-invariant effective potential, Eq.~(\ref{Vfull}), we will
now examine in more details the behavior of its different contributions with respect
to $m_{1/2}$. By inspection of the RGE, it is
possible to infer that, for generic boundary conditions with a linear 
dependence upon the soft breaking parameters as given in (\ref{softbc}), all the resulting RG-evolved parameters at 
the EW scale, in the one-loop RG approximation, will have a similar linear behavior,
\be
B_{EW}= b\, m_{1/2},\;\; m^2_{H_u}= u \,m^2_{1/2},\;\; m^2_{H_d}= d\, m^2_{1/2},\;\; A_i=a_i\, m_{1/2}\;.
\label{softevol}
\ee 
The unspecified scale-dependent $b$, $u, d$ and $a_i$ parameters in Eq.~(\ref{softevol}) have 
of course complicated expression in terms of the 
original ones in Eq. (\ref{softbc}), but entirely determined numerically by the RGEs. 
Note indeed that the linear behavior in Eq. (\ref{softevol}) is obtained even for the strict no-scale
boundary conditions in (\ref{strict}), because an extra $m_{1/2}$ linear dependence is induced
from the RGE of the gaugino mass parameters $M_i$. 

Although the order of the three different minimizations in Eq.~(\ref{ewsb}) and (\ref{ns}) is in principle
irrelevant, before examining the $m_{1/2}$ minimization 
it is much more convenient to sit first at the EW minima, which greatly simplifies the procedure. 
(This is also because we do not consider the full expression of the one-loop effective potential
for any $H_u$, $H_d$ field values, since it is equivalent and more convenient to put the tadpole 
contributions when the effective potential is evaluated at the EW minimum). 
This is just the familiar way of expressing the EW minimization Eqs. (\ref{ewsb}) as
constraints to express $B_{EW}$ and $\mu_{EW}$ in terms of the other parameters:
\be
B \mu = (\hat m^2_{H_u}+ \hat m^2_{H_d} +2\mu^2) \frac{\sin 2\beta}{2} 
\label{bmuew}
\ee
\be
\mu^2 = \frac{\hat m^2_{H_d} -\hat m^2_{H_u}\tan^2\beta}{\tan^2\beta-1}-\frac{g^2+g^{'2}}{4}v^2
\label{muew}
\ee
where $\tan\beta\equiv v_u/v_d$, $v^2=v^2_u+v^2_d$ in our conventions and 
$\hat m^2_{H_i}\equiv m^2_{H_i}+\delta H_i $ where
$\delta H_{u,d}$ denotes the corrections implied by one-loop tadpoles.
The latter have the generic form
\be
\delta H_i =\frac{1}{16\pi^2}\sum_n c_n (-1)^{2n} M^2_n (\ln \frac{M^2_n}{Q^2}-1) 
\label{tad}
\ee
where $c_n$ are the different couplings to the respective Higgs fields $H_u$, $H_d$ of 
all relevant particles in the sum over $n$. 
Notice that, for reasons that will become clear in section \ref{procede-A-C}, 
we did not yet use in Eq.~(\ref{muew}) the additional constraint
that the Z mass should be reproduced at the EW minimum, i.e. the {\em extra} constraint:
\be
\frac{g^2+g^{'2}}{2}v^2 \equiv m^2_Z\;.
\label{mz}
\ee  
As discussed above, one expects the soft-SUSY breaking parameters $m^2_{H_i}$ and $B$ in no-scale scenarios 
to be directly related to the single source of SUSY-breaking, $m_{3/2}$ (or equivalently $m_{1/2}$). 
Concerning the supersymmetric $\mu$ parameter, as stated at the end of section \ref{basics}, it may 
either be considered as an independent parameter or else related to $m_{1/2}$ at high scale. But it is in both 
cases entirely determined, at the EW scale, via the constraints (\ref{muew}). \\
From (\ref{bmuew} -- \ref{tad}), one obtains, after straightforward
algebra, the effective potential at the EW minimum in the form,
\be
V_{full}^{EW min}= -\frac{g^2+g^{'2}}{8} v^4 (1-2s^2_\beta)^2 -v^2 (s^2_\beta\, \delta H_u +c^2_\beta\, \delta H_d)
+V_{loop}+\tilde{\Lambda}_{vac}
\label{Vft}
\ee
with $c_\beta\equiv \cos\beta$, $s_\beta\equiv\sin\beta$, and 
all terms have implicitly a scale dependence here omitted for simplicity of notation. 
Requiring further the constraint of correct physical Z mass (\ref{mz}),
one obtains
 \be
V_{full}^{EW min}(m_Z \, {\rm fixed})= -\frac{m^4_Z}{2(g^2+g^{'2})} (1-2s^2_\beta)^2 -2 \frac{m^2_Z}{(g^2+g^{'2})} 
(c^2_\beta \,\delta H_u +s^2_\beta \,\delta H_d)
+V_{loop}+\tilde{\Lambda}_{vac}
\label{Vftz}
\ee
which is formally different from (\ref{Vft}), in particular as far as the functional $m_{1/2}$ dependence is concerned.
We will come back to this point in more detail in the next section. 
Note that this difference is strenghthend by the fact that the dependence on $m_{H_u}$, $m_{H_d}$ has 
disappeared from the tree-level term of (\ref{Vft}):  indeed, away from the EW minimum 
the (tree-level) potential in Eq.~(\ref{Vtree})
depends on the five parameters $v_u$, $v_d$, $m^2_1$, $m^2_2$, $B\mu$ (let alone the two gauge couplings). Now $B$ and 
$\mu$ can be eliminated upon use of the EW minimum constraints (\ref{bmuew})-(\ref{muew}), but the very structure
of (\ref{Vtree}) implies that, at the EW minimum, a third parameter disappears so that
(\ref{Vft}) in fact depends only on two independent parameters, that we may choose here for convenience
to be $v_u$ and $v_d$.

\section{Looking for minima of the RG invariant effective potential}
In this section, and before entering a more phenomenological discussion, we examine generically 
the possible existence and fate of $V_{full}(m_{1/2})$ minima for representative input parameters.  
We will also specify for this purpose some important aspects of our minimization procedure.   
\subsection{$\mathbf B_0$ input}\label{B0in} 
We shall first consider an important feature concerning the choice of the input parameters.
Most scenarios in no-scale models imply a fixed $B_0$ high scale value, in particular 
in the strict no-scale, $B_0=0$ as in Eq. (\ref{strict}), or in the string inspired case (\ref{stringbc}). 
However, the by now standard MSSM model-independent procedure is to determine $B_{EW}$ together with $\mu_{EW}$
from the EWSB minimization conditions (\ref{ewsb}), not caring usually for high scale values of $B_0$.
Even in a more phenomenological framework, it is of interest to 
perform the minimization rather with $B_0$ input, considering $\tb$ as dynamically determined rather than an input.
Essentially one has to consider Eq. (\ref{bmuew}) as determining $\tan\beta$ from $B_{EW}$ and
the other relevant parameters at the EW scale. 
It turns out to be a rather non-trivial exercise to make such a consistent algorithm. Actually 
the EW minimization condition (\ref{ewsb}) together with $\tb(B)$ determination turns out to give
a fourth order equation for $\tb$ (with not always real solutions), but this
is a rather straightforward part of the derivation. More problematic is that 
the dynamical $\tb$ values thus determined at the EW scale 
is a sensible parameter in all subsequent calculations, and in particular it drastically affects  
the influential top Yukawa coupling (due to the low energy matching relations Eq.(\ref{Ythresh})), 
which in turn is driving strongly the RGE of the $B$ parameter. Therefore, the algorithmically non trivial 
feature is to get consistent values of $B_0$ and $\tb$, matching both high and low energy boundary 
conditions, satisfying the EWSB constraints
etc, because of the induced effects on the RGE. 
This has to be solved iteratively and a new algorithm was introduced in
SuSpect for this purpose.\\
\begin{figure}[htb!]
\begin{center}
\epsfig{figure=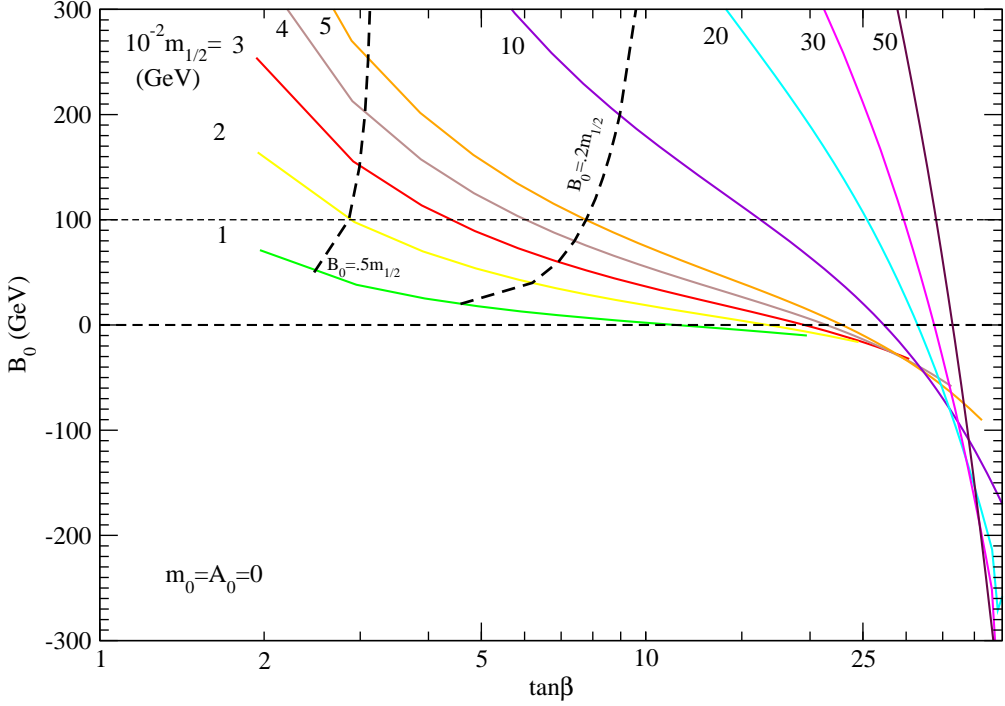,width=12cm,angle=-90}
\caption[long]{\label{Btb} The $B_0-\tb$ connection for various choices of $m_{1/2}$ (with $m_0=A_0=0$).}

\end{center}
\end{figure}
We illustrate in Fig.~\ref{Btb} the connection between $B_0$ and $\tb$
for different representative values of $m_{1/2}$. As one can see,
rather low $\tb$ values are favored, especially for increasingly large $B_0$. 
Note that Fig.~\ref{Btb} 
illustrates only the $m_0=0$ case, but the influence of $m_0 \ne 0$ is not very drastic, only pushing the 
curves for a given $B_0 \ne 0$ to slightly smaller $\tb$, in an almost parallel way.
The obtained values of $\tb$ are essentially located in the range $2-25$ (increasing for decreasing $B_0$
and increasing $m_0$) for $100 \mbox{GeV}  \lsim m_{1/2}\lsim {\cal O}(1 \mbox{TeV})$. 
It is still possible to reach larger $\tb$ values, but for very large $m_{1/2}$ and small $B_0\sim 0$ values.
\subsection{Renormalization scale prescriptions and naive RG-improved effective potential}\label{reno+pot}
Since the full effective potential Eq.~(\ref{Vfull}) is approximately scale-independent
(up to two-loop order), it is convenient to go a step further, following 
\cite{Cremmer:1983bf,*Ellis:1983sf,*Ellis:1983ei}, 
and define a naive `RG-improved' effective potential by choosing the 
arbitrary scale $Q_0$ such that
\be
 V_{1-loop}(Q_0) \equiv 0
\label{Vlzero}
\ee 
This choice simplifies largely the analysis since one can then consider only the minimization 
of $V_{tree}(Q_0)+\tilde{\Lambda}_{vac}(Q_0)$ which has a relatively simpler dependence on $m_{1/2}$.
The latter combination still embeds one-loop corrections, since the
dependence $Q_0(Q)$ at one-loop is implicit and thus consistently absorbed within $V_{tree}(Q_0)+\tilde{\Lambda}_{vac}(Q_0)$.
This prescription is partly inspired by the construction of RG-improved effective
potential~\cite{Kastening:1991gv,Bando:1992wz,Bando:1992wy,rginv2}, where RG-resummation properties have been 
rigorously established 
for simpler effective potentials with a single mass scale. In the present multi-scale MSSM case, a
rigorous generalization is not available at present, and choosing simply the scale cancelling the loop term
is certainly not sufficient to establish correct resummation properties, but at least this prescription is 
well-defined and greatly simplifies the analysis.  
An expression for $Q_0$ is obtained from Eqs.~(\ref{Veff}), (\ref{V1loop}) in the form
\be
Q_0 = e^{-3/4} \; \exp \frac{\sum_n (2s_n+1) M^4_n(Q_0)\ln M^2_n(Q_0) }
{2 \, \sum_n (2s_n+1) M^4_n(Q_0) } 
\label{RGI}
\ee 
which, at the one-loop level, gives explicitly $Q_0$. [The $Q_0$ dependence through $M_n(Q)$ on the RHS of 
Eq. (\ref{RGI}) gives contributions formally of higher (two-loop) order. 
To include these effects one can still solve (\ref{RGI}) numerically by iterating on $Q_0$.]
Nevertheless, the occurrence and location of possible no-scale $m_{1/2}$ 
minima are not expected to depend too much on this prescription,
up to higher order small effects, due to the overall approximate RG-invariance. 
We emphasize that (\ref{RGI}) is only one convenient choice among many possible prescriptions,  
and in our numerical analysis, as will be more explicit later, we also 
perform and compare the minimization results when using other well-defined 
renormalization scale prescriptions to quantify the residual scale dependence.  
In those expressions is also implicit the field-dependence of the masses: 
as mentioned above, 
 minimizing with respect to the $H_u, H_d$ field 
is more conveniently done by canceling the corresponding one-loop 
tadpoles~\cite{tad}, which is completely equivalent to
minimizing explicitly with respect to the $H_u, H_d$ fields~\footnote{
The equivalence only holds when sitting at the EW minimum, as the cancellation of the tadpoles
does not give the field dependence {\em away} from the EW minimum.}.   
All that is performed numerically within SuSpect~\cite{suspect}, involving a number of subtleties
like necessary iterations etc.
\subsection{Different $\mathbf m_{1/2}$ minimization procedures} \label{procede-A-C}

Having determined the EW direction from Eq. (\ref{ewsb}), one obtains the 
effective potential (\ref{Vft}) in the EW valley, $V_{full}^{EW val}(m_{1/2})$, and then perform 
the extra minimization Eq. (\ref{ns}) along this valley. In this
subsection we examine in some details different approaches to this minimization procedure, having in mind 
what can be most easily implementable within the present tools for the computation of the MSSM spectra. 
Though we will carry out a numerical analysis with SuSpect 
at the full one-loop level (including dominant two-loop effects as well),
it is useful at a first stage to examine some approximations 
to appreciate the main features of the minimization results. 

At first sight it could be very tempting to simply evaluate the full effective potential
(\ref{Vfull}) from the output of one's preferred SUSY spectrum code, and to perform the extra
minimization with respect  to $m_{1/2}$ numerically. 
However, there is an important subtlety:
in the no-scale approach, $v_u, v_d$ and $m_{1/2}$ should be treated as three
independent dynamical variables with respect to which we seek a minimum of $V_{full}$. It is only at such
a minimum that one requires the physical quantities, such as the pole masses of the
Z boson, the top and bottom quarks, the $\tau$ lepton (as well as the ones of all other quarks and leptons),
 to be properly reproduced. Now the point is that in most publicly available SUSY spectra 
codes~\cite{isajet,suspect,softsusy, spheno} only the $v_u,v_d$ two-parameter EW minimization is performed.
Accordingly, the $m_Z, m^{pole}_{top}$ and other mass constraints are hard-coded everywhere, which is 
perfectly consistent in so far as the EW minimization conditions (\ref{bmuew}, \ref{muew}) are by definition valid 
only at the minimum. In contrast, if one is determining possible 
$m_{1/2}$ minima through a numerical analysis of the shape of $V_{full}^{EW val}(m_{1/2})$, while requiring 
{\sl beforehand} the above-mentioned mass constraints, which is the most straightforward way
of using the existing codes, then this amounts to 
requiring the correct physical mass values even when not sitting at the physical vacuum. This induces 
an artificial dependence of the soft parameters on $m_{1/2}$
distorting the shape of $V_{full}^{EW val}(m_{1/2})$ and thus the location or even the possible existence of
the minima. More specifically, assuming (\ref{softbc}) for 
definiteness and fixing $m_Z$ 
away from the $m_{1/2}$ minimum, induces through (\ref{muew}) and (\ref{mz}) a functional dependence of $\mu$ on 
$m_{1/2}$ which is incompatible with an (otherwise perfectly acceptable) $m_{1/2}$-independent boundary condition for 
$\mu$, or even with a boundary condition having an $m_{1/2}$ functional dependence of the type (\ref{muscale}). 
Feeding back in (\ref{Vft}) this fixed-$m_Z$ induced $\mu(m_{1/2})$, would lead to an EW valley potential 
$V_{full}^{EW val}(m_{1/2})$ that is incompatible with a 
large class of possible boundary conditions for $\mu$, and to the false conclusion that they are forbidden. 
In practice this would show up in a modification of $\frac{\partial \mu}{\partial m_{1/2}}$ 
in the identity
\be
\frac{d}{d m_{1/2}}\,V_{full}^{EW val}(m_{1/2},\mu)  = 
\frac{\partial}{\partial m_{1/2}} V_{full}^{EW val} + 
\frac{\partial V_{full}^{EW val}}{\partial \mu} \, \frac{\partial \mu}{\partial m_{1/2}}\;,  
\label{diff}
\ee
that would displace the solutions of $\frac{d}{d m_{1/2}}\,V_{full}^{EW val}(m_{1/2},\mu)=0$.
A similar rationale holds for the physical value of $m^{pole}_{top}$ when fixed 
beforehand. In this case equation (\ref{Ythresh}) induces a spurious dependence of $Y_t(Q_{in})$ on $m_{1/2}$ 
through $v_u$ as well as the non-RGE loop corrections $\delta^{RC}_y(Q_{in})$. When fed back in the potential along the EW 
valley, such a dependence will again distort unphysically its shape. 

In order to avoid all these difficulties one can adopt two possible procedures:
\begin{itemize}
\item[(A)] derive an explicit form for (\ref{ns}), akin to the explicit forms (\ref{bmuew}, \ref{muew}) derived 
from (\ref{ewsb0}), and solve numerically this system of three explicit equations. This permits to use consistently  
the fixed mass constraints (as they stand in the public codes without modification), since one is now sitting at the 
actual physical vacuum. 
\item[(B)] deactivate the physical mass constraints in the codes, determine numerically the minimum of 
$V_{full}^{EW val}(m_{1/2})$ along the EW valley, then impose
these mass constraints once the minimum is found, to check for consistency. 
\end{itemize}

Implementation of procedure (B) including the full loop-level effective potential, can be very involved and  
would necessitate a highly non-trivial extension of the standard procedures of the various public codes.
Although the EW minimization part would not change\footnote{the explicit forms at the full one-loop level, 
(\ref{bmuew}), (\ref{muew}), should still be solved numerically  due to the tadpoles
involving a highly non-linear dependence on all parameters.},  
the further minimization with respect to $m_{1/2}$, (\ref{ns}), if performed purely numerically, becomes in general
a rather involved task; in the EW valley, one will have to scan a two-parameter space 
$(m_{1/2},\mu)$ , requiring the correct 
$\mu(m_Z)$ relationship via (\ref{muew}), (\ref{mz}), as well as consistency of $m^{pole}_{top}$ via
(\ref{Ythresh}),  only once the minimum is found. This holds irrespective of whether we assume (\ref{muscale}) or not.
However, as we will discuss below, the boundary condition (\ref{muscale}) will have the benefit that
requiring the mass constraints {\sl before} determining the minimum, although in principle wrong, can become a good 
approximation. This allows a simplified version of (B) as follows: 
\begin{itemize}
\item[(C)] in this procedure the minimum of $V_{full}^{EW val}(m_{1/2})$ is determined fully numerically along the EW
 valley as in (B), but the $m_Z$ and $m^{pole}_{top}$ constraints are applied from the outset as in (A).  
\end{itemize}
Procedure (C) is thus desirable as it provides a simpler picture and is easy to use, not needing 
drastic modification of the present codes. 
The boundary condition (\ref{muscale}) has also some practical benefits in the context of procedure (A), 
as it allows to obtain an explicit analytical form for (\ref{ns}), which otherwise would not be
easily tractable if $\mu$ were taken as a free parameter. In the sequel we will thus stick to (\ref{muscale}) in 
all our study when comparing the outcome of procedures (A) and (C).
Furthermore, as a consequence of the RGE of $\mu$ which is of the form $d\mu/d\ln Q \sim \mu$,
 the boundary condition assumption (\ref{muscale}) leads to 
\begin{equation}
\mu(Q) =c_\mu(Q) m_{1/2}
\label{muscalevol}
\end{equation}
that is valid at any scale $Q$.
Now since $\mu_{EW} \equiv \mu(Q_{EW}^{default})$ is anyway fixed a posteriori  
by the EW minimization and the $m_Z$ constraint, Eqs.~(\ref{muew}), (\ref{mz}),
its high scale value $\mu_0$ is entirely determined through the RGE from 
$\mu_{EW}$. Yet, one has still to verify that $\mu_0$ remains of order $m_{1/2}$ in the spirit of the no-scale 
framework, as emphasized at the end of section II.
Indeed, this is generically the case (except possibly for $\tan \beta$ very close to $1$ which is anyway 
phenomenologically largely excluded), given the structure of (\ref{muew}) and the fact that $\mu(Q)$ does not run much.

We now discuss in more detail the implementation of procedures (A) and (C) as well as the approximate validity 
of the latter. Using  (\ref{softbc}) and  (\ref{muscale}) and neglecting the running of all masses occurring in the 
one-loop part of  (\ref{Vfull}) which is consistent at one-loop level strictly,   
one can derive from (\ref{Vfull}) an `explicit' equation for (\ref{ns}) to be used in procedure (A):
\be
m_{1/2} \frac{\partial}{\partial m_{1/2}}V_{full}(m_{1/2}) = 0 \Rightarrow  
V_{full}(m_{1/2}) +\frac{1}{128\pi^2} \sum_n (-1)^{2n} M^4_n(m_{1/2}) 
+ \frac{1}{4} m_{1/2}^5 \frac{d \tilde{\eta}_0}{d m_{1/2}} =0
\label{nsapprox}
\ee
where we also used (\ref{Lambdaeta}). Various comments are in order concerning the validity of 
this equation. The assumptions (\ref{softevol}, \ref{muscale}) are crucial 
to derive this compact expression, since the first term $V_{full}$  on the RHS of 
(\ref{nsapprox}) originates (upon use of $m_{1/2}\partial_{m_{1/2}} (m^4_{1/2}) = 4m^4_{1/2}$) 
from all terms within $V_{full}$ that scale trivially as $m^4_{1/2}$,  
while the supertrace term comes from the explicit $\ln (m^2_{1/2}/Q^2)$ dependence within (\ref{V1loop}),
see e.g. refs.~\cite{Kounnas:1994fr,Kelley:1994ab,nsvac2}. There is however more to it. 
Since the squared mass eigenvalues $M_n^2$ in (\ref{Veff}) are $H_u$-, $H_d$-field dependent, then strictly speaking
the scaling $M_n^2 \sim m_{1/2}^2$ assumed in deriving (\ref{nsapprox}) does not hold in general. 
There should thus be extra terms in
(\ref{nsapprox}) due to the breakdown of the $m_{1/2}$ scaling. Such terms are indeed present in general, but they 
actually vanish when the conditions (\ref{ewsb0}) are taken into account, i.e. at the true electroweak 
vacuum.\footnote{
This is not specific to our boundary condition assumptions. For any function $V(\chi, \phi)$ with $\chi$ and $\phi$ 
two independent fields, one can always artificially treat $\phi$ as being linear in 
$\chi$, $\phi \equiv \chi \hat{\phi}$, to recast $\partial_\chi V(\chi, \phi)$ in the form of a {\sl total} derivative 
$\frac{d}{d \chi} V(\chi, \chi \hat{\phi})$, provided one considers only the points where  
$\partial_\chi V(\chi, \phi) = \partial_\phi V(\chi, \phi) = 0$.}
Obviously, such a scaling holds as well at the point $H_u=H_d=0$, in particular
for the RGE of $\tilde{\Lambda}_{vac}$, (\ref{rgevac}). It then immediately follows from (\ref{Lambdaeta}), 
(\ref{softevol}) and (\ref{muscalevol}) that the running of 
$\tilde{\eta}$ does not depend on $m_{1/2}$. Thus, the only possible dependence of $\tilde{\eta}$ on 
this parameter would come from the boundary condition, whence the last term on the RHS of (\ref{nsapprox}). This term, 
usually not considered in the literature, will be important when discussing the alternative subtraction prescription in 
section \ref{sub-presc}.
At this point equations (\ref{bmuew}), (\ref{muew}) and (\ref{nsapprox}) form a consistent set of EWSB conditions 
given the assumptions (\ref{softbc}), (\ref{muscale}).
In a more general context, one would still want to consider models 
where the (supersymmetric) parameter $\mu_0$ is an independent one
 at high scale so that the scaling (\ref{muscale})
does not hold, (reflecting the well-known $\mu$-parameter
problem). In this case, Eq.~(\ref{nsapprox}) will have to be modifed in a non-trivial way including additional terms 
involving essentially
$\partial V_{full}/\partial \mu$ that are not straightforward to evaluate for $V_{loop}$.\footnote{
The exploration of this more general case will be pursued elsewhere.}  
 
Equation ~(\ref{nsapprox}), together with (\ref{bmuew}) and (\ref{muew}), complete the ingredients of procedure (A). 
They will have to be solved numerically as a non-trivial system of equations giving the values of 
$m_{1/2}, v_u$ and $v_d$ at the minimum of the potential.
     
We turn now to procedure (C) and discuss the degree of validity of the approximation involved when $m_Z$ is fixed 
before minimization. To illustrate the case we focus first on the tree-level part of the potential (\ref{Vft}). 
Before fixing $m_Z$, there is evidently a non-trivial $m_{1/2}$ dependence in this part, in the EW valley; from
equations (\ref{bmuew}) and (\ref{muew}), 
$v_u$ and $v_d$ (or $v$ and $s_\beta^2$) can be re-expressed in terms of the soft parameters and $\mu$
 as,
\be
v^2= -\frac{2}{(g^2+g^{'2})} \left(\frac{m^2_{H_u} -m^2_{H_d}}{|\cos 2\beta|} +m^2_A\right)
\ee
with
\be
|\cos 2\beta| = \left( 1- 4\frac{B^2 \mu^2}{m^4_A}\right)^{1/2}
\ee
where we defined for convenience $m^2_A \equiv m^2_{H_u} +m^2_{H_d}+2\mu^2 $, so that
\be
V_{tree, EW min} = -\frac{1}{2(g^2+g^{'2})} \left(m^2_{H_d} -m^2_{H_u} -m^2_A
|\cos 2\beta| \right)^2
\label{vthuhd}
\ee
which therefore exhibits an explicit non-trivial $m_{1/2}$ dependence via the relations in 
(\ref{softevol}). 
Thus one easily infers that $V_{tree, EW min} \sim m^4_{1/2}$.
In contrast,  if one imposes first $m_Z$ via 
Eq. (\ref{mz}) (as is hard-coded in the public version of SuSpect and in other public 
codes~\cite{isajet,softsusy,spheno}), 
clearly the effective potential 
has then a different $m_{1/2}$-dependence as is explicit from (\ref{Vftz}), whose tree-level part
only depends on $m_Z$ and $\tb$: it has accordingly practically
no more dependence on $m_{1/2}$ (at least a very mild one as compared to the $m^4_{1/2}$ dependence
inferred from the previous analysis not fixing $m_Z$). This illustrates the point we stressed previously
in this section: fixing $m_Z$ via Eq.~(\ref{muew}) beforehand induces a spurious functional dependence,
$\mu \equiv \mu(m_{1/2},m_Z)$, which overwrites any initial assumption about the $\mu$ boundary condition and 
thus modifies unphysically the structure of the minima. This remains of course true beyond the tree-level. 
More generally, it is clear from (\ref{diff}) that the reliability of procedure (C) will depend on how far or close  
is the spurious dependence $\mu(m_{1/2},m_Z)$ from the initially assumed model-dependence $\mu(m_{1/2})$.
For instance, if $\mu$ is taken to be a free $m_{1/2}$-independent parameter, then obviously
the second term on the RHS of (\ref{diff}) should be vanishing, while within  procedure (C) this
is not the case and can even lead to substantial differences for large $m_{1/2}$, thus degrading the
quality of the approximation in determining the actual minimum. In contrast, the situation is much more favorable
when the boundary condition (\ref{muscale}) is assumed. As can be easily seen from (\ref{muew}), (\ref{softevol})
and (\ref{mz}), the spurious dependence is of the form 
$\mu(m_{1/2},m_Z) = ( \frac{u \tan^2 \beta- d}{1 - \tan^2 \beta}m_{1/2}^2 - \frac{m_Z^2}{2})^{1/2} \sim m_{1/2}$ in 
the limit $m_{1/2} \gg m_Z$, and the functional dependence in (\ref{muscalevol}) is properly reproduced at 
the electroweak scale in this limit. 
Moreover, relatively large $m_{1/2}$ values are favored by the most recent experimental exclusion limits.    
From this rather crude analysis, it is expected that as far as the effect of fixing $m_Z$ is concerned,  
the (relative) difference between the output of procedures (A) and (C)
should decrease like $\sim m^2_Z/m^2_{1/2}$ for increasing $m_{1/2}$. 
We have performed explicit minimizations for rather generic mSUGRA input 
with $B_0, m_0,\cdots \ne 0$ to compare both procedures, and checked that the above described qualitative
behavior is indeed essentially observed. The corresponding $m_{1/2}$ minima that can be obtained from the naive 
(incorrect) procedure can differ substantially for $m_{1/2}\sim m_Z$, while for $m_{1/2}\gsim 300$ GeV  
the difference between the two procedures decreases to reach about 10\% for larger $m_{1/2}$, as will be illustrated 
with concrete examples. The residual 10\% difference is not due to the fixing of $m_Z$ but actually essentially to
the effect of fixing $m^{pole}_{top}$ through (\ref{Ythresh}) which was not taken into account in the
above discussion. It is, furthermore, always positive in a large part of the parameter space.
More precisely, the leading non-RGE one-loop SUSY corrections in  (\ref{Ythresh}) 
have the generic form,
\begin{eqnarray}
\delta^{QCD}_{SUSY}&=& \frac{\alpha_s}{3 \pi} ( r + 2 \ln \frac{m_{1/2}}{m_Z} ) \label{deltaQCDSUSY}
\end{eqnarray} 
where we used (\ref{softevol}) and  (\ref{muscalevol}) and where $r$ is a complicated
function  of $b_0, x_0, a_0$ of order $1$ -- $2$ depending on the input values. The $\ln \frac{m_{1/2}}{m_Z}$
dependence translates through the induced $m_{1/2}$ dependence in $Y_t$ into a shift of the form
$\zeta m_{1/2}^4$ in (\ref{nsapprox}) which {\sl in fine} accounts for the above mentioned constant relative 
difference in the values of $m_{1/2}$ minima.

With these features in mind, we will illustrate in most of our subsequent numerical analysis, the results of 
both procedures (A) and (C).

%

\subsection{No-scale scenarios and vacuum energy} \label{ns+vc}
We now illustrate minimization results for a representative set of parameter values, adopting
the previously described prescriptions and minimization procedures, as well as the following three different 
choices of EW scale: 
\begin{itemize}
\item[1)] the `default' scale, \be Q_{EW}^{default}=(m_{\t t_1} m_{\t t_2})^{1/2} \label{QEWSB} \ee
largely adopted in nowadays SUSY spectrum calculations.

\item[2)] the scale $Q_{EW}$ such that 
\be
V_{1-loop}(Q_{EW})=0
\label{vloop0}
\ee
as motivated previously in subsection IV B, so that the expression being minimized is $V_{tree}+\tilde{\Lambda}_{vac}$. 
This scale is to be determined dynamically according to Eq.~(\ref{RGI}). 
 
\item[3)] the scale $Q_{EW}$ such that 
\be
\tilde{\Lambda}_{vac}(Q_{EW})=0
\label{vac0}
\ee
motivated by the requirement of a vanishingly small `vacuum energy' at the EW scale.\footnote{
Note that this prescription is just a convenient choice. 
It is by no means intended as a cheap  solution to the notorious `cosmological constant problem'. 
For one thing, at the electroweak scale the true vacuum energy is not given by $\Lambda$, but by the value of 
$V_{full}$ at the minimum, which has various tree-level and loop contributions for non-vanishing $v_u, v_d$ and 
$m_{1/2}$. Then one could rather consider a scale prescription such as $V_{full} (Q)=0$. But again this is nothing
but adjusting the minimum, requiring possibly a proper choice of the boundary condition 
$\Lambda^0 = \Lambda(Q_{GUT})$, and certainly not more a solution to the `cosmological constant problem'.  
More generally, the presently measured (small and positive) cosmological constant is a very large distance 
observable, and its relation to the very short distance vacuum energy computed from a well-defined 
quantum field theory is another side of the unsolved problem.}
Eq.~(\ref{vac0}) has also to be solved iteratively, since 
a different choice of $Q_{EW}$ affects the whole spectrum.
\end{itemize}

As it turns out, these three scales are all quite different numerically, so that
the comparison of the ensuing results is expected to be a reasonably good cross-check of the
(necessarily approximate) scale invariance of our minimization results. Note moreover that the three choices
correspond to dynamically determined scales, being all non trivial functions of $m_{1/2}$ rather than fixed  
values. 
 
In Table \ref{tabpresc} we summarize, for $m_0=A_0=0, B_0=0.2 m_{1/2}$ and $\eta_0=10$, the resulting $m_{1/2}$ minima
and the corresponding values of the $Q_{EW}$ scale for these three scale prescriptions, 
using both the (A) and (C) procedures described in section \ref{procede-A-C}.
\begin{table}
\begin{center}
\caption[long]{\label{tabpresc} Values of $m_{1/2}$ minima and the corresponding values
of $Q_{EW}(m_{1/2})$, for $m_0=A_0=0, B_0=0.2 m_{1/2}, \eta_0=10$ and three 
different scales, using the two minimization procedures (A) and (C). { A conservative intrinsic numerical error of
about 1\% is to be added, taking into account uncertainties in the RGE and spectrum calculations}.   
We also indicate for comparison the value of $Q$ at the EW border of (\ref{ewsbc}).} 
\begin{tabular}{|c||c|c|c||c|}
\hline
 & $Q_{EW}=(m_{\t t_1} m_{\t t_2})^{1/2}$  & $V_{1-loop}(Q_{EW})=0$ & $\Lambda_{vac}(Q_{EW})=0$ & EW border\\ 
\hline\hline
1) procedure (A): &      &      &     &   \\
$Q_{EW}$(GeV) &   610  & 307    &  500 &  700 \\
 $m_{1/2}(min)(GeV)$ & 335 & 332 & 334 & -- \\
$\eta_{EW}$      &   1.1    & -0.6    &  0    &  \\     
\hline 
2) procedure (C): &      &      &     &   \\
$Q_{EW}$(GeV) &   544  & 277  & 430 & 580 \\
 $m_{1/2}(min)(GeV)$ & 297 & 299 & 300& -- \\
$\eta_{EW}$          & 0.6     & -1.15    &  0   &  \\     
\hline 
\end{tabular}
\end{center}
\end{table}
%
A first welcome feature is that the existence and values of the minima do not depend
much on the choice of scale, at one-loop order, as expected if RG-invariance is consistently implemented.
We have checked that the same rather generic properties are observed for other values
of the input parameters. (We comment more on those checks later in this section.) 
Another feature is that there is a definite, but rather moderate, difference 
between the results of the two minimization procedures, procedure (A) 
giving generically slightly higher values of $m_{1/2}$ than procedure (C) (by about $10\%$ here
for $m_{1/2}\sim 300 $ GeV). This confirms the discussion at the end of the previous section where
the origin of this difference was traced back to the effect of fixing the physical top-quark mass
prior to minimization in procedure (C).

But perhaps the most important feature is that there exists a somewhat narrow window for 
$\eta(Q_{EW})$ where the vacuum energy contribution
is neither too large nor too small, compensating efficiently the behavior of the tree-level and loop
contributions such as to produce non-trivial $m_{1/2}$ minima. For each value of $\eta_0$, the precise values of these 
upper and lower critical values of $\eta_{EW}$ depend on $B_0$ (mildly) and on $m_0$, and also on the choice of 
renormalization/EW scale. For example the case illustrated in Table \ref{tabpresc} leads approximately to:
\be
-1.5 \lsim \eta(Q_{EW}) \lsim 1.2\;.
\label{etacritEW}
\ee 
with the GUT boundary condition $\eta_0=10$.
Specific values of $\eta(Q_{EW})$ within this range, corresponding to different $Q_{EW}$ scales, are given in the table. 
In fact the lower and upper bounds in (\ref{etacritEW}) can be  respectively associated to the 
lowest and highest values of $Q_{EW}$ consistent with REWSB.  For instance, $\eta \gsim 1.2$ could still lead to a 
minimum of the potential in the $m_{1/2}$ direction, but would require  $Q_{EW} > 580 {\rm GeV}$, which is beyond 
the EW border (see Table \ref{tabpresc} ) so that there is no minimum in the $v_u, v_d$ direction. As for the lower 
bound $ \simeq -1.5$, although related to the same physical feature, its value is simply dictated by the lower bound 
$Q_{EW} \gsim {\cal O}(m_Z)$ which is a natural requirement. We stress that, although corresponding to the 
same value of $\eta_0$ at the GUT scale, the various $\eta(Q_{EW})$ in the range (\ref{etacritEW}) do not lie on 
one and the same trajectory of the running $\eta(Q)$. The reason is as follows: when 
the EWSB is fulfilled, the values of $Y_t(Q_{EW})$ determined from (\ref{Ythresh}) with the physical $m_{top}^{pole}$ 
imposed, amount to different and incompatible boundary conditions for $Y_t$ when varying the $Q_{EW}$ prescription.  
This leads to a modification of the running of the MSSM parameters, in particular $m_{H_u}^2(Q)$, and so to a 
modification of the values taken by the beta function in (\ref{rgevac}) implying different running $\eta(Q)$ 
trajectories for different $Q_{EW}$ prescriptions.
  
\begin{figure}[h!]
\begin{center}
\epsfig{figure=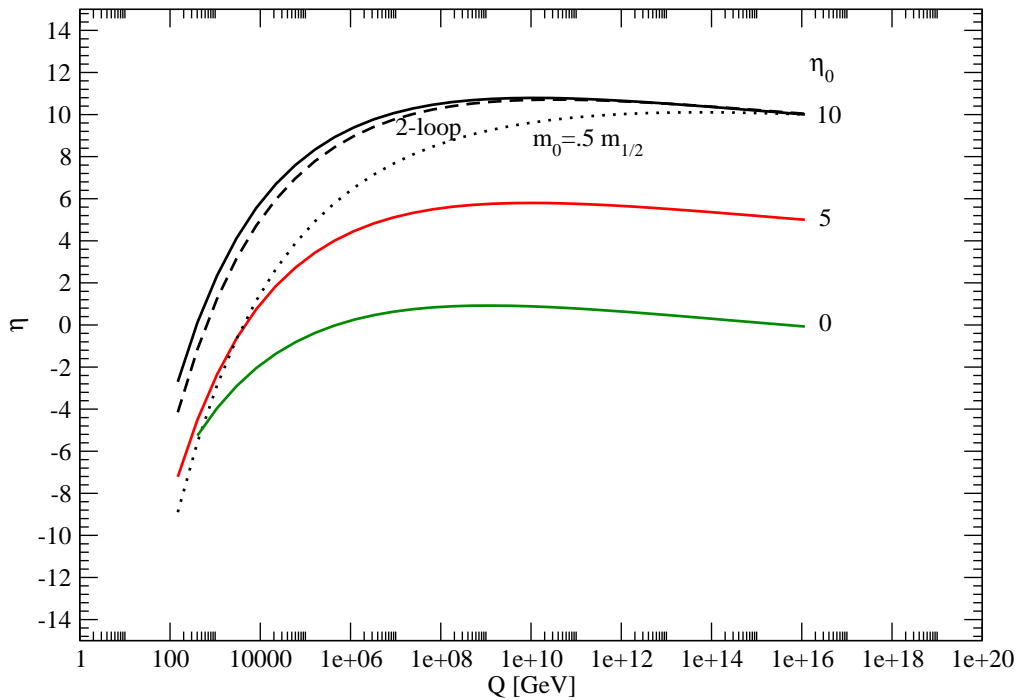,width=12cm,angle=-90}
\caption[long]{\label{etaevol} The one-loop RG evolution of the vacuum energy between the EW 
and GUT scales, for 
$m_0=A_0=0$, $B_0=.2 m_{1/2}$, $Q_{EW}=(m_{\t t_1} m_{\t t_2})^{1/2}$ and three different sets of 
$(\eta_{EW}, m_{1/2})$ as determined by the minimum of 
the potential; the corresponding $\eta_0$ values at the GUT scale are explicitly indicated.
For $\eta_0=10$, we also show the RG evolution 
including two-loop effects (dashed line) and the $m_0=.5 m_{1/2}$ case (dotted line), for comparison.
See text for further comments.}
\end{center}
\end{figure}

This point is important to keep in mind when discussing how allowed ranges 
for $\eta(Q_{EW})$ such as (\ref{etacritEW}) are mapped on allowed ranges of $\eta_0$ at the GUT scale.
We illustrate in Fig~\ref{etaevol} the connection between low and high energy $\eta$
values as dictated by the RG evolution supplemented with the $m_{top}^{pole}$ constraint,
but now for a unique EW scale prescription $Q_{EW}^{default}(m_{1/2})$ given by (\ref{QEWSB}). In this case the 
induced difference in the boundary conditions for $Y_t$ comes from the different values of $m_{1/2}$ minima. Thus 
each full-line curve in the figure corresponds simultaneously to a different $\eta_0$, a different $m_{1/2}$ minimum  
and different numerical values for the beta function in (\ref{rgevac}). It is then clear why, in contrary to what 
(\ref{rgevac}) would naively dictate, these curves are not globally shifted with respect to each other and can even 
intersect. Moreover, there is in general no simple one-to-one correspondence between the low energy 
and high energy values of $\eta$ when the physical EWSB constraints are taken into account, since
varying the $Q_{EW}$ prescription would result in a beam of trajectories for each value of $\eta_0$, which 
can overlap at low $Q$. It follows that the lower 
and upper bounds on $\eta_0$ are somewhat tricky to determine precisely, as they correspond to an envelop
deduced from the largest allowed range for $\eta(Q_{EW})$ at the EW scale. Hereafter we only give a qualitative 
discussion.
An approximate range of $\eta_0$ for which $m_{1/2}$ minima exist, is given by:
\be
0 \lsim \eta_0(Q_{GUT}) \lsim 15 
\label{etarange}
\ee
and the corresponding largest allowed range at the EW scale is found to be 
\be
-3 \lsim \eta(Q_{EW}) \lsim 1.7 \;.
\label{etarangeEW}
\ee
Actually, this EW range is obtained from the beam of RG trajectories of $\eta_0=15$ alone, 
when spanning all possible $Q_{EW}$ choices. It encompasses all other ranges corresponding to lower $\eta_0$ values. 
Fig.~\ref{etaevol} is also instructive as regards the connection between $\eta_0$ and $m_{1/2}$, given that 
whatever the precise choice of $Q_{EW}$ prescription the latter remains of the same order of magnitude as $m_{1/2}$.
Accordingly, the location of the $m_{1/2}$ minimum is quite sensitive to $\eta_0$, as illustrated
in Table \ref{tabtbh} for three representative values of $\eta_0$. 
In (\ref{etarange}) the largest $10\lsim \eta_0\lsim 15$ values correspond to $m_{1/2}\lsim 330 $ GeV approximately. 
Though it  will depend on the precise values of the other input parameters $B_0$ and $m_0$, one may anticipate
 that $\eta_0\gsim 10$ is at the verge of being excluded in a substantial part of the other mSUGRA parameter space 
by present collider (LEP,Tevatron and LHC) constraints, as we shall investigate in more details in section 
\ref{collider}. 
On the lower side of the $\eta_0$ range, $\eta_0 \lesssim 3$ in (\ref{etarange}) corresponds already 
to $m_{1/2}(min) \gsim 4$TeV, and $m_{1/2}(min)$
increases very rapidly to extremely large $m_{1/2}(\gg 1 \rm TeV)$ for lower values of $\eta_0$.
  
This behavior is rather generic and not very 
strongly dependent on $B_0$. Varying $m_0$ has a more substantial effect as illustrated on Fig.~\ref{etaevol}:
for a given $\eta_0$, increasing the ratio $m_0/m_{1/2}$ will result in larger values of possible
$m_{1/2}$ minima. Including two-loop RGE effects for the vacuum energy, induces some systematic large shifts of 
about 20\% of the $m_{1/2}$ minima found in Table~\ref{tabpresc}, but the results
remain qualitatively similar. Pushing it to an extreme, one can extrapolate this behavior 
down to $\eta_0 \simeq 0$ which would lead to $m_{1/2}$ minima of order $50$--$80$TeV,
corresponding respectively to large ($\simeq 50$) and small ($\simeq 10$) $\tan \beta$ values. All sparticles have 
then tens of TeV masses while the lightest Higgs mass remains of order $135$GeV, a not very exciting scenario
for SUSY searches at the LHC.\footnote{ It is amusing to note 
here that such a configuration, with an obviously severe electroweak fine-tuning problem, comes along with a 
vanishing vacuum energy at the GUT scale.}
We also observe that the sensitivity to the different scale choices 
tends to increase as $\eta_0$ decreases ($m_{1/2}$ increases), 
in other words the precise determination of $m_{1/2}(min)$ somewhat degrades, reflecting the increasing influence of 
ignored higher order terms.

In Table \ref{tabtbh} we also give, for each $m_{1/2}$  minimum that we found, 
the corresponding range 
of values for $\tb$ and for some of the most relevant sparticle masses, when different scale prescriptions are applied.
We indicate as well the corresponding values of $c_\mu^0$, cf. Eq.~(\ref{muscale}); the latter is entirely determined by 
the RGE and the EW constraints (\ref{muew}), (\ref{mz}) and comes out of order $1$, as expected from
the discussion following (\ref{muscalevol}). \\

\begin{table}
\begin{center}
\caption[long]{\label{tabtbh} $m_{1/2}$ minima values for the two
minimization procedures and different
$\eta_0$, with
corresponding values of $\tb$. Other input are $B_0=.2 m_{3/2}$,
$m_0=A_0=0$. The variation of $m_{1/2}$
 minima values
corresponds to the various renormalization/EW scales, similarly to
Table~\ref{tabpresc}.
We also give some of the phenomenologically most
 relevant sparticle masses ($m_{\t q min}$ designates the lightest squark
of the first two generations)}
\begin{tabular}{|c|c||c|c||c|c|c|c|c|c|}
\hline
\rm min. procedure &$\eta_0$ & $m_{1/2}(min)$ & $\tb(Q_{EW})$ & $c_\mu^0$ & $m_h$ & $m_{\chi^0_1} $ &
$m_{\t \tau}$ &$m_{\t q min}$ &$m_{\t g}$\\
\hline\hline
(A) & 10 &  332 ~-~ 335 & 6.9 ~-~ 7.1 & $\sim$ .82 & $\sim$ 111  &  132 ~-~ 133  & 125 ~-~ 126 &  692 ~-~ 698  &
785 ~-~ 791 \\
(C) & 10 &  297 ~-~ 300 & 6.9 ~-~ 7.1 & $\sim$ .8 & $\sim$ 110  &  117 ~-~ 118  &  112 ~-~ 114 & 626 ~-~ 632 &
708 ~-~ 715  \\
(A) & 8 & 550 ~-~ 570 & 7.7 ~-~ 7.9   & $\sim$ .86  & $\sim$ 115 & 227 ~-~ 235 & 202 ~-~ 209  & 1093 ~-~ 1129&
1250 ~-~ 1293\\
(C) & 8 & 490 ~-~ 505 & 7.7 ~-~ 7.9   & $\sim$ .85  & $\sim$ 114 & 200 ~-~ 207 & 180 ~-~ 185  & 984 ~-~ 1012   &
1124 ~-~ 1156  \\
(A) & 5 & 1540 ~-~ 1560 & 9.5 ~-~ 9.7 & $\sim$ .99  & $\sim$ 121 & 670 ~-~ 679 & 552 ~-~ 559 & 2789 ~-~ 2822 &
3240 ~-~ 3288 \\
(C) & 5 & 1340 ~-~ 1360 & 9.5 ~-~ 9.7 & $\sim$ .97  & $\sim$ 121 & 579 ~-~ 588 & 481 ~-~ 488 & 2457 ~-~ 2490  &
2854 ~-~ 2894 \\
\hline
\end{tabular}
\end{center}
\end{table}

\begin{figure}[t]
\begin{picture}(200,200)
\hspace{-6cm}
\begin{minipage}[t]{0.5\linewidth}
\vspace{-7cm}
\epsfig{figure=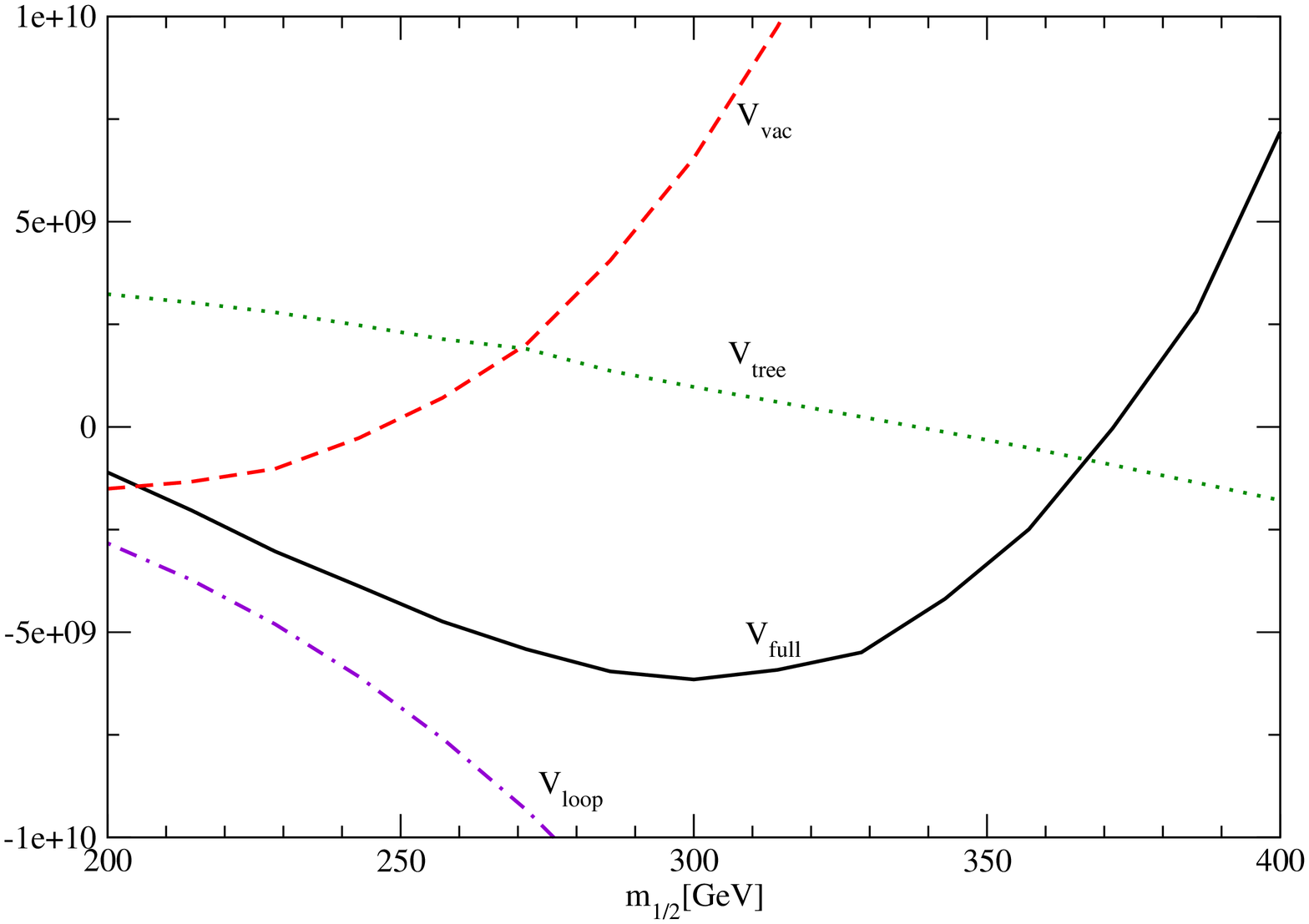,width=7.5cm,angle=-90}
\put(-145, -20){(a)}
\end{minipage}

\begin{minipage}[t]{0.5\linewidth}
\vspace{-7cm}
\epsfig{figure=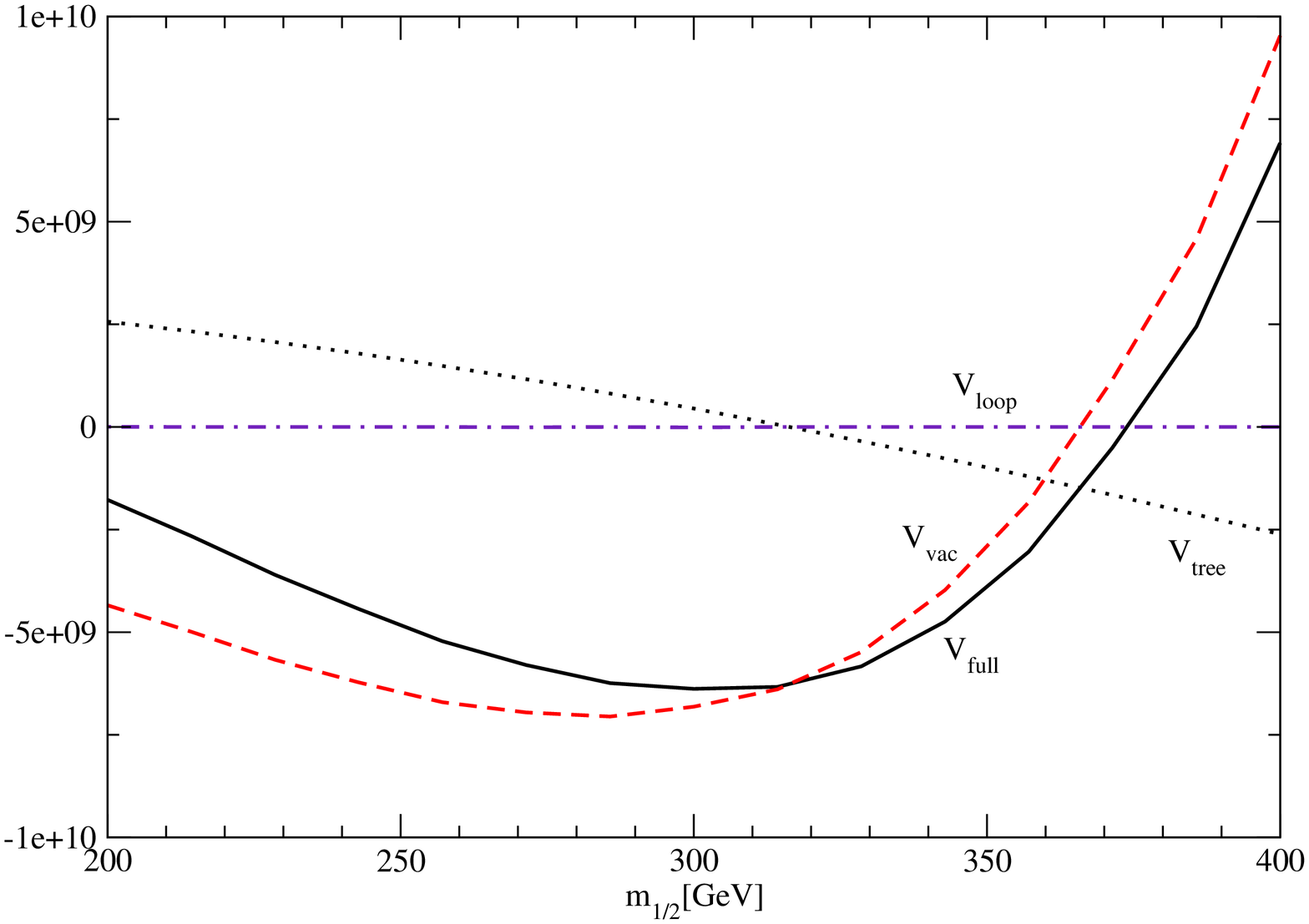,width=7.5cm,angle=-90}
\put(-145, -20){(b)}
\end{minipage}
\end{picture}
\caption[long]{\label{V2diff} relative contributions of $V_{tree}$, $\tilde{\Lambda}_{vac}$ and $V_{loop}$ to $V_{full}$ as functions of 
$m_{1/2}$ (for $m_0=A_0=0$), for two different choices of the renormalization/EW scale, using (\ref{QEWSB}) in 
figure (a) , and (\ref{vloop0}) in figure (b).}
\end{figure}

An important consequence of the analysis is that the contribution of $\tilde{\Lambda}_{vac}$ to the vacuum energy, 
necessary to define an RG-invariant effective potential making it an inseparable part of $V_{full}$,  
plays also a crucial role in the occurrence of the $m_{1/2}$ minima. In the early studies of no-scale 
minima~\cite{Cremmer:1983bf,*Ellis:1983sf,
*Ellis:1983ei} (where this contribution was not included), the parameter choices were 
mostly such that $m_{1/2}(m_{3/2})$ minima were of order $m_Z$, so that these minima resulted from a fair balance of 
tree-level and one-loop terms in the effective potential. But given the present phenomenological 
constraints on $m_{1/2}$ and the behavior of the tree and loop contributions (including a much heavier
top quark mass than assumed in the early days) the situation has now notably changed. For instance, using the scale 
prescription (\ref{QEWSB}), 
one finds that for larger $m_{1/2}$, say $m_{1/2} \gsim 300$GeV (largely favored by the latest LHC results), the 
occurrence of these minima is essentially driven
by the balance between  $\tilde{\Lambda}_{vac}$  and loop contributions. The latter have a clear $\sim m^4_{1/2}$ 
behavior, while the tree-level contributions tend to be relatively suppressed for large $m_{1/2}$; see 
Fig.~\ref{V2diff}(a) for an illustration of this case. However, one should keep in mind that a comparison of the
relative tree-level versus loop level and/or vacuum energy contributions does not make much sense physically 
since they are not separately RG-invariant. Varying the scale $Q$ shifts
parts of $V_{1-loop}$ into $V_{tree}$ or $\tilde{\Lambda}_{vac}$ and vice-versa. It is even possible to choose the 
renormalization/EW scale $Q_{EW}=Q_0(m_{1/2})$ such that one of these contributions, or some combination of them,
vanishes identically for  arbitrary $m_{1/2}$. 
In the example illustrated in Figure~\ref{V2diff}(b), all of $V_{1-loop}$ is 
absorbed in $\tilde{\Lambda}_{vac}$ and tree-level contributions modifying consistently their individual shapes 
in $m_{1/2}$. 
However, once combined, they lead to essentially the same value of $m_{1/2}$ at the minimum irrespective of 
the scale prescription, as expected from the RG-invariance of 
$V_{full}$ and $d V_{full}/d m_{1/2}$.  Comparing  Figs.~\ref{V2diff}(a) and  \ref{V2diff}(b) 
shows indeed a sufficiently good numerical stability  
 of $V_{full}$ and of the value of $m_{1/2}$ at its minimum, against a variation of the renormalization/EW scale.

The dominance of the loop and vacuum contributions with respect to the `tree-level' ones, 
as manifest for instance in Fig.~\ref{V2diff}(a) for large $m_{1/2}$ and the choice (\ref{QEWSB}) for the EW scale, 
should not by itself question the validity of the perturbative expansion. For one thing, 
(\ref{QEWSB}) is supposed to partly re-sum the dominant higher order leading logs, and is thus in some sense 
safe (notwithstanding the multi-mass scale difficulties mentioned in section \ref{reno+pot}). 
For another, the one-loop and vacuum energy relative contributions are further enhanced due to a large cancellation 
in the tree-level contribution
(specially for sufficiently large $\tb$), which makes the latter a very shallow function of $m_{1/2}$
for a large class of $Q_{EW}$ prescriptions. 
In that sense it is not so much a problem of uncontrolled higher perturbative orders, but a rather 
accidental very mild dependence of the tree-level on the relevant minimization parameter. 

Finally, the two-loop contributions to the effective potential, 
though certainly non-negligible in practice, are known to remain well under control~\cite{V2Martin, Martin:2001vx} for the rather 
moderate values of  $ m_{1/2} \lsim 1~\mbox{TeV}$ that are most phenomenologically interesting, and the perturbative 
validity is not endangered. We have included their dominant ${\cal O}(\alpha_s Y_t^2)$ contributions 
\cite{Zhang:1998bm, *Espinosa:1999zm},~\cite{V2Martin} and found that 
they stay at a reasonable level so that the resulting minima are rather stable.
For very large values of $m_{1/2}$ well beyond the 
TeV range, there may be a true perturbativity problem, so that one may not trust too much the $m_{1/2}$ minimization 
results.

\subsection{Alternative subtraction prescription} 
\label{sub-presc}

\begin{figure}[t]
\vspace{1cm}
\begin{picture}(200,200)
\hspace{-3cm}
\begin{minipage}[t]{0.5\linewidth}
\vspace{-9cm}
\epsfig{figure=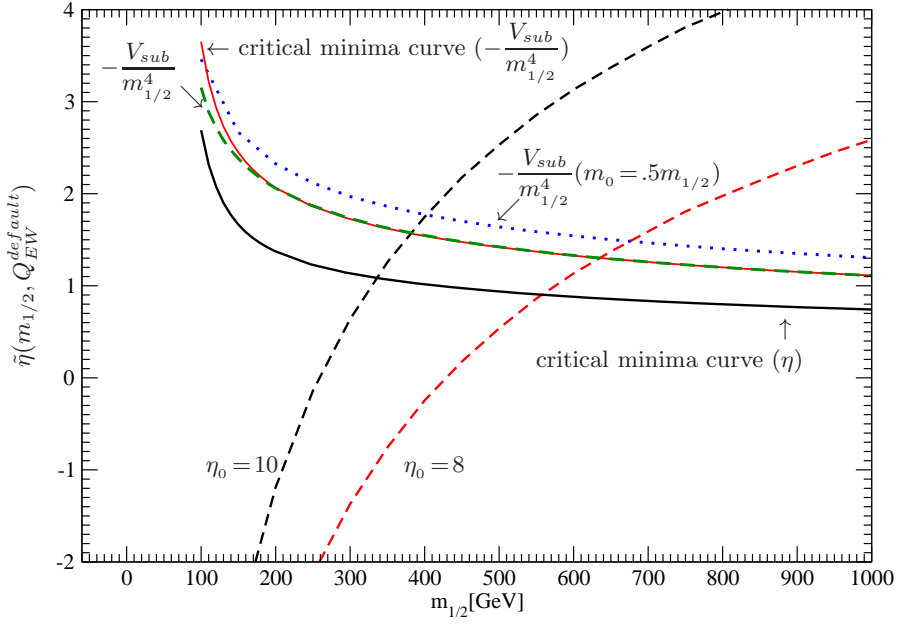,width=10.5cm, keepaspectratio=true, angle=-90}
\put(-370,-180){\makebox(0,0)[bl]{\rotatebox{90}{{$\tilde{\eta}(m_{1/2}, Q_{EW}^{default})$}}}}
\put(-295, -220){$\eta_{{}_0}\!=\!10$}
\put(-220, -220){$\eta_{{}_0}\!=\!8$}
\put(-185, -120){\makebox(0,0)[bl]{$\displaystyle { - \frac{V_{sub}}{m^{4}_{{}_{1/2}}}(m_{{}_0} \!= \!.5 m_{{}_{1/2}})}$} }
\put(-185, -123){$\swarrow$}
\put(-335, -80){\makebox(0,0)[bl]{$\displaystyle { - \frac{V_{sub}}{m^{4}_{{}_{1/2}}}} $ }}
\put(-305, -80){$\searrow$}
\put(-78, -168){$\uparrow$}
\put(-170, -180){critical minima curve $(\eta)$}
\put(-298, -62){ $\leftarrow$ critical minima curve $(\displaystyle { - \frac{V_{sub}}{m^{4}_{{}_{1/2}}}})$}
\end{minipage}
\end{picture}
\vspace{.5cm}
\caption[long]{ \label{etavssub} the $m_{1/2}$~--~$\tilde{\eta}(Q_{EW}^{default})$ correlations. 
All curves (apart from the blue dotted curve) 
correspond to $m_0=A_0=0$ and $B_0=.2 m_{1/2}$ and
$Q_{EW}^{default} \equiv (m_{\t t_1} m_{\t t_2})^{1/2}$. 
The two `critical minima' full-line curves give the locations of
the $m_{1/2}$ minima for a given $\tilde{\eta}(Q_{EW}^{default})$ as determined by procedure (A), Eq.~(\ref{nsapprox}).
(The red curve corresponds to $\tilde{\eta} \equiv -V_{sub}/m^{4}_{1/2}$ and the black curve to 
$\tilde{\eta} \equiv \eta$.) The black and red dashed-line curves give $\eta(m_{1/2}, Q=Q_{EW}^{default})$ for different
boundary $\eta_0$, as induced by the RGE. Similarly, the green dashed-line curve corresponds to the subtraction
prescription.}
\end{figure}

We have also analyzed for completeness the structure of the $m_{1/2}$ minima when  
adopting the subtraction prescription (\ref{sub}) rather than the untwiddled $\eta$ prescription. In this section
we compare the outcome of the two prescriptions for a representative set of boundary values for $B_0$, $m_0$ and
$A_0$. For the subtraction prescription, Eq.~(\ref{nsapprox}) can be recast in the more convenient form, 
\be
V_{full}(m_{1/2}) +\frac{1}{128\pi^2} \sum_n (-1)^{2n} \, [ M^4_n(m_{1/2}) 
-  M^4_n(m_{1/2}, v_u=v_d=0) \,] =0.
\label{nsapproxsub}
\ee
Since this prescription provides a special case of the $m_{1/2}$-dependent $\tilde{\eta}_0$ boundary
conditions, viz (\ref{initialsub}, \ref{solsub}), one expects it  to lead to a different shape of 
$V_{full}(m_{1/2})$ than the untwiddled $\eta$ prescription, i.e. not just a constant shift, 
but rather a different structure of the minima in the $m_{1/2}$ direction. For the same reason the comparison between 
the minimization procedures (A) and (C) of section \ref{procede-A-C}, carried out in section \ref{ns+vc} for the 
untwiddled $\eta$ prescription, does not 
necessarily hold for the subtraction prescription. In particular, as we will see, the very existence of $m_{1/2}$ 
minima may now not be necessarily guaranteed simultaneously in both  procedures (A) and (C), in contrast to what was 
found in section \ref{ns+vc}. We show in Fig.~\ref{etavssub} the various dependencies of $\tilde{\eta}$ on $m_{1/2}$ 
taken {\sl at the electroweak scale} with $Q$ fixed to $Q_{EW}^{default}$ as defined by (\ref{QEWSB}) and boundary conditions
$m_0=A_0=0, B_0=.2 m_{1/2}$. Following our previously defined conventional notations, $\tilde{\eta}$ represents here 
either $ -V_{sub}/m^{4}_{1/2}$ or the (untwiddled) $\eta$.
In order to understand better the meaning of this plot, we stress that the $m_{1/2}$-dependence in 
$\tilde{\eta}$ has three different sources---the boundary condition $\tilde{\eta}_0$---the EW scale
$Q_{EW}^{default}(m_{1/2})$---and the fixing of $m_Z$ and $m^{pole}_{top}$ leading to $\mu_{EW}(m_{1/2})$ and 
$Y_{top}(m_{1/2})$. 
The two full line curves indicated as `critical minima' on Fig.~\ref{etavssub} are actually the locations of
the $m_{1/2}$ minima as determined by procedure (A), that is where Eq.~(\ref{nsapprox}) is satisfied; the red curve
corresponds to $\tilde{\eta} \equiv -V_{sub}/m^{4}_{1/2}$ and the black curve to $\tilde{\eta} \equiv \eta$, 
taken at the electroweak scale.\footnote{Each of these two cases satisfies (\ref{nsapprox}), where, as emphasized
previously, the latter equation has been derived
{\sl before} imposing the physical mass constraints. This is encoded in the fact that $d\tilde{\eta_0}/dm_{1/2}$
appears in (\ref{nsapprox}), rather than $d\tilde{\eta}/dm_{1/2}$, so that this term is non-vanishing for
$\tilde{\eta} \equiv -V_{sub}/m^{4}_{1/2}$ but vanishes for $\tilde{\eta} \equiv \eta$. This is not in contradiction
with the fact that $\tilde{\eta}$ has in both cases a non-trivial dependence on $m_{1/2}$ but only
{\sl at the electroweak scale}, as visible on the plot.} The outcome of procedure (C) is always below those critical 
curves, due to the shift to lower $m_{1/2}$ minima by about 10\% induced mainly by the fixing of 
$m_{top}^{pole}$, as discussed 
at the end of section \ref{procede-A-C} and in section \ref{ns+vc}, Tables \ref{tabpresc}, \ref{tabtbh}. It is then 
easy to trace the structure and existence of $m_{1/2}$ minima by overlaying various curves of $-V_{sub}/m^{4}_{1/2}$ 
and $\eta$, corresponding to different boundary values of $\eta_0$ and (\ref{softbc}), 
and looking for intersections with the two critical curves. For instance the intersections between the $\eta$ critical 
curve and the set of curves with different 
$m_{1/2}$-independent values of $\eta_0$ give the values of $m_{1/2}$ and $\eta_{EW}$ at the minimum
of the potential, as obtained from procedure (A). E.g. $\eta_0=8, 10$ depicted on Fig.~\ref{etavssub}, lead to 
$m_{1/2} \simeq 550$GeV, $\eta_{EW} \simeq .9$  and $m_{1/2} \simeq 335$GeV, $\eta_{EW} \simeq 1.1$ respectively,
see also Table \ref{tabtbh}.
Varying $\eta_0$, one can easily read out from the general trend of the monotonically increasing $\eta$ curves as 
compared to the monotonically decreasing critical $\eta$ curve, 
that there are basically always intersections (i.e. existence of $m_{1/2}$ minima from procedure (A)), except for too 
large or too small values of $\eta_0$. This reproduces features similar to (\ref{etacritEW}) and (\ref{etarangeEW})
albeit here for the specific prescription $Q_{EW}^{default}$ for which the lower critical value of $\eta_{EW}$ is 
around $0.7$ for an $m_{1/2}(min)$ around $1$TeV, as can be seen from  Fig.~\ref{etavssub}. Moreover, since each 
$\eta$ curve sweeps out all the region
below the critical curve, it follows that one will always find a $m_{1/2}$ minimum solution from procedure (C)
whenever there exists one from procedure (A). This is good news as it guarantees the qualitative
equivalence of the two procedures even if they differ quantitatively. The situation is drastically different
for the subtraction prescription. The reason is that the $\tilde{\eta}$ curves of
$ -V_{sub}/m^{4}_{1/2}$ have a shape very similar  to that of the corresponding critical curve, rendering the
intersections rather scarce. One can see this in Fig.~\ref{etavssub} where the green (dashed line) curve remains
very close to the critical curve (red full line) but in fact intersects it at $m_{1/2} \simeq 210$~GeV which is thus
the outcome of procedure (A). But after intersecting, the $-V_{sub}/m^{4}_{1/2}$ does not go enough below
the critical curve to match the 10\% difference that ensures solutions from procedure (C). Indeed, we found that 
in this case the latter procedure does not yield any $m_{1/2}$ minimum for this specific $Q_{EW}$ prescription.
In fact, even for procedure (A), the minimum  is found in the relatively restricted range 
$300\,{\rm GeV}  \lesssim Q \lesssim 500\, {\rm GeV}$. This is not surprising in view of the almost parallel
red full-line and green dashed curves, which makes the intersection more sensitive to spurious scale dependence
from higher order corrections in this prescription.  
Other  $-m^{-4}_{1/2} V_{sub}$ curves with 
different boundary conditions are expected to have the same behavior (see e.g. the blue curve in Fig.~\ref{etavssub}) 
so that one expects generically that, within the subtraction prescription, procedure (C) does
not yield solutions when procedure (A) does. Thus, in contrast with the untwiddled prescription where a mere 
10\% effect on the minima is obtained between procedures (A) and (C), the subtraction prescription clearly provides an 
example where the flaws of procedure (C) become severe enough to make it qualitatively unreliable.\\

In summary, we learned from this section that:
\begin{itemize}
\item[1)] the subtraction prescription, although theoretically sound, does suffer from an increased sensitivity
to higher order effects making it in practice perhaps less reliable.

\item[2)] the subtraction prescription might seem more predictive than
the untwiddled prescription with the extra $\eta_0$ parameter, but this is an artifact of the specific subtraction
at $H_u=H_d=0$. In general one can subtract at other arbitrary values of the Higgs fields with presumably different
values of the minimization, thus recovering more freedom than naively expected.
\item[3)] although, strictly speaking, only procedure (A) is correct, the comparison between the (A) and (C) 
procedures
allowed us to assess better the impact of the various contributions (radiative corrections, physical constraints) on
the determination of the minima. For instance the effect of fixing $m_Z$ prior to minimization  in procedure
(C) becomes quickly mild for increasing $m_{1/2}$, while the radiative corrections when fixing similarly 
$m_{top}^{pole}$ lead to moderate but incompressible relative differences of order 10\%. 
\item[4)] The different choices of scale prescription we considered allowed to check 
(within numerical uncertainties) the 
expected approximate one-loop scale independence of $V_{full}$ and of the resulting $m_{1/2}$ minima.
Also each one of these choices can have its own practical benefit as stated previously, $Q_{EW}^{default}$ being the 
simplest in practice since readily implemented in most public codes. 
\end{itemize}

In the following we will rely exclusively on the untwiddled $\eta$ prescription, namely with $m_{1/2}$-independent
$\eta_0$ at the GUT scale. 
 
\section{Collider and other Phenomenological Constraints} \label{collider}

\begin{figure}[h!]
\begin{center}
\epsfig{figure=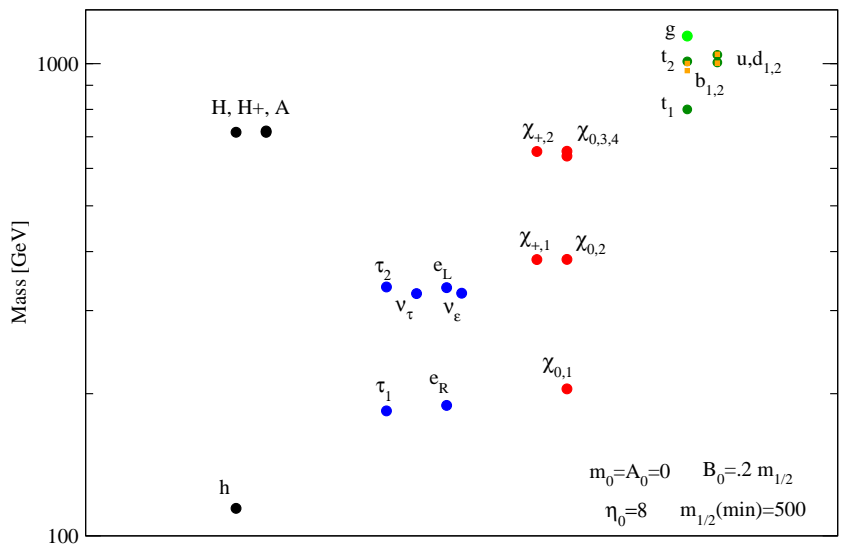,width=10cm,angle=-90}
\caption[long]{\label{spect} A typical representative spectrum with a consistent $m_{1/2}$ minimum.}
\end{center}
\end{figure}

In this section we examine the present collider and other phenomenological constraints
combined with the theoretical constraints from the requirement of non trivial $m_{1/2}$ minima. 
Before going into more details let us start with a first lap regarding the constraints on the relevant high scale 
parameters $B_0$ (or equivalently $\tb$) and $\eta_0$. Qualitatively, as $B_0$ increases (here for fixed $A_0=0$), 
the occurrence of $m_{1/2}$ minima is not  changing drastically, being mostly sensitive to $\eta_0$ values. 
But higher $B_0$ implies (for fixed $A_0=0$) higher $B_{EW}$, and correspondingly smaller $\tb$ (see the discussion
in subsection \ref{B0in} and Fig.~\ref{Btb}). Therefore the lightest Higgs mass tends to decrease, for virtually the same
$m_{1/2}$ values, with consequently a larger exclusion range in the $(m_{1/2}, \eta_0)$ parameter space. 
But the present lightest Higgs bound can be accommodated even for $m_0 = 0$, provided that the $m_{1/2}$ minima 
are sufficiently large as can be found for specific $\eta_0$ values. More precisely, taking $\eta_0 \lsim 8$ and 
$B_0 =0.2 m_{1/2}$ one finds approximately $m_h \gsim 114$ GeV in agreement with present bounds \cite{Nakamura:2010zzi}.
Note also that a consistent $B_0$ input tends to favor $5 \lsim \tb\lsim 25 $, as long as $m_{1/2}\lsim 1$ TeV. 
We give for illustration in Fig.~\ref{spect} the full sparticle spectrum
obtained for the case $\eta_0=8$, $B_0=.2m_{1/2}$ (and $m_0=A_0=0$) with a minimum at about $m_{1/2}\sim 500$ GeV.
The lightest Higgs mass is $m_h\simeq 114.3$ GeV and this spectrum passes all other present 
constraints (including $b \to s \gamma$, $g_\mu-2$, as well as the recent LHC constraints 
~\cite{Khachatryan:2011tk, *Chatrchyan:2011bz, *Collaboration:2011bj}, \cite{Aad:2011xm, *Aad:2011hh, *Aad:2011ks} ), 
except for the important fact that
the $\tilde \tau$ is the LSP. This is the case more generally in a large part of the parameter space. 
We shall discuss in the next section how to evade this dark matter issue by assuming that the gravitino
is the true LSP and the stau the NLSP. The longstanding cosmological issues related to the gravitino
including gravitino LSP as a possible
dark matter candidate \cite{Pagels:1981ke,*Weinberg:1982zq,*Nanopoulos:1983up,*Khlopov:1984pf,
*Ellis:1984eq,*Juszkiewicz:1985gg,*Ellis:1984er,*Kawasaki:1986my,*Berezinsky:1991kf},
 were also considered in the  no-scale framework~\cite{Moroi:1993mb, Ellis:1984kd}.

%

\subsection{The constrained MSSM and the LSP issue}
In standard mSUGRA, substantial parts of the parameter space  
where the lightest neutralino is the LSP and assumed to be the dark matter, can be excluded by the previously 
established collider and relic density combined constraints~\cite{Drees:1992am, *Baer:1995nc, *Ellis:1997wva,
*Ellis:1998kh, *Djouadi:2001yk,*Baer:2002gm,*Baer:2003yh,
*Chattopadhyay:2003xi, *Ellis:2003cw,*Battaglia:2003ab,
*Arnowitt:2003vw,*Ellis:2003si,*Gomez:2004eka,*Ellis:2004tc,*Belanger:2005jk}, \cite{Djouadi:2006be}.  
In such a scenario, most of these results may be roughly applied
in our case, provided that one superimposes on those constraints the specific $(m_{1/2},\tb)$
values theoretically constrained by the no-scale $B_0$ input, together with the extra
constraints on the vacuum energy via $\eta_0$. 
This nevertheless deserves a specific study and update, as we illustrate for 
a few representative cases below. 
Now in fact, for a large part of the relevant no-scale parameter space, the LSP is the charged stau,  
 which is in conflict with the requirement of an electrically neutral dark matter.
The neutralino mass grows faster with $\mhalf$ than the stau, which consequently 
tends to be lighter. Having a non vanishing $m_0$ raises the initial value of the scalars 
and delays the moment where radiative contributions raise the neutralino mass, giving more easily 
a neutralino LSP. For rather small values of $\mhalf$, say less than about 300 GeV,
 and rather low values of $\tb$, there is a window in which the neutralino and stau neutrino are 
lighter than the stau, but this part of the parameter space is largely excluded at present 
by other lower sparticle mass 
bounds from the LEP~\cite{Nakamura:2010zzi}, Tevatron\cite{D0lim}, and the latest LHC limits
~\cite{Khachatryan:2011tk, *Chatrchyan:2011bz, *Collaboration:2011bj}, \cite{Aad:2011xm, *Aad:2011hh, *Aad:2011ks}
. 
In particular both the right selectron, the lightest chargino, and the gluino recently constrained by the LHC,
can easily be too light. 

This issue can be solved by assuming that the gravitino is the true LSP, thus lighter than the stau,
so that the latter is decaying to a gravitino plus a $\tau$ lepton in the early universe. 
In our generalized no-scale framework the gravitino mass is required to be ${\cal O}(\mhalf)$ but otherwise 
essentially free.
For this alternative scenario new and specific constraints arise mainly from relic density confronted 
with WMAP results, including also in this case the very relevant 
gravitino thermal contributions to the relic density.     

We thus assume that the gauginos and the gravitino are linked in a manner that allows a light gravitino,
though not exhibiting a precise relation. Note that the gravitino does not need to be {\em very } light,
but just slightly below the stau and neutralino masses still
with $m_{3/2}/m_{1/2}\sim {\cal O}(1)$.  
On purely phenomenological ground we assume the simple 
relation (\ref{mgravmhalf}) and explore the different constraints for representative choices $0.1 \lsim c_{3/2} \lsim 1$.
More elaborate model-dependent relations are also possible~\cite{noscalerev}, in particular 
in superstring derived models. 

From the point of view of dark matter relic density constraints, assuming a gravitino LSP opens up a whole 
new area of parameters that is otherwise excluded for a neutralino LSP.
Requiring in addition non-trivial no-scale $m_{1/2}$ minima and consistent $B_0$ input, 
gives tighter constraints on $m_{1/2}$ and also on $\eta_0$ compatible values as we have illustrated above. 

In addition other more indirect phenomenological constraints on supersymmetric models, such as  
those obtained from the muon anomalous moment and the B decay observables, will be taken into account. But
indirect constraints are less drastic in general, since they can always be fulfilled by additional 
contributions or slightly modified scenarios. 
While relic density constraints are the most important from our perspective (on top of direct collider limits), 
since it is the only way to  
put constraints on the gravitino mass in this general no-scale scenario.
\subsubsection{Sparticle mass limits}
We use present collider limits on sparticle masses some of which are, however, model-dependent. For instance many 
available bounds assume a neutralino LSP, or when assuming the decay to a gravitino, that $m_{3/2}$ is very small
($m_{3/2} \lesssim 1\unit{keV}$). 
It is thus difficult to read out general bounds on the masses from the existing limits. 
For simplicity we may apply conservatively the pre-LHC limits for the colorless MSSM particles that are strictly
speaking valid only for a mSUGRA scenario with neutralino LSP~\cite{Nakamura:2010zzi}:

\begin{itemize}
\item neutralino mass: $m_{\chi_{1}^0} > 46\unit{GeV}$; 
this bound could be easily evaded, however, by relaxing the gaugino unification assumption, that we took
for simplicity but which is by no means mandatory in a no-scale framework. Indeed there are little 
theoretical constraints on the form of the gauge kinetic function which generates a non-zero 
$m_{1/2}$ value, so that it may not need to be universal~\cite{noscalerev}. 
\item chargino mass : $m_{\chi_{1}^\pm} > 104 \unit{GeV}$ for $m_{\tilde \nu} > 200$GeV; 
this lower bound can be somehow evaded  in the case of a very light sneutrino 
(which contributes to the t-channel with destructive interference). 
\item stau : $m_{\tilde{\tau}} > 86-95\unit{GeV}$, valid as long as 
($|m_{\tilde{\tau}} - m_\chi| \gsim 7 \unit{GeV}$) 
\item smuon : $m_{\tilde{\mu}} > 95 \unit{GeV}$ 

\item squarks and gauginos: 

Recently CMS~\cite{Khachatryan:2011tk, *Chatrchyan:2011bz, *Collaboration:2011bj} 
and ATLAS~\cite{Aad:2011xm, *Aad:2011hh, *Aad:2011ks} have put new exclusion 
limits on squark and gluino masses 
from the LHC run at 7 GeV center-of-mass energy with
integrated luminosity of $35 \mbox{pb}^{-1}$.  
In the mSUGRA model this translates into
lower bounds in the $m_{1/2}$--$m_0$ plane that are more severe from searches for multi-jet events than
for events containing two or more leptons in the final state. Furthermore, these constraints are not very sensitive
to $A_0$ and $\tan \beta$ as far as the latter is not very large (a regime not relevant to our case). 
These limits lead to a lower $m_{1/2}$ bound of about $300$ GeV for moderate $m_0$ values $0$--$200$ GeV. 
While finalizing this paper a very recent
ATLAS note~\cite{ATLAS-CONF-2011-086} appeared, extending the study of jet
events with missing energy
for a higher luminosity of $165 {\rm pb}^{-1}$. This appears to exclude $m_{1/2}
\lsim 450$GeV for
low $m_0$ values and $A_0=0$ $\tb=10$, thus ruling out a priori our
first benchmark study in Table \ref{tabpresc}, if applying conservatively
the limits valid for a neutralino LSP.
In any case this is not much of a problem
in so far as our different benchmarks are only theoretical examples for the
occurrence of no-scale minima.
Note that the spectrum e.g. shown in Fig. \ref{spect} is still (slightly)
above the border of present exclusions.
\end{itemize}

However, it
should be emphasized that all those constraints do not necessarily apply in a large part of the
parameter space considered here, where $m_{\tilde \tau_1} < m_{\chi^0_1}$, with a  
gravitino LSP and $\tilde \tau_1$ NLSP. Indeed, for
the range of gravitino masses that we will consider, the stau NLSP is sufficiently
long-lived not to produce a signal with missing energy in the detector, so that limits obtained from searches for such
signals do not apply.  Very recent results by ATLAS~\cite{ATLAS_stau_NLSP} dedicated to searches for long charged 
tracks give bounds on stable staus NLSP, $m_{\tilde \tau} > 136$GeV, for a representative
gauge-mediated SUSY-breaking (GMSB) model. Since this limit concerns a different model, its implication
on no-scale would necessitate a detailed study. Nevertheless, naively fitting this ${\tilde \tau}$ mass
bound for $m_0=A_0=0$ and an appropriate value of $\tan \beta (B_0)$, leads to a quite similar
MSSM spectrum  corresponding approximately to $m_{1/2}\gsim 360$GeV.

\subsubsection{Higgs boson mass}
Strong limits on the lightest Higgs mass are obtained from LEP and the Tevatron~\cite{Nakamura:2010zzi}. 
In fact, apart from a very small window for relatively small $m_{1/2}$, which is  
now essentially excluded by Tevatron and the above mentioned very recent LHC limits, 
in most of the no-scale parameter space we are in the Higgs-decoupling scenario 
$m_A \gg m_h$ such that the standard-model like
limit on $m_h$ essentially applies. Thus in most of the parameter space we are considering here
the LEP II constraint \cite{LEPsusy, Nakamura:2010zzi} should hold :
\begin{equation}
m_h \geq 114.4 \unit{GeV}
\label{mhbound}
\end{equation}
where the limit depends on $\tan \beta$ to some extent. 
This implies rather stringent lower bounds
on $m_{1/2}$ prior to LHC squark/gluino mass limits, specially for low
$\tb$ (i.e. large $B_0$).
Even for $B_0=0$, requiring Eq.~(\ref{mhbound}) corresponds to $m_{1/2}
\gsim 400$GeV e.g. for $m_0=A_0=0$.
Otherwise, the most conservative limit is $m_h \geq 92.8 \unit{GeV}$ 
\cite{Nakamura:2010zzi}. 
To take into account theoretical uncertainties, 
we will use a conservative limit allowing for about 3-4 GeV of theoretical
uncertainties as is customary.

\subsubsection{Muon anomalous moment}
Supersymmetric particles can contribute at the loop level to the 
muon anomalous moment $a_\mu =(g-2)_\mu/2$ \cite{Lopez:1993vi, 
Chattopadhyay:1995ae, Moroi:1995yh, Carena:1996qa, Goto:1999mk, Drees:2000bs}. 
It could explain the deviation measured \cite{Bennett:2006fi}:
\be
\Delta a_\mu = a_{\mu}^{exp} - a_{\mu}^{th} = (22\pm 10 \;\mbox{to}\; 26\pm 9)\times 10^{-10}
\label{gmu}
\ee
which is about 2 to 3 standard deviations
 from the theoretical prediction of the standard
model (for a review see \cite{g-2_th_rev}).

The contributions to $a_\mu$ in no-scale models is of course a particular case
of mSUGRA general contributions. These can be quite important when slepton masses are light. 
Corrections in a general MSSM come dominantly from loops with a chargino and a muon sneutrino and loops with a
neutralino and a smuon \cite{Moroi:1995yh}. The MSSM correction is proportional to $\tan \beta$ and 
its sign follows the sign of $\mu$, thus favoring a positive sign for the $\mu$ parameter in view
of (\ref{gmu}).
For not too large $\tb$ values and choosing $\mu>0$, one can accomodate the preferred range
(\ref{gmu}) for some regions of the mSUGRA parameter values compatible 
with no-scale models (provided of course that $m_{1/2}$ is not too large), 
 as we shall illustrate in the next section.
%
\subsubsection{${\rm b\rightarrow s \, \gamma}$}
Another largely studied probe for supersymmetry is B-meson physics, in
particular the
decay $b\rightarrow s\gamma$ which has been extensively measured with good
accuracy and is theoretically
well under control. Indeed,
theoretical calculations from standard model contributions
have been now performed at the next-to-next-to-leading logarithm (NNLL)
order~\cite{bsg_sm_nlo}, including also nonperturbative
corrections~\cite{bsg_sm_np}. Confronted
with recent experimental measurements~\cite{bsg_exp, Nakamura:2010zzi}:
\be
Br(b\rightarrow s\gamma) = (3.55 \pm 0.24 \pm 0.09) \times 10^{-4}
\label{bsg_exp}
\ee
it results in a discrepancy with the standard model slightly above one
standard deviation, therefore
potentially very constraining for new physics.   
The possible contributions from the MSSM are
dominated by one-loop effects from chargino plus stops, and top plus
charged
Higgses. NLL SUSY-QCD corrections have also been
calculated~\cite{Degrassi:2000qf, *bsg_susy_nlo,*bsg_susy_nlo2,*bsg_susy_nlo3}.
In mSUGRA, contributions can become sizeable for relatively large $\tb$
and
sufficiently small $m_{1/2}$, which is not much favored in no-scale
scenarios, due to the $m_{1/2}-B_{EW}-\tb$
correlations inducing rather moderate $\tb \lsim 10$ values for not too
large $m_{1/2}$ (see the discussion
in subsection \ref{B0in}).   
We therefore anticipate that $b\to s \gamma$ constraints are relatively
marginal for a large part
of our no-scale inspired scenarios, specially for the strict no-scale with
$m_0=0$.
 
In practice we have used in our analysis the bounds (\ref{bsg_exp})
conservatively augmented by
theoretical uncertainties as quoted e.g. in ~\cite{bsg_sm_nlo, bsg_sm_np}.
The precise limits are, however, not very crucial
for our analysis, since as we will illustrate they give anyway (mild)
constraints only for rather 
small $m_{1/2} \lsim 250-300$ GeV which, for
the range of $\tb$ values under consideration, are largely superseded
both by the lightest
Higgs mass limit from LEP and by the Tevatron and the recent LHC limits on
$m_{1/2}$. (In fact
for $A_0=0$ and $m_0 \ne 0$ moderately small, the lightest Higgs mass
bound $m_h > 114.4$ GeV
generally supersedes the $b\rightarrow s \gamma$ constraints even
for large $\tb$ values up to $\tb \simeq 40-50$, anyway unreachable in the
no-scale framework. 
$b\rightarrow s \gamma$ can be much more constraining if $-A_0$ is large
enough, $|A_0| \sim 1-2$TeV
such that the stops could be light enough,
see e.g the discussion in \cite{Djouadi:2006be}).

\subsubsection{Dark matter relic density}
The lightest neutralino as a candidate for dark matter has been extensively studied in 
many scenarios~\cite{Drees:1992am, *Baer:1995nc, *Ellis:1997wva, *Ellis:1998kh, *Djouadi:2001yk,*Baer:2002gm,
*Baer:2003yh,*Chattopadhyay:2003xi, *Ellis:2003cw,*Battaglia:2003ab,
*Arnowitt:2003vw,*Ellis:2003si,*Gomez:2004eka,*Ellis:2004tc, *Belanger:2005jk}, \cite{Djouadi:2006be}. 
The part of mSUGRA parameter space giving a stau LSP should be normally excluded.
This has been one argument 
advocating against the viability of the strict no-scale scenarios. But if one considers a gravitino 
lighter than the stau then this part of the parameter space regains interest. 
The gravitino dark matter candidate has also been quite studied 
in the past decade. In the case of gauge mediation supersymmetry breaking scenarios, this particle is naturally the LSP 
and was considered both for cosmological issues and in colliders signatures \cite{Ambrosanio:2000ik, Baltz:2001rq, 
Fujii:2002fv, Kawagoe:2003jv, Lemoine:2005hu, Jedamzik:2005ir, Bailly:2008yy}. The gravitino can also be the LSP and a 
very interesting dark matter candidate in the context of mSUGRA scenarios \cite{Feng:2003xh, Feng:2003uy, Ellis:2003dn, 
Feng:2004zu, Feng:2004mt, Roszkowski:2004jd, Cerdeno:2005eu, Steffen:2006hw, Kawasaki:2007xb, Pradler:2007is, 
Kawasaki:2008qe, Bailly:2008yy, Bailly:2009pe, Bailly:2010hh}. 
 
In the present analysis we will illustrate a few scenarios
for the most representative no-scale cases as studied above. 
A more complete study of the constraints obtained for the full mSUGRA parameter space
will be done in a forthcoming analysis, where we also consider in detail some implications
and constraints from Big Bang Nucleosynthesis on such LSP gravitino in no-scale scenario.

For the relic density, we use micrOMEGAs 2.0 \cite{Belanger:2006is}, to compute the relic density
 of the neutralino or stau MSSM LSP. For scenarios with a gravitino being the real LSP 
(neutralino or stau being the NLSP), all supersymmetric particles decay to the
 NLSP well before the latter has decayed to a gravitino, because all interactions to the gravitino are suppressed by 
the Planck mass. We first compute the relic density $\Omega_{\rm NLSP}h^2$ the NLSP would have if it did not decay to 
the gravitino. Then 
assuming that each NLSP with mass $m_{\rm NLSP}$ decays to one gravitino, leads to the non-thermal contribution to the 
gravitino relic density
\begin{eqnarray}
\Omega_{3/2}^{\rm NTP}h^2={m_{3/2} \over m_{\rm NLSP}} \Omega_{\rm NLSP}h^2
\label{omntp}
\end{eqnarray}
with $h=0.73^{+0.04}_{-0.03}$ the Hubble constant in units of $100\unit{km Mpc}^{-1} \unit{s}^{-1}$.

The gravitino can also be produced in scattering processes during reheating after
 inflation \cite{Moroi:1993mb, Bolz:2000fu, Pradler:2006qh, Pradler:2006hh, Rychkov:2007uq}. 
Following \cite{Pradler:2006qh, Pradler:2006hh}, the resulting gravitino yield from thermal production is
controlled by the reheat temperature $T_R$ as follows,
\begin{eqnarray}
Y_{3/2}^{\rm TP}(T\ll T_R) = \sum_{i=1}^{3} y_i g_i^2(T_R) \left( 1+\frac{M_i^{2}(T_R)}{3m_{3/2}^2}\right) 
\ln \pfrac{k_i}{g_i(T_R)}\pfrac{T_R}{10^{10} \unit{GeV}} \label{TPyield}
\end{eqnarray}
where i sums over gauge groups,  
$y_i/10^{-12}=(0.653, 1.604, 4.276)$, $k_i=(1.266, 1.312, 1.271)$  and gauge couplings and gaugino masses are 
calculated using the one-loop RGE. Assuming standard thermal history without release of entropy, the gravitino relic 
density from thermal production is
\begin{eqnarray}
\Omega_{3/2}^{\rm TP}h^2 = m_{3/2} Y_{3/2}^{\rm TP}(T_0)s(T_0)h^2/\rho_c \label{TPrelic}
\end{eqnarray}
with $\rho_c/[s(T_0)h^2]=3.6\times10^{-9} \unit{GeV}$ and $T_0$ the background temperature.

Comparing the total gravitino relic density
\begin{eqnarray}
\Omega_{3/2}h^2=\Omega_{3/2}^{\rm TP}h^2+\Omega_{3/2}^{\rm NTP}h^2
\label{omtot}
\end{eqnarray}
to the one inferred from the measurements of the CMB anisotropies will constrain $T_R$ and the MSSM parameter space. 
The 3-year WMAP satellite survey has given 
at $3\sigma$ confidence level \cite{Spergel:2006hy}
\begin{eqnarray}
\Omega_{\rm DM}^{3\sigma} h^2 = 0.105^{+0.021}_{-0.030}
\label{wmap}
\end{eqnarray}
\subsection{Combined Constraints}

We can now combine both the theoretical no-scale constraints (i.e., the existence of non-trivial $m_{1/2}$ minima
and the $B_0-\tb$ relationship) and phenomenological
 constraints on $\mhalf$ and other parameters from direct and indirect search limits at colliders and from 
other observables.
LEP, Tevatron, and latest LHC constraints tell us that rather  
low $\mhalf$  are now essentially excluded, also to avoid the 114 GeV lightest Higgs 
limit and other indirect phenomenological and theoretical exclusions.
WMAP will also give stringent constraints excluding essentially all low $m_0$ values 
(in particular the pure no-scale $m_0=0$ case) if the neutralino is the true LSP, because the corresponding 
relic density comes out below the observational bound (\ref{wmap}).

\subsubsection{Collider and other phenomenological constraints}
\begin{figure}[h!]
\begin{center}
\epsfig{figure=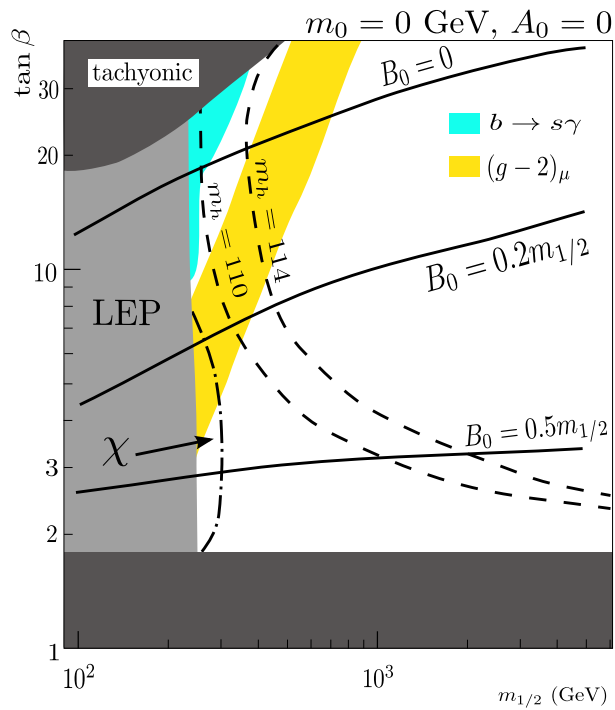,width=8cm}
\caption[long]{\label{consm00} Direct collider and other indirect constraints in the $(m_{1/2},\tb)$ plane 
for $m_0=A_0=0$. The lines for three different representative $B_0$ input  give  the $m_{1/2}$ -- $\tb$ 
no-scale correlations.
Dark and light grey colors indicate respectively the areas excluded by inconsistent EWSB (`tachyonic'), and by 
direct sparticle mass limits from LEP and Tevatron.
The light blue zone is excluded by $b\to s\gamma$ constraints (\ref{bsg_exp}), while the yellow zone
corresponds to values falling into the measured $g_\mu-2$ range of (\ref{gmu}). The dashed lines give
the lightest Higgs mass limits and the dot-dashed line the $m_{\tilde \tau} =m_{\tilde{N}_1}$ border. 
}
\end{center}
\end{figure}

In Fig.~\ref{consm00} we give the present constraints from direct sparticle search limits and
the low energy constraints from (\ref{bsg_exp}), (\ref{gmu}),  
in the $(m_{1/2},\tb)$ plane for $m_0=A_0=0$, most relevant to the true no-scale scenario. 
We do not put explicitly the above mentioned latest LHC exclusions~\cite{Khachatryan:2011tk, Aad:2011xm} 
on this and other subsequent plots, since these are anyway debatable given that in most of our scenarios
the neutralino is not the LSP as discussed previously.  
A number of conclusions may be easily drawn from this figure:
\begin{itemize}
 \item Due to the $B_0$ no-scale input, one obtains for each $B_0$ value specific
$m_{1/2},\tb$ correlations. In particular for the strict no-scale (\ref{strict}) the values of $\tb$
are restricted to be $\gsim 20$ when taking into account other constraints.
This can be consistent with the Higgs mass lower bound of $\sim 114$ GeV and falls into the preferred
$b\to s\gamma$ range, but in all this region of parameter space the stau is the MSSM LSP as indicated,
so that a gravitino true LSP becomes a very appealing scenario. 
\item For larger $B_0 $ values, there can be regions where the neutralino is again the LSP,
typically for $B_0\gsim .15 m_{1/2}$ and sufficiently small $m_{1/2}$, see the figure. But
this is generally not compatible with the light Higgs mass limit, even when allowing a large
theoretical uncertainty. In principle there could be a tiny region for such $B_0\sim 0.2-0.5 m_{1/2}$ 
values, where for sufficiently small $m_{1/2}$ one is no longer in the decoupling limit, 
i.e such that $m_A$ is light enough and the bound in (\ref{mhbound}) no longer applies. However
in that case the very recent direct limits from the LHC~\cite{Khachatryan:2011tk, *Chatrchyan:2011bz, 
*Collaboration:2011bj}, \cite{Aad:2011xm, *Aad:2011hh, *Aad:2011ks} exclude virtually all of this small corner.  
\item The requirement of non-trivial $m_{1/2}$ 
minima leads to constraints e.g. in the plane $(B_0, \eta_0)$ or
equivalently $(\tb, \eta_0)$ for $m_0=0$, or more general ones for $m_0\ne 0$, that we do not give
explicitly. Suffice it to say that for any $\eta_0$ in the range (\ref{etarange}) one can find $m_{1/2}$
minima, but present lower bounds on $m_{1/2}$ exclude accordingly $\eta_0 \gsim 8$--$10$, approximately. 
The most phenomenologically interesting range, obtained for not too large $m_{1/2} \lsim 1$ TeV, 
corresponds to $5 \lsim \eta_0 \lsim 8$, while for smaller $\eta_0$ the corresponding $m_{1/2}$ minima
increase very fast as discussed before.
\end{itemize}

\subsubsection{Relic density constraints and gravitino LSP}
\begin{figure}[h!]
\begin{center}
\epsfig{figure=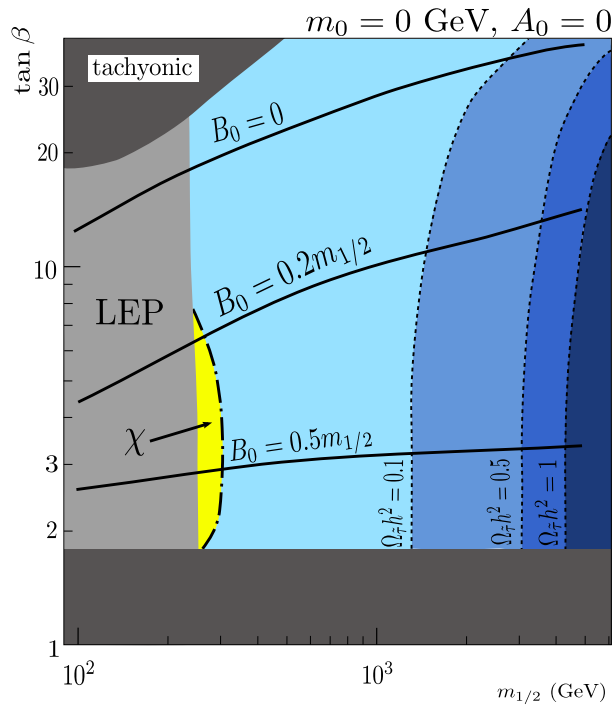,width=8cm}
\caption[long]{\label{DMm00} $\tilde \tau$ relic density values at decoupling, in the $(m_{1/2},\tb)$ plane
 with $m_0=A_0=0$. In all blue regions $\tilde \tau$ is the MSSM LSP. Different levels of blue correspond to 
different values of the $\tilde \tau$ abundance, as indicated; in the yellow region $\tilde N_1$ becomes
ligher than $\tilde \tau$ but with very small relic abundance (see text for more comments). 
other captions as in Fig.~\ref{consm00}. }
\end{center}
\end{figure}

\begin{figure}[h!]
\begin{center}
\epsfig{figure=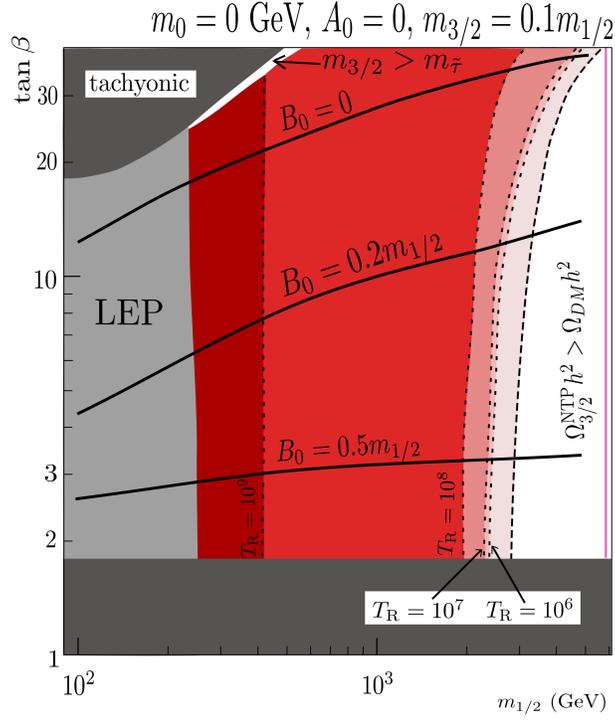,width=8cm}
\caption[long]{\label{DMGm00} Gravitino DM relic density values consistent with the 
WMAP constraints (\ref{wmap}) in the $(m_{1/2},\tb)$ plane, assuming $m_{3/2}=0.1 m_{1/2}$ and
 $m_0=A_0=0$.
The different levels of red correspond to different reheat
temperature values. The small region where the gravitino is not the LSP is also indicated; 
other captions as in Fig.~\ref{consm00}.
}
\end{center}
\end{figure}

\begin{figure}[h!]
\begin{center}
\epsfig{figure=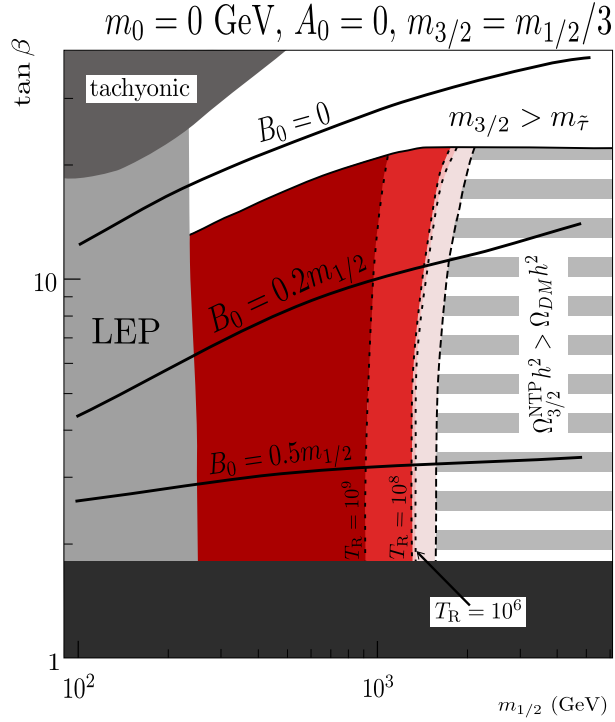,width=8cm}
\caption[long]{\label{DMGm00m32_2} Same as Fig~\ref{DMGm00} but for $m_{3/2}=\frac{1}{3} m_{1/2}$. 
}
\end{center}
\end{figure}

In Fig.~\ref{DMm00} we show the relic density values in the $(m_{1/2},\tb)$ plane corresponding
to Fig.~\ref{consm00}, calculated for $\tilde \tau$ MSSM LSP after it has decoupled from the thermal bath. 
We also show, in yellow, the small region 
where $\tilde N_1$ becomes the MSSM LSP. In the latter region the $\tilde N_1$ abundance remains too small to be
consistent with WMAP, thus excluding $\tilde N_1$ as a dark matter candidate but still allowing it as an NLSP 
with, in this case, a rather small non-thermal contribution to the gravitino relic density.
The relic density values obtained in the (largely dominant) 
region where the $\tilde \tau$ is 
the MSSM LSP only make sense if the gravitino is the true LSP, and will lead in some parts of the parameter space
 to a substantial non-thermal contribution (\ref{omntp}) to the total gravitino relic density.
We illustrate the gravitino total relic density, obtained from (\ref{omntp} --\ref{omtot}), 
in Fig.~\ref{DMGm00} for $m_{3/2}=0.1 m_{1/2}$ and in 
Fig.~\ref{DMGm00m32_2} for a higher $m_{3/2}/m_{1/2}$
ratio, for different values of the reheating temperature $T_R$. 
The first value is such that the gravitino is the true LSP in most of the parameter space,
while in the second case the uncacceptable region where $\mgrav >m_{\tilde \tau}$ is enlarged for 
 large $\tb$. Now one can see that
it is easy to recover consistency with the WMAP relic density constraint in a large part 
of the parameter space, provided that $T_R$ is sufficiently large, $T_R\gsim 10^6$ GeV. 
There is not much qualitative differences for the two illustrated $\mgrav$ masses. 
In fact we observe that a phenomenologically most interesting case for potential 
early discovery at the LHC, namely for not too large $m_{1/2}\gsim 400-500$, and 
consistency with the $g_\mu-2$, $b \to s \gamma$ and WMAP
constraints, implies a relatively large $T_R\gsim 10^8$--$10^9$ GeV, the darker red region on the figure.
A large $T_R$ is also welcome by other independent issues such as thermal leptogenesis scenarios\cite{lepgen}. 
Indeed, a comparison of Fig~\ref{consm00} and Fig.~\ref{DMGm00} shows the interesting fact that
a part of the strict no-scale model $B_0=m_0=A_0=0$ is not excluded: there is a range,
for $m_{1/2}\sim 400-800$ GeV, $\tb\sim 20-25$ compatible
with $m_h \gsim 114$ GeV, the $g_\mu-2$ deviation (Fig.~\ref{consm00}) and other constraints, 
provided the reheating temperature is $10^8-10^9$ GeV.  

\subsubsection{Generalized no-scale scenarios}
\begin{figure}[h!]
\begin{center}
\epsfig{figure=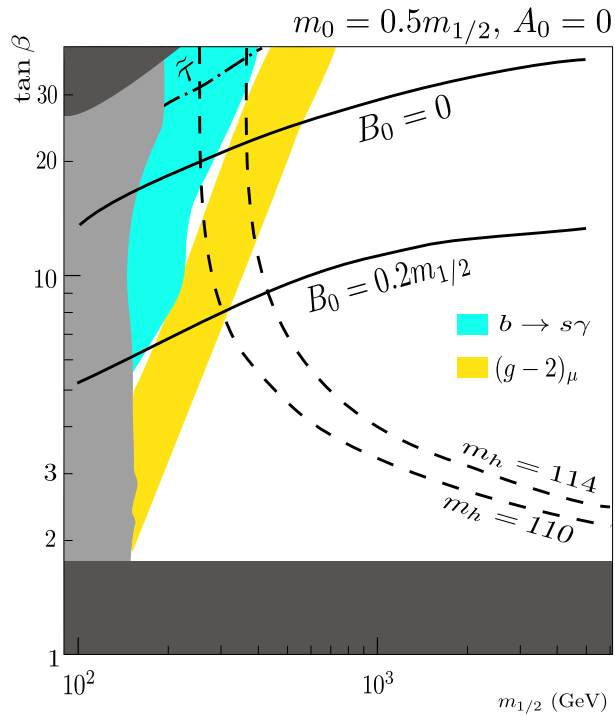,width=8cm}
\caption[long]{\label{consm05} Direct collider and other indirect constraints in the $(m_{1/2},\tb)$ plane 
for $m_0=.5 m_{1/2}$, $A_0=0$. See Fig.~\ref{consm00} for captions. }
\end{center}
\end{figure}

\begin{figure}[h!]
\begin{center}
\epsfig{figure=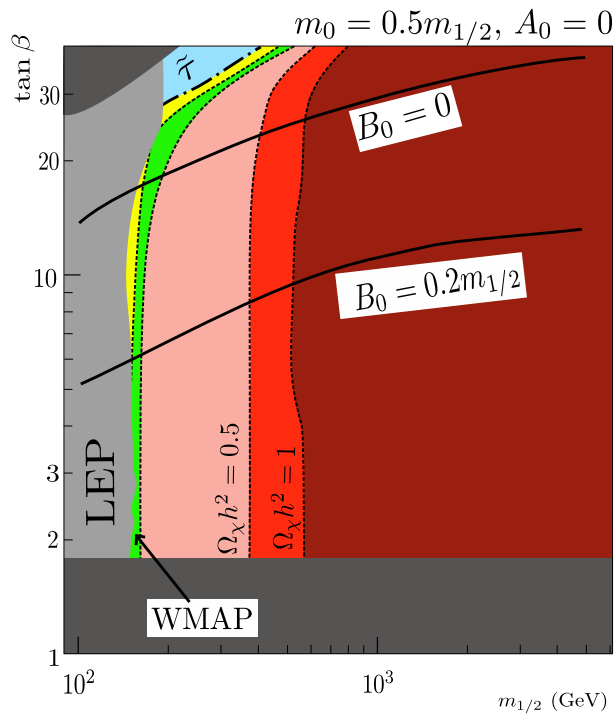,width=8cm}
\caption[long]{\label{DMm05} Non thermal relic density in the $(m_{1/2},\tb)$ plane 
for $m_0=.5 m_{1/2}$, $A_0=0$. Same captions as in Fig.~\ref{DMm00}. }
\end{center}
\end{figure}

\begin{figure}[h!]
\begin{center}
\epsfig{figure=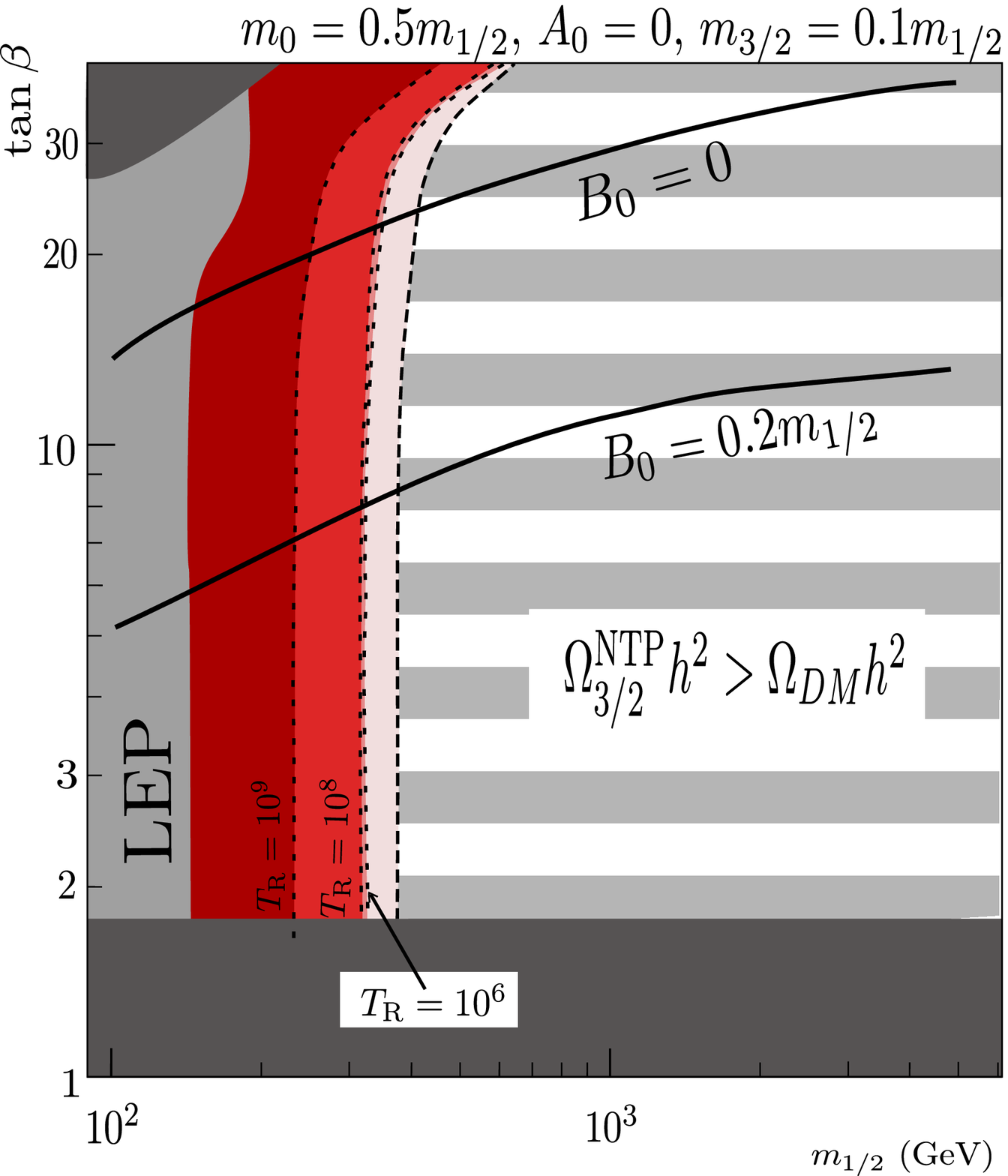,width=8cm}
\caption[long]{\label{DMGm05} Dark matter relic density values consistent with WMAP constraints(\ref{wmap})
 in the $(m_{1/2},\tb)$ plane, 
calculated assuming a gravitino $\tilde G$ LSP in most of the parameter space, or otherwise indicated. 
$m_0=.5 m_{1/2}$, $A_0=0$ and 
other captions as in Fig.~\ref{consm00}. The different levels of red corresponds to different reheating
temperatures, as indicated. }
\end{center}
\end{figure}

In this subsection we consider one typical example illustrating more general cases with $A_0, m_0 \ne 0$.
Since $A_0$ has little effect on the existence of non-trivial no-scale minima, we illustrate for 
simplicity here only the case $A_0=0$. 
In Fig~\ref{consm05} we show the results of the same analysis as before but for $m_0=.5 \mhalf$. An important 
difference with the $m_0=0$ case in Fig~\ref{consm00} 
is a larger region excluded by $b\to s \gamma$, which includes a substantial part
of the $B_0=0$ line, for rather small $m_{1/2}$.  Now coming to the relic density constraints on 
Fig~\ref{DMm05} there exists an interesting region (in green) where the {\em neutralino}, if it is the true LSP,
is compatible with WMAP. Note however that this region is almost entirely excluded
by $b\to s\gamma$ for the $B_0 =0$ line as can be seen on Fig.~ \ref{consm05} and also basically excluded
by the latest LHC limits on $m_{1/2}$. But one can easily find appropriate $m_0$ and
$B_0$ values such that the $b\to s \gamma$ and LHC constraints are compatible with a neutralino LSP. 
A full scan of the parameter space will be explored elsewhere. 
Alternatively, the gravitino LSP case with its relic density is illustrated in 
Fig~\ref{DMGm05}, where the main difference with the $m_0=0$ pure no-scale case is that the consistent
region with $\Omega_{3/2}^{\rm NTP}h^2 < \Omega_{DM} h^2$ is shrinked to much smaller $\mhalf$ values
for which there is accordingly a tension with the latest LHC $m_{1/2}$ lower limits. 
 For 
sufficiently small (but not yet all excluded) $\mhalf$, one can have the right relic density
with a high reheating temperature almost independently of $\tb$ similarly to $m_0=0$.   
%

\section{Conclusions}
We have re-examined generalized no-scale supergravity-inspired scenarios, 
 in which the gravitino mass
and all other connected soft supersymmetry-breaking parameters can be dynamically
determined  through radiative corrections, triggering a non-trivial minimum of the RG-improved
potential. For representative 
high scale boundary conditions on the minimal supergravity model, we have examined critically the theoretical 
and phenomenological viability of such a mechanism in view of up-to-date calculations of the low energy 
supersymmetric spectrum, taking into account all important one-loop radiative corrections. 
We also have investigated the impact of different prescriptions and possible 
variants of the minimization procedure, paying attention to the extra $m_{1/2}$-dependence,
genuine or fake, induced by the implementation of physical mass constraints with 
various sources of radiative corrections.

We emphasize the importance of using
a RG-invariant effective potential including consistenly a scale dependent vacuum energy contribution. 
We find that the occurrence of phenomenologically interesting minima restrict the vacuum energy to lie
within a rather restricted range at the EW scale, translating into a corresponding 
restricted range $\Lambda^0_{vac}\sim (3-10)
m^4_{1/2}$, at the GUT scale, when taking into account present LHC and other phenomenological constraints.  
The main practical consequences for phenomenology is to provide additional
 constraints on top of standard
mSUGRA parameter constraints, due to the tight connexion between $\eta_0$ and 
non trivial $m_{1/2}$ minima, as well the $m_{1/2}-B_0-\tb$ correlations from $B_0$ input within the no-scale
framework.
Allowed regions are very restricted when considering the strict no-scale boundary conditions with
$m_0=A_0=0$, providing theoretical exclusion domain prior to any additional 
experimental constraints.

Concerning the dark matter relic density, a considerably enlarged allowed region of the mSUGRA
parameter space can be obtained 
provided one assumes the gravitino to be the true LSP, accounting for the observed relic density
with important thermal contributions.
Perhaps of particular interest is the fact the strict no-scale model $B_0=m_0=A_0=0$ is not 
excluded by present LHC and other experimental constraints; 
in particular there is a range for $m_{1/2}\sim 400-800$ GeV, $\tb\sim 20-25$ compatible
with $m_h \gsim 114$ GeV, the $g_\mu-2$ deviation (Fig.~\ref{consm00}), provided the reheating temperature is 
$10^8-10^9$ GeV as illustrated in Fig.~\ref{DMGm00}. Incidentally
this is rather close to the $(m_{1/2}, \tb)$ range also preferred in recent analysis of the flipped-$SU(5)$
no-scale scenario~\cite{Li:2010ws,*Li:2011dw}. A direct comparison of our results with the ones in these papers is 
however limited, since the flipped $SU(5)$ model is quite different, with modified RGEs affecting in particular the runnings 
in the gaugino sector, allowing for a neutralino LSP even for larger $m_{1/2}$ than found in our study.\\

Even if departing slightly from the original no-scale scenarios, 
the idea of dynamically fixing the soft breaking masses from extra minimization at the EW scale remains very
attractive, even more so as it emphasizes the role of the vacuum energy being crucial  
for the occurrence of non-trivial no-scale minima. Any future experimental determination or exclusion of
$m_{1/2}$ interpreted within no-scale supergravity framework will thus help pinpoint information related to
the vacuum energy contribution at the EW and possibly at the GUT scale.

\bibliography{biblio-FIN}

\ifx\mcitethebibliography\mciteundefinedmacro
\PackageError{unsrtM.bst}{mciteplus.sty has not been loaded}
{This bibstyle requires the use of the mciteplus package.}\fi
\begin{mcitethebibliography}{100}

\bibitem{Witten:1981nf}
E.~Witten,
\newblock Nucl.Phys. {\bf B188}, 513 (1981)\relax
\mciteBstWouldAddEndPuncttrue
\mciteSetBstMidEndSepPunct{\mcitedefaultmidpunct}
{\mcitedefaultendpunct}{\mcitedefaultseppunct}\relax
\EndOfBibitem
\bibitem{Sakai:1981gr}
N.~Sakai,
\newblock Z.Phys. {\bf C11}, 153 (1981)\relax
\mciteBstWouldAddEndPuncttrue
\mciteSetBstMidEndSepPunct{\mcitedefaultmidpunct}
{\mcitedefaultendpunct}{\mcitedefaultseppunct}\relax
\EndOfBibitem
\bibitem{Dimopoulos:1981zb}
S.~Dimopoulos and H.~Georgi,
\newblock Nucl.Phys. {\bf B193}, 150 (1981)\relax
\mciteBstWouldAddEndPuncttrue
\mciteSetBstMidEndSepPunct{\mcitedefaultmidpunct}
{\mcitedefaultendpunct}{\mcitedefaultseppunct}\relax
\EndOfBibitem
\bibitem{Kaul:1981hi}
R.~K. Kaul and P.~Majumdar,
\newblock Nucl.Phys. {\bf B199}, 36 (1982)\relax
\mciteBstWouldAddEndPuncttrue
\mciteSetBstMidEndSepPunct{\mcitedefaultmidpunct}
{\mcitedefaultendpunct}{\mcitedefaultseppunct}\relax
\EndOfBibitem
\bibitem{Ellis:1990wk}
J.~R. Ellis, S.~Kelley, and D.~V. Nanopoulos,
\newblock Phys.Lett. {\bf B260}, 131 (1991)\relax
\mciteBstWouldAddEndPuncttrue
\mciteSetBstMidEndSepPunct{\mcitedefaultmidpunct}
{\mcitedefaultendpunct}{\mcitedefaultseppunct}\relax
\EndOfBibitem
\bibitem{Amaldi:1991cn}
U.~Amaldi, W.~de~Boer, and H.~Furstenau,
\newblock Phys.Lett. {\bf B260}, 447 (1991)\relax
\mciteBstWouldAddEndPuncttrue
\mciteSetBstMidEndSepPunct{\mcitedefaultmidpunct}
{\mcitedefaultendpunct}{\mcitedefaultseppunct}\relax
\EndOfBibitem
\bibitem{Langacker:1991an}
P.~Langacker and M.-x. Luo,
\newblock Phys.Rev. {\bf D44}, 817 (1991)\relax
\mciteBstWouldAddEndPuncttrue
\mciteSetBstMidEndSepPunct{\mcitedefaultmidpunct}
{\mcitedefaultendpunct}{\mcitedefaultseppunct}\relax
\EndOfBibitem
\bibitem{Giunti:1991ta}
C.~Giunti, C.~Kim, and U.~Lee,
\newblock Mod.Phys.Lett. {\bf A6}, 1745 (1991)\relax
\mciteBstWouldAddEndPuncttrue
\mciteSetBstMidEndSepPunct{\mcitedefaultmidpunct}
{\mcitedefaultendpunct}{\mcitedefaultseppunct}\relax
\EndOfBibitem
\bibitem{Goldberg:1983nd}
H.~Goldberg,
\newblock Phys.Rev.Lett. {\bf 50}, 1419 (1983)\relax
\mciteBstWouldAddEndPuncttrue
\mciteSetBstMidEndSepPunct{\mcitedefaultmidpunct}
{\mcitedefaultendpunct}{\mcitedefaultseppunct}\relax
\EndOfBibitem
\bibitem{Ellis:1983ew}
J.~R. Ellis, J.~Hagelin, D.~V. Nanopoulos, K.~A. Olive, and M.~Srednicki,
\newblock Nucl.Phys. {\bf B238}, 453 (1984)\relax
\mciteBstWouldAddEndPuncttrue
\mciteSetBstMidEndSepPunct{\mcitedefaultmidpunct}
{\mcitedefaultendpunct}{\mcitedefaultseppunct}\relax
\EndOfBibitem
\bibitem{dmrev}
G.~Jungman, M.~Kamionkowski, and K.~Griest,
\newblock Phys.Rept. {\bf 267}, 195 (1996), hep-ph/9506380\relax
\mciteBstWouldAddEndPuncttrue
\mciteSetBstMidEndSepPunct{\mcitedefaultmidpunct}
{\mcitedefaultendpunct}{\mcitedefaultseppunct}\relax
\EndOfBibitem
\bibitem{Ibanez:1982fr}
L.~E. Ibanez and G.~G. Ross,
\newblock Phys.Lett. {\bf B110}, 215 (1982)\relax
\mciteBstWouldAddEndPuncttrue
\mciteSetBstMidEndSepPunct{\mcitedefaultmidpunct}
{\mcitedefaultendpunct}{\mcitedefaultseppunct}\relax
\EndOfBibitem
\bibitem{Ibanez:1982ee}
L.~E. Ibanez,
\newblock Phys.Lett. {\bf B118}, 73 (1982)\relax
\mciteBstWouldAddEndPuncttrue
\mciteSetBstMidEndSepPunct{\mcitedefaultmidpunct}
{\mcitedefaultendpunct}{\mcitedefaultseppunct}\relax
\EndOfBibitem
\bibitem{Ellis:1982wr}
J.~R. Ellis, D.~V. Nanopoulos, and K.~Tamvakis,
\newblock Phys.Lett. {\bf B121}, 123 (1983)\relax
\mciteBstWouldAddEndPuncttrue
\mciteSetBstMidEndSepPunct{\mcitedefaultmidpunct}
{\mcitedefaultendpunct}{\mcitedefaultseppunct}\relax
\EndOfBibitem
\bibitem{AlvarezGaume:1983gj}
L.~Alvarez-Gaume, J.~Polchinski, and M.~B. Wise,
\newblock Nucl.Phys. {\bf B221}, 495 (1983)\relax
\mciteBstWouldAddEndPuncttrue
\mciteSetBstMidEndSepPunct{\mcitedefaultmidpunct}
{\mcitedefaultendpunct}{\mcitedefaultseppunct}\relax
\EndOfBibitem
\bibitem{Chamseddine:1982jx}
A.~H. Chamseddine, R.~L. Arnowitt, and P.~Nath,
\newblock Phys.Rev.Lett. {\bf 49}, 970 (1982)\relax
\mciteBstWouldAddEndPuncttrue
\mciteSetBstMidEndSepPunct{\mcitedefaultmidpunct}
{\mcitedefaultendpunct}{\mcitedefaultseppunct}\relax
\EndOfBibitem
\bibitem{Barbieri:1982eh}
R.~Barbieri, S.~Ferrara, and C.~A. Savoy,
\newblock Phys.Lett. {\bf B119}, 343 (1982)\relax
\mciteBstWouldAddEndPuncttrue
\mciteSetBstMidEndSepPunct{\mcitedefaultmidpunct}
{\mcitedefaultendpunct}{\mcitedefaultseppunct}\relax
\EndOfBibitem
\bibitem{Hall:1983iz}
L.~J. Hall, J.~D. Lykken, and S.~Weinberg,
\newblock Phys.Rev. {\bf D27}, 2359 (1983)\relax
\mciteBstWouldAddEndPuncttrue
\mciteSetBstMidEndSepPunct{\mcitedefaultmidpunct}
{\mcitedefaultendpunct}{\mcitedefaultseppunct}\relax
\EndOfBibitem
\bibitem{Cremmer:1982vy}
E.~Cremmer, P.~Fayet, and L.~Girardello,
\newblock Phys.Lett. {\bf B122}, 41 (1983)\relax
\mciteBstWouldAddEndPuncttrue
\mciteSetBstMidEndSepPunct{\mcitedefaultmidpunct}
{\mcitedefaultendpunct}{\mcitedefaultseppunct}\relax
\EndOfBibitem
\bibitem{Ohta:1982wn}
N.~Ohta,
\newblock Prog.Theor.Phys. {\bf 70}, 542 (1983)\relax
\mciteBstWouldAddEndPuncttrue
\mciteSetBstMidEndSepPunct{\mcitedefaultmidpunct}
{\mcitedefaultendpunct}{\mcitedefaultseppunct}\relax
\EndOfBibitem
\bibitem{Cremmer:1983bf}
E.~Cremmer, S.~Ferrara, C.~Kounnas, and D.~V. Nanopoulos,
\newblock Phys.Lett. {\bf B133}, 61 (1983)\relax
\mciteBstWouldAddEndPuncttrue
\mciteSetBstMidEndSepPunct{\mcitedefaultmidpunct}
{\mcitedefaultendpunct}{\mcitedefaultseppunct}\relax
\EndOfBibitem
\bibitem{Ellis:1983sf}
J.~R. Ellis, A.~Lahanas, D.~V. Nanopoulos, and K.~Tamvakis,
\newblock Phys.Lett. {\bf B134}, 429 (1984)\relax
\mciteBstWouldAddEndPuncttrue
\mciteSetBstMidEndSepPunct{\mcitedefaultmidpunct}
{\mcitedefaultendpunct}{\mcitedefaultseppunct}\relax
\EndOfBibitem
\bibitem{Ellis:1983ei}
J.~R. Ellis, C.~Kounnas, and D.~V. Nanopoulos,
\newblock Nucl.Phys. {\bf B241}, 406 (1984)\relax
\mciteBstWouldAddEndPuncttrue
\mciteSetBstMidEndSepPunct{\mcitedefaultmidpunct}
{\mcitedefaultendpunct}{\mcitedefaultseppunct}\relax
\EndOfBibitem
\bibitem{noscalerev}
A.~Lahanas and D.~V. Nanopoulos,
\newblock Phys.Rept. {\bf 145}, 1 (1987)\relax
\mciteBstWouldAddEndPuncttrue
\mciteSetBstMidEndSepPunct{\mcitedefaultmidpunct}
{\mcitedefaultendpunct}{\mcitedefaultseppunct}\relax
\EndOfBibitem
\bibitem{vacplanck}
S.~Ferrara, C.~Kounnas, and F.~Zwirner,
\newblock Nucl.Phys. {\bf B429}, 589 (1994), hep-th/9405188\relax
\mciteBstWouldAddEndPuncttrue
\mciteSetBstMidEndSepPunct{\mcitedefaultmidpunct}
{\mcitedefaultendpunct}{\mcitedefaultseppunct}\relax
\EndOfBibitem
\bibitem{Dutta:2007xr}
B.~Dutta, Y.~Mimura, and D.~V. Nanopoulos,
\newblock Phys.Lett. {\bf B656}, 199 (2007), 0705.4317\relax
\mciteBstWouldAddEndPuncttrue
\mciteSetBstMidEndSepPunct{\mcitedefaultmidpunct}
{\mcitedefaultendpunct}{\mcitedefaultseppunct}\relax
\EndOfBibitem
\bibitem{Maxin:2008kp}
J.~A. Maxin, V.~E. Mayes, and D.~V. Nanopoulos,
\newblock Phys.Rev. {\bf D79}, 066010 (2009), 0809.3200\relax
\mciteBstWouldAddEndPuncttrue
\mciteSetBstMidEndSepPunct{\mcitedefaultmidpunct}
{\mcitedefaultendpunct}{\mcitedefaultseppunct}\relax
\EndOfBibitem
\bibitem{Ellis:2010jb}
J.~Ellis, A.~Mustafayev, and K.~A. Olive,
\newblock Eur.Phys.J. {\bf C69}, 219 (2010), 1004.5399\relax
\mciteBstWouldAddEndPuncttrue
\mciteSetBstMidEndSepPunct{\mcitedefaultmidpunct}
{\mcitedefaultendpunct}{\mcitedefaultseppunct}\relax
\EndOfBibitem
\bibitem{Li:2010mi}
T.~Li, J.~A. Maxin, D.~V. Nanopoulos, and J.~W. Walker,
\newblock Phys. Lett. {\bf B699}, 164 (2011), 1009.2981\relax
\mciteBstWouldAddEndPuncttrue
\mciteSetBstMidEndSepPunct{\mcitedefaultmidpunct}
{\mcitedefaultendpunct}{\mcitedefaultseppunct}\relax
\EndOfBibitem
\bibitem{Li:2011xu}
T.~Li, J.~A. Maxin, D.~V. Nanopoulos, and J.~W. Walker,
\newblock (2011), 1105.3988\relax
\mciteBstWouldAddEndPuncttrue
\mciteSetBstMidEndSepPunct{\mcitedefaultmidpunct}
{\mcitedefaultendpunct}{\mcitedefaultseppunct}\relax
\EndOfBibitem
\bibitem{Li:2010ws}
T.~Li, J.~A. Maxin, D.~V. Nanopoulos, and J.~W. Walker,
\newblock Phys. Rev. {\bf D83}, 056015 (2011), 1007.5100\relax
\mciteBstWouldAddEndPuncttrue
\mciteSetBstMidEndSepPunct{\mcitedefaultmidpunct}
{\mcitedefaultendpunct}{\mcitedefaultseppunct}\relax
\EndOfBibitem
\bibitem{Li:2011dw}
T.~Li, J.~A. Maxin, D.~V. Nanopoulos, and J.~W. Walker,
\newblock (2011), 1101.2197\relax
\mciteBstWouldAddEndPuncttrue
\mciteSetBstMidEndSepPunct{\mcitedefaultmidpunct}
{\mcitedefaultendpunct}{\mcitedefaultseppunct}\relax
\EndOfBibitem
\bibitem{Kounnas:1994fr}
C.~Kounnas, F.~Zwirner, and I.~Pavel,
\newblock Phys.Lett. {\bf B335}, 403 (1994), hep-ph/9406256\relax
\mciteBstWouldAddEndPuncttrue
\mciteSetBstMidEndSepPunct{\mcitedefaultmidpunct}
{\mcitedefaultendpunct}{\mcitedefaultseppunct}\relax
\EndOfBibitem
\bibitem{Kelley:1994ab}
S.~Kelley, J.~L. Lopez, D.~V. Nanopoulos, and A.~Zichichi,
\newblock (1994), hep-ph/9409223\relax
\mciteBstWouldAddEndPuncttrue
\mciteSetBstMidEndSepPunct{\mcitedefaultmidpunct}
{\mcitedefaultendpunct}{\mcitedefaultseppunct}\relax
\EndOfBibitem
\bibitem{nsvac2}
G.~Leontaris and N.~Tracas,
\newblock Phys.Lett. {\bf B351}, 487 (1995), hep-ph/9502246\relax
\mciteBstWouldAddEndPuncttrue
\mciteSetBstMidEndSepPunct{\mcitedefaultmidpunct}
{\mcitedefaultendpunct}{\mcitedefaultseppunct}\relax
\EndOfBibitem
\bibitem{Ellis:1984kd}
J.~R. Ellis, K.~Enqvist, and D.~V. Nanopoulos,
\newblock Phys.Lett. {\bf B147}, 99 (1984)\relax
\mciteBstWouldAddEndPuncttrue
\mciteSetBstMidEndSepPunct{\mcitedefaultmidpunct}
{\mcitedefaultendpunct}{\mcitedefaultseppunct}\relax
\EndOfBibitem
\bibitem{Ellis:1984bs}
J.~R. Ellis, C.~Kounnas, and D.~V. Nanopoulos,
\newblock Phys.Lett. {\bf B143}, 410 (1984)\relax
\mciteBstWouldAddEndPuncttrue
\mciteSetBstMidEndSepPunct{\mcitedefaultmidpunct}
{\mcitedefaultendpunct}{\mcitedefaultseppunct}\relax
\EndOfBibitem
\bibitem{Ellis:1984bm}
J.~R. Ellis, C.~Kounnas, and D.~V. Nanopoulos,
\newblock Nucl.Phys. {\bf B247}, 373 (1984)\relax
\mciteBstWouldAddEndPuncttrue
\mciteSetBstMidEndSepPunct{\mcitedefaultmidpunct}
{\mcitedefaultendpunct}{\mcitedefaultseppunct}\relax
\EndOfBibitem
\bibitem{Brignole:1993dj}
A.~Brignole, L.~E. Ibanez, and C.~Munoz,
\newblock Nucl.Phys. {\bf B422}, 125 (1994), hep-ph/9308271\relax
\mciteBstWouldAddEndPuncttrue
\mciteSetBstMidEndSepPunct{\mcitedefaultmidpunct}
{\mcitedefaultendpunct}{\mcitedefaultseppunct}\relax
\EndOfBibitem
\bibitem{Munoz:1995bi}
C.~Munoz,
\newblock (1995), hep-ph/9509290\relax
\mciteBstWouldAddEndPuncttrue
\mciteSetBstMidEndSepPunct{\mcitedefaultmidpunct}
{\mcitedefaultendpunct}{\mcitedefaultseppunct}\relax
\EndOfBibitem
\bibitem{Giudice:1988yz}
G.~Giudice and A.~Masiero,
\newblock Phys.Lett. {\bf B206}, 480 (1988)\relax
\mciteBstWouldAddEndPuncttrue
\mciteSetBstMidEndSepPunct{\mcitedefaultmidpunct}
{\mcitedefaultendpunct}{\mcitedefaultseppunct}\relax
\EndOfBibitem
\bibitem{Soni:1983rm}
S.~K. Soni and H.~Weldon,
\newblock Phys.Lett. {\bf B126}, 215 (1983)\relax
\mciteBstWouldAddEndPuncttrue
\mciteSetBstMidEndSepPunct{\mcitedefaultmidpunct}
{\mcitedefaultendpunct}{\mcitedefaultseppunct}\relax
\EndOfBibitem
\bibitem{Bagger:1994hh}
J.~Bagger, E.~Poppitz, and L.~Randall,
\newblock Nucl.Phys. {\bf B426}, 3 (1994), hep-ph/9405345\relax
\mciteBstWouldAddEndPuncttrue
\mciteSetBstMidEndSepPunct{\mcitedefaultmidpunct}
{\mcitedefaultendpunct}{\mcitedefaultseppunct}\relax
\EndOfBibitem
\bibitem{Munoz:1995yp}
C.~Munoz,
\newblock (1995), hep-th/9507108\relax
\mciteBstWouldAddEndPuncttrue
\mciteSetBstMidEndSepPunct{\mcitedefaultmidpunct}
{\mcitedefaultendpunct}{\mcitedefaultseppunct}\relax
\EndOfBibitem
\bibitem{LeMouel:1997tk}
C.~Le~Mouel and G.~Moultaka,
\newblock Nucl.Phys. {\bf B518}, 3 (1998), hep-ph/9711356\relax
\mciteBstWouldAddEndPuncttrue
\mciteSetBstMidEndSepPunct{\mcitedefaultmidpunct}
{\mcitedefaultendpunct}{\mcitedefaultseppunct}\relax
\EndOfBibitem
\bibitem{Kastening:1991gv}
B.~M. Kastening,
\newblock Phys.Lett. {\bf B283}, 287 (1992)\relax
\mciteBstWouldAddEndPuncttrue
\mciteSetBstMidEndSepPunct{\mcitedefaultmidpunct}
{\mcitedefaultendpunct}{\mcitedefaultseppunct}\relax
\EndOfBibitem
\bibitem{Bando:1992wz}
M.~Bando, T.~Kugo, N.~Maekawa, and H.~Nakano,
\newblock Phys.Lett. {\bf B301}, 83 (1993), hep-ph/9210228\relax
\mciteBstWouldAddEndPuncttrue
\mciteSetBstMidEndSepPunct{\mcitedefaultmidpunct}
{\mcitedefaultendpunct}{\mcitedefaultseppunct}\relax
\EndOfBibitem
\bibitem{Bando:1992wy}
M.~Bando, T.~Kugo, N.~Maekawa, and H.~Nakano,
\newblock Prog.Theor.Phys. {\bf 90}, 405 (1993), hep-ph/9210229\relax
\mciteBstWouldAddEndPuncttrue
\mciteSetBstMidEndSepPunct{\mcitedefaultmidpunct}
{\mcitedefaultendpunct}{\mcitedefaultseppunct}\relax
\EndOfBibitem
\bibitem{rginv2}
C.~Ford, D.~Jones, P.~Stephenson, and M.~Einhorn,
\newblock Nucl.Phys. {\bf B395}, 17 (1993), hep-lat/9210033\relax
\mciteBstWouldAddEndPuncttrue
\mciteSetBstMidEndSepPunct{\mcitedefaultmidpunct}
{\mcitedefaultendpunct}{\mcitedefaultseppunct}\relax
\EndOfBibitem
\bibitem{tad}
M.~Sher,
\newblock Phys.Rept. {\bf 179}, 273 (1989)\relax
\mciteBstWouldAddEndPuncttrue
\mciteSetBstMidEndSepPunct{\mcitedefaultmidpunct}
{\mcitedefaultendpunct}{\mcitedefaultseppunct}\relax
\EndOfBibitem
\bibitem{CWV}
S.~R. Coleman and E.~J. Weinberg,
\newblock Phys.Rev. {\bf D7}, 1888 (1973)\relax
\mciteBstWouldAddEndPuncttrue
\mciteSetBstMidEndSepPunct{\mcitedefaultmidpunct}
{\mcitedefaultendpunct}{\mcitedefaultseppunct}\relax
\EndOfBibitem
\bibitem{V2Martin}
S.~P. Martin,
\newblock Phys.Rev. {\bf D66}, 096001 (2002), hep-ph/0206136\relax
\mciteBstWouldAddEndPuncttrue
\mciteSetBstMidEndSepPunct{\mcitedefaultmidpunct}
{\mcitedefaultendpunct}{\mcitedefaultseppunct}\relax
\EndOfBibitem
\bibitem{Martin:2001vx}
S.~P. Martin,
\newblock Phys.Rev. {\bf D65}, 116003 (2002), hep-ph/0111209\relax
\mciteBstWouldAddEndPuncttrue
\mciteSetBstMidEndSepPunct{\mcitedefaultmidpunct}
{\mcitedefaultendpunct}{\mcitedefaultseppunct}\relax
\EndOfBibitem
\bibitem{rginv0}
M.~Einhorn and D.~Jones,
\newblock Nucl.Phys. {\bf B211}, 29 (1983)\relax
\mciteBstWouldAddEndPuncttrue
\mciteSetBstMidEndSepPunct{\mcitedefaultmidpunct}
{\mcitedefaultendpunct}{\mcitedefaultseppunct}\relax
\EndOfBibitem
\bibitem{rginv3}
S.~Kelley, J.~L. Lopez, D.~V. Nanopoulos, H.~Pois, and K.-j. Yuan,
\newblock Nucl.Phys. {\bf B398}, 3 (1993), hep-ph/9206218\relax
\mciteBstWouldAddEndPuncttrue
\mciteSetBstMidEndSepPunct{\mcitedefaultmidpunct}
{\mcitedefaultendpunct}{\mcitedefaultseppunct}\relax
\EndOfBibitem
\bibitem{suspect}
A.~Djouadi, J.-L. Kneur, and G.~Moultaka,
\newblock Comput.Phys.Commun. {\bf 176}, 426 (2007), hep-ph/0211331\relax
\mciteBstWouldAddEndPuncttrue
\mciteSetBstMidEndSepPunct{\mcitedefaultmidpunct}
{\mcitedefaultendpunct}{\mcitedefaultseppunct}\relax
\EndOfBibitem
\bibitem{isajet}
F.~E. Paige, S.~D. Protopopescu, H.~Baer, and X.~Tata,
\newblock (2003), hep-ph/0312045\relax
\mciteBstWouldAddEndPuncttrue
\mciteSetBstMidEndSepPunct{\mcitedefaultmidpunct}
{\mcitedefaultendpunct}{\mcitedefaultseppunct}\relax
\EndOfBibitem
\bibitem{softsusy}
B.~Allanach,
\newblock Comput.Phys.Commun. {\bf 143}, 305 (2002), hep-ph/0104145\relax
\mciteBstWouldAddEndPuncttrue
\mciteSetBstMidEndSepPunct{\mcitedefaultmidpunct}
{\mcitedefaultendpunct}{\mcitedefaultseppunct}\relax
\EndOfBibitem
\bibitem{spheno}
W.~Porod,
\newblock Comput.Phys.Commun. {\bf 153}, 275 (2003), hep-ph/0301101\relax
\mciteBstWouldAddEndPuncttrue
\mciteSetBstMidEndSepPunct{\mcitedefaultmidpunct}
{\mcitedefaultendpunct}{\mcitedefaultseppunct}\relax
\EndOfBibitem
\bibitem{Zhang:1998bm}
R.-J. Zhang,
\newblock Phys.Lett. {\bf B447}, 89 (1999), hep-ph/9808299\relax
\mciteBstWouldAddEndPuncttrue
\mciteSetBstMidEndSepPunct{\mcitedefaultmidpunct}
{\mcitedefaultendpunct}{\mcitedefaultseppunct}\relax
\EndOfBibitem
\bibitem{Espinosa:1999zm}
J.~R. Espinosa and R.-J. Zhang,
\newblock JHEP {\bf 0003}, 026 (2000), hep-ph/9912236\relax
\mciteBstWouldAddEndPuncttrue
\mciteSetBstMidEndSepPunct{\mcitedefaultmidpunct}
{\mcitedefaultendpunct}{\mcitedefaultseppunct}\relax
\EndOfBibitem
\bibitem{Nakamura:2010zzi}
Particle Data Group, K.~Nakamura {\em et~al.},
\newblock J.Phys.G {\bf G37}, 075021 (2010)\relax
\mciteBstWouldAddEndPuncttrue
\mciteSetBstMidEndSepPunct{\mcitedefaultmidpunct}
{\mcitedefaultendpunct}{\mcitedefaultseppunct}\relax
\EndOfBibitem
\bibitem{Khachatryan:2011tk}
CMS Collaboration, V.~Khachatryan {\em et~al.},
\newblock Phys.Lett. {\bf B698}, 196 (2011), 1101.1628\relax
\mciteBstWouldAddEndPuncttrue
\mciteSetBstMidEndSepPunct{\mcitedefaultmidpunct}
{\mcitedefaultendpunct}{\mcitedefaultseppunct}\relax
\EndOfBibitem
\bibitem{Chatrchyan:2011bz}
CMS Collaboration, S.~Chatrchyan {\em et~al.},
\newblock (2011), 1103.1348\relax
\mciteBstWouldAddEndPuncttrue
\mciteSetBstMidEndSepPunct{\mcitedefaultmidpunct}
{\mcitedefaultendpunct}{\mcitedefaultseppunct}\relax
\EndOfBibitem
\bibitem{Collaboration:2011bj}
CMS Collaboration, C.~Collaboration,
\newblock (2011), 1106.3272\relax
\mciteBstWouldAddEndPuncttrue
\mciteSetBstMidEndSepPunct{\mcitedefaultmidpunct}
{\mcitedefaultendpunct}{\mcitedefaultseppunct}\relax
\EndOfBibitem
\bibitem{Aad:2011xm}
ATLAS Collaboration, G.~Aad {\em et~al.},
\newblock (2011), 1103.6214\relax
\mciteBstWouldAddEndPuncttrue
\mciteSetBstMidEndSepPunct{\mcitedefaultmidpunct}
{\mcitedefaultendpunct}{\mcitedefaultseppunct}\relax
\EndOfBibitem
\bibitem{Aad:2011hh}
Atlas Collaboration, G.~Aad {\em et~al.},
\newblock Phys.Rev.Lett. {\bf 106}, 131802 (2011), 1102.2357\relax
\mciteBstWouldAddEndPuncttrue
\mciteSetBstMidEndSepPunct{\mcitedefaultmidpunct}
{\mcitedefaultendpunct}{\mcitedefaultseppunct}\relax
\EndOfBibitem
\bibitem{Aad:2011ks}
ATLAS, G.~Aad {\em et~al.},
\newblock (2011), 1103.4344\relax
\mciteBstWouldAddEndPuncttrue
\mciteSetBstMidEndSepPunct{\mcitedefaultmidpunct}
{\mcitedefaultendpunct}{\mcitedefaultseppunct}\relax
\EndOfBibitem
\bibitem{Pagels:1981ke}
H.~Pagels and J.~R. Primack,
\newblock Phys.Rev.Lett. {\bf 48}, 223 (1982)\relax
\mciteBstWouldAddEndPuncttrue
\mciteSetBstMidEndSepPunct{\mcitedefaultmidpunct}
{\mcitedefaultendpunct}{\mcitedefaultseppunct}\relax
\EndOfBibitem
\bibitem{Weinberg:1982zq}
S.~Weinberg,
\newblock Phys.Rev.Lett. {\bf 48}, 1303 (1982)\relax
\mciteBstWouldAddEndPuncttrue
\mciteSetBstMidEndSepPunct{\mcitedefaultmidpunct}
{\mcitedefaultendpunct}{\mcitedefaultseppunct}\relax
\EndOfBibitem
\bibitem{Nanopoulos:1983up}
D.~V. Nanopoulos, K.~A. Olive, and M.~Srednicki,
\newblock Phys.Lett. {\bf B127}, 30 (1983)\relax
\mciteBstWouldAddEndPuncttrue
\mciteSetBstMidEndSepPunct{\mcitedefaultmidpunct}
{\mcitedefaultendpunct}{\mcitedefaultseppunct}\relax
\EndOfBibitem
\bibitem{Khlopov:1984pf}
M.~Khlopov and A.~D. Linde,
\newblock Phys.Lett. {\bf B138}, 265 (1984)\relax
\mciteBstWouldAddEndPuncttrue
\mciteSetBstMidEndSepPunct{\mcitedefaultmidpunct}
{\mcitedefaultendpunct}{\mcitedefaultseppunct}\relax
\EndOfBibitem
\bibitem{Ellis:1984eq}
J.~R. Ellis, J.~E. Kim, and D.~V. Nanopoulos,
\newblock Phys.Lett. {\bf B145}, 181 (1984)\relax
\mciteBstWouldAddEndPuncttrue
\mciteSetBstMidEndSepPunct{\mcitedefaultmidpunct}
{\mcitedefaultendpunct}{\mcitedefaultseppunct}\relax
\EndOfBibitem
\bibitem{Juszkiewicz:1985gg}
R.~Juszkiewicz, J.~Silk, and A.~Stebbins,
\newblock Phys.Lett. {\bf B158}, 463 (1985)\relax
\mciteBstWouldAddEndPuncttrue
\mciteSetBstMidEndSepPunct{\mcitedefaultmidpunct}
{\mcitedefaultendpunct}{\mcitedefaultseppunct}\relax
\EndOfBibitem
\bibitem{Ellis:1984er}
J.~R. Ellis, D.~V. Nanopoulos, and S.~Sarkar,
\newblock Nucl.Phys. {\bf B259}, 175 (1985)\relax
\mciteBstWouldAddEndPuncttrue
\mciteSetBstMidEndSepPunct{\mcitedefaultmidpunct}
{\mcitedefaultendpunct}{\mcitedefaultseppunct}\relax
\EndOfBibitem
\bibitem{Kawasaki:1986my}
M.~Kawasaki and K.~Sato,
\newblock Phys.Lett. {\bf B189}, 23 (1987)\relax
\mciteBstWouldAddEndPuncttrue
\mciteSetBstMidEndSepPunct{\mcitedefaultmidpunct}
{\mcitedefaultendpunct}{\mcitedefaultseppunct}\relax
\EndOfBibitem
\bibitem{Berezinsky:1991kf}
V.~Berezinsky,
\newblock Phys.Lett. {\bf B261}, 71 (1991)\relax
\mciteBstWouldAddEndPuncttrue
\mciteSetBstMidEndSepPunct{\mcitedefaultmidpunct}
{\mcitedefaultendpunct}{\mcitedefaultseppunct}\relax
\EndOfBibitem
\bibitem{Moroi:1993mb}
T.~Moroi, H.~Murayama, and M.~Yamaguchi,
\newblock Phys.Lett. {\bf B303}, 289 (1993)\relax
\mciteBstWouldAddEndPuncttrue
\mciteSetBstMidEndSepPunct{\mcitedefaultmidpunct}
{\mcitedefaultendpunct}{\mcitedefaultseppunct}\relax
\EndOfBibitem
\bibitem{Drees:1992am}
M.~Drees and M.~M. Nojiri,
\newblock Phys.Rev. {\bf D47}, 376 (1993), hep-ph/9207234\relax
\mciteBstWouldAddEndPuncttrue
\mciteSetBstMidEndSepPunct{\mcitedefaultmidpunct}
{\mcitedefaultendpunct}{\mcitedefaultseppunct}\relax
\EndOfBibitem
\bibitem{Baer:1995nc}
H.~Baer and M.~Brhlik,
\newblock Phys.Rev. {\bf D53}, 597 (1996), hep-ph/9508321\relax
\mciteBstWouldAddEndPuncttrue
\mciteSetBstMidEndSepPunct{\mcitedefaultmidpunct}
{\mcitedefaultendpunct}{\mcitedefaultseppunct}\relax
\EndOfBibitem
\bibitem{Ellis:1997wva}
J.~R. Ellis, T.~Falk, K.~A. Olive, and M.~Schmitt,
\newblock Phys.Lett. {\bf B413}, 355 (1997), hep-ph/9705444\relax
\mciteBstWouldAddEndPuncttrue
\mciteSetBstMidEndSepPunct{\mcitedefaultmidpunct}
{\mcitedefaultendpunct}{\mcitedefaultseppunct}\relax
\EndOfBibitem
\bibitem{Ellis:1998kh}
J.~R. Ellis, T.~Falk, and K.~A. Olive,
\newblock Phys.Lett. {\bf B444}, 367 (1998), hep-ph/9810360\relax
\mciteBstWouldAddEndPuncttrue
\mciteSetBstMidEndSepPunct{\mcitedefaultmidpunct}
{\mcitedefaultendpunct}{\mcitedefaultseppunct}\relax
\EndOfBibitem
\bibitem{Djouadi:2001yk}
A.~Djouadi, M.~Drees, and J.~Kneur,
\newblock JHEP {\bf 0108}, 055 (2001), hep-ph/0107316\relax
\mciteBstWouldAddEndPuncttrue
\mciteSetBstMidEndSepPunct{\mcitedefaultmidpunct}
{\mcitedefaultendpunct}{\mcitedefaultseppunct}\relax
\EndOfBibitem
\bibitem{Baer:2002gm}
H.~Baer {\em et~al.},
\newblock JHEP {\bf 0207}, 050 (2002), hep-ph/0205325\relax
\mciteBstWouldAddEndPuncttrue
\mciteSetBstMidEndSepPunct{\mcitedefaultmidpunct}
{\mcitedefaultendpunct}{\mcitedefaultseppunct}\relax
\EndOfBibitem
\bibitem{Baer:2003yh}
H.~Baer and C.~Balazs,
\newblock JCAP {\bf 0305}, 006 (2003), hep-ph/0303114\relax
\mciteBstWouldAddEndPuncttrue
\mciteSetBstMidEndSepPunct{\mcitedefaultmidpunct}
{\mcitedefaultendpunct}{\mcitedefaultseppunct}\relax
\EndOfBibitem
\bibitem{Chattopadhyay:2003xi}
U.~Chattopadhyay, A.~Corsetti, and P.~Nath,
\newblock Phys.Rev. {\bf D68}, 035005 (2003), hep-ph/0303201\relax
\mciteBstWouldAddEndPuncttrue
\mciteSetBstMidEndSepPunct{\mcitedefaultmidpunct}
{\mcitedefaultendpunct}{\mcitedefaultseppunct}\relax
\EndOfBibitem
\bibitem{Ellis:2003cw}
J.~R. Ellis, K.~A. Olive, Y.~Santoso, and V.~C. Spanos,
\newblock Phys.Lett. {\bf B565}, 176 (2003), hep-ph/0303043\relax
\mciteBstWouldAddEndPuncttrue
\mciteSetBstMidEndSepPunct{\mcitedefaultmidpunct}
{\mcitedefaultendpunct}{\mcitedefaultseppunct}\relax
\EndOfBibitem
\bibitem{Battaglia:2003ab}
M.~Battaglia {\em et~al.},
\newblock Eur.Phys.J. {\bf C33}, 273 (2004), hep-ph/0306219\relax
\mciteBstWouldAddEndPuncttrue
\mciteSetBstMidEndSepPunct{\mcitedefaultmidpunct}
{\mcitedefaultendpunct}{\mcitedefaultseppunct}\relax
\EndOfBibitem
\bibitem{Arnowitt:2003vw}
R.~L. Arnowitt, B.~Dutta, and B.~Hu,
\newblock p.~25 (2003), hep-ph/0310103\relax
\mciteBstWouldAddEndPuncttrue
\mciteSetBstMidEndSepPunct{\mcitedefaultmidpunct}
{\mcitedefaultendpunct}{\mcitedefaultseppunct}\relax
\EndOfBibitem
\bibitem{Ellis:2003si}
J.~R. Ellis, K.~A. Olive, Y.~Santoso, and V.~C. Spanos,
\newblock Phys.Rev. {\bf D69}, 095004 (2004), hep-ph/0310356\relax
\mciteBstWouldAddEndPuncttrue
\mciteSetBstMidEndSepPunct{\mcitedefaultmidpunct}
{\mcitedefaultendpunct}{\mcitedefaultseppunct}\relax
\EndOfBibitem
\bibitem{Gomez:2004eka}
M.~E. Gomez, T.~Ibrahim, P.~Nath, and S.~Skadhauge,
\newblock Phys.Rev. {\bf D70}, 035014 (2004), hep-ph/0404025\relax
\mciteBstWouldAddEndPuncttrue
\mciteSetBstMidEndSepPunct{\mcitedefaultmidpunct}
{\mcitedefaultendpunct}{\mcitedefaultseppunct}\relax
\EndOfBibitem
\bibitem{Ellis:2004tc}
J.~R. Ellis, S.~Heinemeyer, K.~A. Olive, and G.~Weiglein,
\newblock JHEP {\bf 0502}, 013 (2005), hep-ph/0411216\relax
\mciteBstWouldAddEndPuncttrue
\mciteSetBstMidEndSepPunct{\mcitedefaultmidpunct}
{\mcitedefaultendpunct}{\mcitedefaultseppunct}\relax
\EndOfBibitem
\bibitem{Belanger:2005jk}
G.~Belanger, S.~Kraml, and A.~Pukhov,
\newblock Phys.Rev. {\bf D72}, 015003 (2005), hep-ph/0502079\relax
\mciteBstWouldAddEndPuncttrue
\mciteSetBstMidEndSepPunct{\mcitedefaultmidpunct}
{\mcitedefaultendpunct}{\mcitedefaultseppunct}\relax
\EndOfBibitem
\bibitem{Djouadi:2006be}
A.~Djouadi, M.~Drees, and J.-L. Kneur,
\newblock JHEP {\bf 0603}, 033 (2006), hep-ph/0602001\relax
\mciteBstWouldAddEndPuncttrue
\mciteSetBstMidEndSepPunct{\mcitedefaultmidpunct}
{\mcitedefaultendpunct}{\mcitedefaultseppunct}\relax
\EndOfBibitem
\bibitem{D0lim}
D0 Collaboration, V.~Abazov {\em et~al.},
\newblock Phys.Lett. {\bf B680}, 34 (2009), 0901.0646\relax
\mciteBstWouldAddEndPuncttrue
\mciteSetBstMidEndSepPunct{\mcitedefaultmidpunct}
{\mcitedefaultendpunct}{\mcitedefaultseppunct}\relax
\EndOfBibitem
\bibitem{ATLAS-CONF-2011-086}
ATLAS Collaboration,
\newblock (2011),
\newblock ATLAS-CONF-2011-086\relax
\mciteBstWouldAddEndPuncttrue
\mciteSetBstMidEndSepPunct{\mcitedefaultmidpunct}
{\mcitedefaultendpunct}{\mcitedefaultseppunct}\relax
\EndOfBibitem
\bibitem{ATLAS_stau_NLSP}
ATLAS Collaboration,
\newblock (2011), 1106.4495,
\newblock * Temporary entry *\relax
\mciteBstWouldAddEndPuncttrue
\mciteSetBstMidEndSepPunct{\mcitedefaultmidpunct}
{\mcitedefaultendpunct}{\mcitedefaultseppunct}\relax
\EndOfBibitem
\bibitem{LEPsusy}
ALEPH, DELPHI, L3, OPAL,
\newblock Notes LEPSUSYWG/01-03.1 and 04-01.1;
  http://lepsusy.web.cern.ch/lepsusy/\relax
\mciteBstWouldAddEndPuncttrue
\mciteSetBstMidEndSepPunct{\mcitedefaultmidpunct}
{\mcitedefaultendpunct}{\mcitedefaultseppunct}\relax
\EndOfBibitem
\bibitem{Lopez:1993vi}
J.~L. Lopez, D.~V. Nanopoulos, and X.~Wang,
\newblock Phys.Rev. {\bf D49}, 366 (1994), hep-ph/9308336\relax
\mciteBstWouldAddEndPuncttrue
\mciteSetBstMidEndSepPunct{\mcitedefaultmidpunct}
{\mcitedefaultendpunct}{\mcitedefaultseppunct}\relax
\EndOfBibitem
\bibitem{Chattopadhyay:1995ae}
U.~Chattopadhyay and P.~Nath,
\newblock Phys.Rev. {\bf D53}, 1648 (1996), hep-ph/9507386\relax
\mciteBstWouldAddEndPuncttrue
\mciteSetBstMidEndSepPunct{\mcitedefaultmidpunct}
{\mcitedefaultendpunct}{\mcitedefaultseppunct}\relax
\EndOfBibitem
\bibitem{Moroi:1995yh}
T.~Moroi,
\newblock Phys.Rev. {\bf D53}, 6565 (1996), hep-ph/9512396\relax
\mciteBstWouldAddEndPuncttrue
\mciteSetBstMidEndSepPunct{\mcitedefaultmidpunct}
{\mcitedefaultendpunct}{\mcitedefaultseppunct}\relax
\EndOfBibitem
\bibitem{Carena:1996qa}
M.~S. Carena, G.~Giudice, and C.~Wagner,
\newblock Phys.Lett. {\bf B390}, 234 (1997), hep-ph/9610233\relax
\mciteBstWouldAddEndPuncttrue
\mciteSetBstMidEndSepPunct{\mcitedefaultmidpunct}
{\mcitedefaultendpunct}{\mcitedefaultseppunct}\relax
\EndOfBibitem
\bibitem{Goto:1999mk}
T.~Goto, Y.~Okada, and Y.~Shimizu,
\newblock Physical Review D  (1999), hep-ph/9908499\relax
\mciteBstWouldAddEndPuncttrue
\mciteSetBstMidEndSepPunct{\mcitedefaultmidpunct}
{\mcitedefaultendpunct}{\mcitedefaultseppunct}\relax
\EndOfBibitem
\bibitem{Drees:2000bs}
M.~Drees, Y.~G. Kim, T.~Kobayashi, and M.~M. Nojiri,
\newblock Phys.Rev. {\bf D63}, 115009 (2001), hep-ph/0011359\relax
\mciteBstWouldAddEndPuncttrue
\mciteSetBstMidEndSepPunct{\mcitedefaultmidpunct}
{\mcitedefaultendpunct}{\mcitedefaultseppunct}\relax
\EndOfBibitem
\bibitem{Bennett:2006fi}
Muon G-2 Collaboration, G.~Bennett {\em et~al.},
\newblock Phys.Rev. {\bf D73}, 072003 (2006), hep-ex/0602035\relax
\mciteBstWouldAddEndPuncttrue
\mciteSetBstMidEndSepPunct{\mcitedefaultmidpunct}
{\mcitedefaultendpunct}{\mcitedefaultseppunct}\relax
\EndOfBibitem
\bibitem{g-2_th_rev}
J.~P. Miller, E.~de~Rafael, and B.~Roberts,
\newblock Rept.Prog.Phys. {\bf 70}, 795 (2007), hep-ph/0703049\relax
\mciteBstWouldAddEndPuncttrue
\mciteSetBstMidEndSepPunct{\mcitedefaultmidpunct}
{\mcitedefaultendpunct}{\mcitedefaultseppunct}\relax
\EndOfBibitem
\bibitem{bsg_sm_nlo}
M.~Misiak {\em et~al.},
\newblock Phys.Rev.Lett. {\bf 98}, 022002 (2007), hep-ph/0609232\relax
\mciteBstWouldAddEndPuncttrue
\mciteSetBstMidEndSepPunct{\mcitedefaultmidpunct}
{\mcitedefaultendpunct}{\mcitedefaultseppunct}\relax
\EndOfBibitem
\bibitem{bsg_sm_np}
M.~Benzke, S.~J. Lee, M.~Neubert, and G.~Paz,
\newblock JHEP {\bf 08}, 099 (2010), 1003.5012\relax
\mciteBstWouldAddEndPuncttrue
\mciteSetBstMidEndSepPunct{\mcitedefaultmidpunct}
{\mcitedefaultendpunct}{\mcitedefaultseppunct}\relax
\EndOfBibitem
\bibitem{bsg_exp}
Heavy Flavor Averaging Group, D.~Asner {\em et~al.},
\newblock (2010), 1010.1589\relax
\mciteBstWouldAddEndPuncttrue
\mciteSetBstMidEndSepPunct{\mcitedefaultmidpunct}
{\mcitedefaultendpunct}{\mcitedefaultseppunct}\relax
\EndOfBibitem
\bibitem{Degrassi:2000qf}
G.~Degrassi, P.~Gambino, and G.~F. Giudice,
\newblock JHEP {\bf 12}, 009 (2000), hep-ph/0009337\relax
\mciteBstWouldAddEndPuncttrue
\mciteSetBstMidEndSepPunct{\mcitedefaultmidpunct}
{\mcitedefaultendpunct}{\mcitedefaultseppunct}\relax
\EndOfBibitem
\bibitem{bsg_susy_nlo}
G.~Degrassi, P.~Gambino, and P.~Slavich,
\newblock Phys. Lett. {\bf B635}, 335 (2006), hep-ph/0601135\relax
\mciteBstWouldAddEndPuncttrue
\mciteSetBstMidEndSepPunct{\mcitedefaultmidpunct}
{\mcitedefaultendpunct}{\mcitedefaultseppunct}\relax
\EndOfBibitem
\bibitem{bsg_susy_nlo2}
G.~Degrassi, P.~Gambino, and P.~Slavich,
\newblock Comput. Phys. Commun. {\bf 179}, 759 (2008), 0712.3265\relax
\mciteBstWouldAddEndPuncttrue
\mciteSetBstMidEndSepPunct{\mcitedefaultmidpunct}
{\mcitedefaultendpunct}{\mcitedefaultseppunct}\relax
\EndOfBibitem
\bibitem{bsg_susy_nlo3}
C.~Greub, T.~Hurth, V.~Pilipp, C.~Schupbach, and M.~Steinhauser,
\newblock (2011), 1105.1330\relax
\mciteBstWouldAddEndPuncttrue
\mciteSetBstMidEndSepPunct{\mcitedefaultmidpunct}
{\mcitedefaultendpunct}{\mcitedefaultseppunct}\relax
\EndOfBibitem
\bibitem{Ambrosanio:2000ik}
S.~Ambrosanio, B.~Mele, S.~Petrarca, G.~Polesello, and A.~Rimoldi,
\newblock JHEP {\bf 01}, 014 (2001), hep-ph/0010081\relax
\mciteBstWouldAddEndPuncttrue
\mciteSetBstMidEndSepPunct{\mcitedefaultmidpunct}
{\mcitedefaultendpunct}{\mcitedefaultseppunct}\relax
\EndOfBibitem
\bibitem{Baltz:2001rq}
E.~A. Baltz and H.~Murayama,
\newblock JHEP {\bf 05}, 067 (2003), astro-ph/0108172\relax
\mciteBstWouldAddEndPuncttrue
\mciteSetBstMidEndSepPunct{\mcitedefaultmidpunct}
{\mcitedefaultendpunct}{\mcitedefaultseppunct}\relax
\EndOfBibitem
\bibitem{Fujii:2002fv}
M.~Fujii and T.~Yanagida,
\newblock Phys. Lett. {\bf B549}, 273 (2002), hep-ph/0208191\relax
\mciteBstWouldAddEndPuncttrue
\mciteSetBstMidEndSepPunct{\mcitedefaultmidpunct}
{\mcitedefaultendpunct}{\mcitedefaultseppunct}\relax
\EndOfBibitem
\bibitem{Kawagoe:2003jv}
K.~Kawagoe, T.~Kobayashi, M.~M. Nojiri, and A.~Ochi,
\newblock Phys. Rev. {\bf D69}, 035003 (2004), hep-ph/0309031\relax
\mciteBstWouldAddEndPuncttrue
\mciteSetBstMidEndSepPunct{\mcitedefaultmidpunct}
{\mcitedefaultendpunct}{\mcitedefaultseppunct}\relax
\EndOfBibitem
\bibitem{Lemoine:2005hu}
M.~Lemoine, G.~Moultaka, and K.~Jedamzik,
\newblock Phys. Lett. {\bf B645}, 222 (2007), hep-ph/0504021\relax
\mciteBstWouldAddEndPuncttrue
\mciteSetBstMidEndSepPunct{\mcitedefaultmidpunct}
{\mcitedefaultendpunct}{\mcitedefaultseppunct}\relax
\EndOfBibitem
\bibitem{Jedamzik:2005ir}
K.~Jedamzik, M.~Lemoine, and G.~Moultaka,
\newblock Phys. Rev. {\bf D73}, 043514 (2006), hep-ph/0506129\relax
\mciteBstWouldAddEndPuncttrue
\mciteSetBstMidEndSepPunct{\mcitedefaultmidpunct}
{\mcitedefaultendpunct}{\mcitedefaultseppunct}\relax
\EndOfBibitem
\bibitem{Bailly:2008yy}
S.~Bailly, K.~Jedamzik, and G.~Moultaka,
\newblock Phys. Rev. {\bf D80}, 063509 (2009), 0812.0788\relax
\mciteBstWouldAddEndPuncttrue
\mciteSetBstMidEndSepPunct{\mcitedefaultmidpunct}
{\mcitedefaultendpunct}{\mcitedefaultseppunct}\relax
\EndOfBibitem
\bibitem{Feng:2003xh}
J.~L. Feng, A.~Rajaraman, and F.~Takayama,
\newblock Phys. Rev. Lett. {\bf 91}, 011302 (2003), hep-ph/0302215\relax
\mciteBstWouldAddEndPuncttrue
\mciteSetBstMidEndSepPunct{\mcitedefaultmidpunct}
{\mcitedefaultendpunct}{\mcitedefaultseppunct}\relax
\EndOfBibitem
\bibitem{Feng:2003uy}
J.~L. Feng, A.~Rajaraman, and F.~Takayama,
\newblock Phys. Rev. {\bf D68}, 063504 (2003), hep-ph/0306024\relax
\mciteBstWouldAddEndPuncttrue
\mciteSetBstMidEndSepPunct{\mcitedefaultmidpunct}
{\mcitedefaultendpunct}{\mcitedefaultseppunct}\relax
\EndOfBibitem
\bibitem{Ellis:2003dn}
J.~R. Ellis, K.~A. Olive, Y.~Santoso, and V.~C. Spanos,
\newblock Phys. Lett. {\bf B588}, 7 (2004), hep-ph/0312262\relax
\mciteBstWouldAddEndPuncttrue
\mciteSetBstMidEndSepPunct{\mcitedefaultmidpunct}
{\mcitedefaultendpunct}{\mcitedefaultseppunct}\relax
\EndOfBibitem
\bibitem{Feng:2004zu}
J.~L. Feng, S.~Su, and F.~Takayama,
\newblock Phys.Rev.D {\bf 70}, 063514 (2004), hep-ph/0404198\relax
\mciteBstWouldAddEndPuncttrue
\mciteSetBstMidEndSepPunct{\mcitedefaultmidpunct}
{\mcitedefaultendpunct}{\mcitedefaultseppunct}\relax
\EndOfBibitem
\bibitem{Feng:2004mt}
J.~L. Feng, S.~Su, and F.~Takayama,
\newblock Phys. Rev. {\bf D70}, 075019 (2004), hep-ph/0404231\relax
\mciteBstWouldAddEndPuncttrue
\mciteSetBstMidEndSepPunct{\mcitedefaultmidpunct}
{\mcitedefaultendpunct}{\mcitedefaultseppunct}\relax
\EndOfBibitem
\bibitem{Roszkowski:2004jd}
L.~Roszkowski, R.~Ruiz~de Austri, and K.-Y. Choi,
\newblock JHEP {\bf 08}, 080 (2005), hep-ph/0408227\relax
\mciteBstWouldAddEndPuncttrue
\mciteSetBstMidEndSepPunct{\mcitedefaultmidpunct}
{\mcitedefaultendpunct}{\mcitedefaultseppunct}\relax
\EndOfBibitem
\bibitem{Cerdeno:2005eu}
D.~G. Cerdeno, K.-Y. Choi, K.~Jedamzik, L.~Roszkowski, and R.~Ruiz~de Austri,
\newblock JCAP {\bf 0606}, 005 (2006), hep-ph/0509275\relax
\mciteBstWouldAddEndPuncttrue
\mciteSetBstMidEndSepPunct{\mcitedefaultmidpunct}
{\mcitedefaultendpunct}{\mcitedefaultseppunct}\relax
\EndOfBibitem
\bibitem{Steffen:2006hw}
F.~D. Steffen,
\newblock JCAP {\bf 0609, 001} (2006), hep-ph/0605306\relax
\mciteBstWouldAddEndPuncttrue
\mciteSetBstMidEndSepPunct{\mcitedefaultmidpunct}
{\mcitedefaultendpunct}{\mcitedefaultseppunct}\relax
\EndOfBibitem
\bibitem{Kawasaki:2007xb}
M.~Kawasaki, K.~Kohri, and T.~Moroi,
\newblock Phys. Lett. {\bf B649}, 436 (2007), hep-ph/0703122\relax
\mciteBstWouldAddEndPuncttrue
\mciteSetBstMidEndSepPunct{\mcitedefaultmidpunct}
{\mcitedefaultendpunct}{\mcitedefaultseppunct}\relax
\EndOfBibitem
\bibitem{Pradler:2007is}
J.~Pradler and F.~D. Steffen,
\newblock Phys. Lett. {\bf B666}, 181 (2008), 0710.2213\relax
\mciteBstWouldAddEndPuncttrue
\mciteSetBstMidEndSepPunct{\mcitedefaultmidpunct}
{\mcitedefaultendpunct}{\mcitedefaultseppunct}\relax
\EndOfBibitem
\bibitem{Kawasaki:2008qe}
M.~Kawasaki, K.~Kohri, T.~Moroi, and A.~Yotsuyanagi,
\newblock Phys. Rev. {\bf D78}, 065011 (2008), 0804.3745\relax
\mciteBstWouldAddEndPuncttrue
\mciteSetBstMidEndSepPunct{\mcitedefaultmidpunct}
{\mcitedefaultendpunct}{\mcitedefaultseppunct}\relax
\EndOfBibitem
\bibitem{Bailly:2009pe}
S.~Bailly, K.-Y. Choi, K.~Jedamzik, and L.~Roszkowski,
\newblock JHEP {\bf 05}, 103 (2009), 0903.3974\relax
\mciteBstWouldAddEndPuncttrue
\mciteSetBstMidEndSepPunct{\mcitedefaultmidpunct}
{\mcitedefaultendpunct}{\mcitedefaultseppunct}\relax
\EndOfBibitem
\bibitem{Bailly:2010hh}
S.~Bailly,
\newblock JCAP {\bf 1103}, 022 (2011), 1008.2858\relax
\mciteBstWouldAddEndPuncttrue
\mciteSetBstMidEndSepPunct{\mcitedefaultmidpunct}
{\mcitedefaultendpunct}{\mcitedefaultseppunct}\relax
\EndOfBibitem
\bibitem{Belanger:2006is}
G.~Belanger, F.~Boudjema, A.~Pukhov, and A.~Semenov,
\newblock Comput.Phys.Commun. {\bf 176}, 367 (2007), hep-ph/0607059\relax
\mciteBstWouldAddEndPuncttrue
\mciteSetBstMidEndSepPunct{\mcitedefaultmidpunct}
{\mcitedefaultendpunct}{\mcitedefaultseppunct}\relax
\EndOfBibitem
\bibitem{Bolz:2000fu}
M.~Bolz, A.~Brandenburg, and W.~Buchmuller,
\newblock Nucl.Phys. {\bf B606}, 518 (2001), hep-ph/0012052\relax
\mciteBstWouldAddEndPuncttrue
\mciteSetBstMidEndSepPunct{\mcitedefaultmidpunct}
{\mcitedefaultendpunct}{\mcitedefaultseppunct}\relax
\EndOfBibitem
\bibitem{Pradler:2006qh}
J.~Pradler and F.~D. Steffen,
\newblock Phys.Rev. {\bf D75}, 023509 (2007), hep-ph/0608344\relax
\mciteBstWouldAddEndPuncttrue
\mciteSetBstMidEndSepPunct{\mcitedefaultmidpunct}
{\mcitedefaultendpunct}{\mcitedefaultseppunct}\relax
\EndOfBibitem
\bibitem{Pradler:2006hh}
J.~Pradler and F.~D. Steffen,
\newblock Phys.Lett. {\bf B648}, 224 (2007), hep-ph/0612291\relax
\mciteBstWouldAddEndPuncttrue
\mciteSetBstMidEndSepPunct{\mcitedefaultmidpunct}
{\mcitedefaultendpunct}{\mcitedefaultseppunct}\relax
\EndOfBibitem
\bibitem{Rychkov:2007uq}
V.~S. Rychkov and A.~Strumia,
\newblock Phys.Rev. {\bf D75}, 075011 (2007), hep-ph/0701104\relax
\mciteBstWouldAddEndPuncttrue
\mciteSetBstMidEndSepPunct{\mcitedefaultmidpunct}
{\mcitedefaultendpunct}{\mcitedefaultseppunct}\relax
\EndOfBibitem
\bibitem{Spergel:2006hy}
WMAP Collaboration, D.~Spergel {\em et~al.},
\newblock Astrophys.J.Suppl. {\bf 170}, 377 (2007), astro-ph/0603449\relax
\mciteBstWouldAddEndPuncttrue
\mciteSetBstMidEndSepPunct{\mcitedefaultmidpunct}
{\mcitedefaultendpunct}{\mcitedefaultseppunct}\relax
\EndOfBibitem
\bibitem{lepgen}
S.~Davidson, E.~Nardi, and Y.~Nir,
\newblock Phys.Rept. {\bf 466}, 105 (2008), 0802.2962\relax
\mciteBstWouldAddEndPuncttrue
\mciteSetBstMidEndSepPunct{\mcitedefaultmidpunct}
{\mcitedefaultendpunct}{\mcitedefaultseppunct}\relax
\EndOfBibitem
\end{mcitethebibliography}
\bibliographystyle{h-physrev2}

\end{document}